\documentclass{article}          
\usepackage{graphicx}
\usepackage[utf8]{inputenc}
\usepackage[T1]{fontenc}
\usepackage{authblk}
\usepackage{bm}
\usepackage{import}
\usepackage{color}
\usepackage{amsmath}

\usepackage{verbatim,amsfonts,amssymb,amsthm} 
\usepackage{amsfonts,here}
\usepackage{float}
\usepackage{caption}
\usepackage{subcaption}
\usepackage{natbib}

\newcommand{\opA}{\mathop{\vphantom{\sum}\mathchoice
  {\vcenter{\hbox{\huge A}}}
  {\vcenter{\hbox{\Large A}}}{\mathrm{A}}{\mathrm{A}}}\displaylimits}

\providecommand{\keywords}[1]{\textbf{\textit{Keywords---}} #1}
\begin{document}

\title{Mixed projection- and density-based topology optimization with applications to structural assemblies}


\author[1]{Nicol\`{o} Pollini\thanks{nicolo@alumni.technion.ac.il}}
\author[2]{Oded Amir\thanks{odedamir@technion.ac.il}}
\affil[1]{\small \textit{Ramboll Group A/S, Copenhagen, Denmark}}
\affil[2]{\small \textit{Faculty of Civil and Environmental Engineering, Technion - Israel Institute of Technology, Haifa, Israel}}
\renewcommand\Authands{ and }

\date{}

\maketitle

\begin{abstract}
In this paper we present a mixed projection- and density-based topology optimization approach.
The aim is to combine the benefits of both parametrizations:
the explicit geometric representation provides specific controls on certain design regions while the implicit density representation provides the ultimate design freedom elsewhere.
This approach is particularly suited for structural assemblies, where the optimization of the structural topology is coupled with the optimization of the shape of the interface between the sub-components in a unified formulation. 
The interface between the assemblies is defined by a segmented profile made of linear geometric entities. 
The geometric coordinates of the nodes connecting the profile segments are used as shape variables in the problem, together with density variables as in conventional topology optimization.
The variable profile is used to locally impose specific geometric constraints or to project particular material properties. Examples of the properties considered herein are a local volume constraint, a local maximum length scale control, a variable Young's modulus for the distributed solid material, and 
spatially variable minimum and maximum length scale.
The resulting optimization approach is general and various geometric entities can be used.
The potential for complex design manipulations is demonstrated through several numerical examples.
\end{abstract}

\section{Introduction}
\label{sec:intro}

Topology optimization has experienced an incredible development since its introduction in the seminal paper by Bends{\o}e and Kikuchi in 1988 \citep{bendsoe1988generating} where homogenization of porous micro-structures was used as the underlying parametrization.  
In general terms, topology optimization addresses the engineering question on how to distribute material within a given domain in order to obtain the best performance of the system considered.
The method allows for a great design freedom and often leads to unexpected material layouts that are far from engineering intuition.
Among the different directions along which topology optimization has developed, the density-based \citep{bendsoe1989optimal,zhou1991coc} and level-set \citep{allaire2002level,allaire2004structural,wang2003level} parametrizations have so far received the most significant attention.
An extensive review of various aspects of density-based and homogenization-based topology optimization can be found in the monograph \cite{bendsoe2003topology}.
For recent thorough comparative reviews on topology optimization approaches, the interested reader is referred also to \cite{sigmund2013topology,deaton2014survey}.

Even though topology optimization has proved to be able to identify innovative design solutions with a high performance compared to more traditional solutions, it has been always an open question how to actually manufacture these designs.
In order to comply with the various manufacturing technologies, certain geometric limitations need to be imposed on the design outcome.   
While density-based and implicit level-set parametrizations offer significant design freedom, they also suffer from a certain drawback---the absence of direct geometrical control. 
Indeed, some geometric properties of the optimized structure can be controlled implicitly. 
Minimum length scale or thickness is the property that attracted most attention of researchers, leading to the formulation of various techniques \citep[][to mention a few]{bruns2001topology,bourdin2001filters,guest2004achieving,wang2011projection,allaire2016thickness}.
It has been shown that also maximum length scale or thickness can be controlled: in density-based procedures using filters and projections \citep{guest2009imposing,lazarov2016length,wu2018infill} and in level-set approaches using the signed distance function \citep{allaire2016thickness}.
Another more recent geometric limitation that has been addressed by various researchers is the overhang constraint in additive manufacturing \citep[e.g.][]{gaynor2016topology,langelaar2017additive,allaire2017structural,qian2017undercut}.
Despite these tremendous achievements, the common density-based and level-set approaches do not offer complete and direct control over the resulting geometry, that can be necessary for adapting the computational design procedure to certain manufacturing scenarios.

As an alternative to the density-based and level-set approaches, another class of topology optimization procedures has emerged recently that is based on explicit parametrization of the design. 
We refer to this class as ``geometric projection'' as defined in \cite{norato2015geometry} and include under this definition a variety of parametrizations that see the topological design as a collection of explicit geometric entities that are projected onto a fixed continuum grid for the purpose of finite element analysis.
Among those methods one can find the following examples:
1) Early contributions on spline-based representations projected onto fixed grids \citep{lee2004continuum,edwards2007smooth};
2) A Heaviside projection approach for optimizing the layout of discrete objects \citep{guest2015optimizing};
3) Direct geometric projection of discrete elements, in particular planar structures composed of bars \citep{norato2015geometry} and three-dimensional structures made of plates \citep{zhang2016geometry}; and
4) The Methods of Moving Morphable Components (MMC) and  Moving Morphable Voids (MMV) that define the topology by projecting explicit geometric entities such as beams and closed splines onto the continuum domain \citep{guo2014doing,zhang2017explicit}. 
From a manufacturing-oriented perspective, these methodologies allow for a more direct control over the geometry compared to traditional topology optimization approaches.
In fact, most of the procedures mentioned above are based on explicit geometric information thus they provide a direct link between topology optimization and CAD representation.
This opens possibilities for direct geometric control---some examples are overhang limitations \citep{guo2017self} and hole area and boundary curvatures \citep{yoely2018topology}.  
At the same time, traditional topology optimization approaches offer a truly free-form design parametrization that results in rich design spaces.
It is not straightforward to obtain this abundance of design possibilities using explicit design parametrizations \citep{seo2010isogeometric}.
Hence the following question is raised: Can one combine the richness of traditional topology optimization with the geometric control of projection methods? 

Thus, this paper presents a design parametrization that enables to get the best of both worlds by mixing projection- and density-based topology optimization. 
The underlying principle is based on three pillars: 
1) An explicit geometric representation provides control in particular regions or for particular design purposes, using shape variables;
2) An implicit density-based representation is used otherwise, using topological variables;
3) Projection functions are used to tie the two parts together, thus coupling the shape and topological variables. 
A particular application that benefits from such shape-topology coupling is the optimization of pre-stressed concrete beams \citep{amir2018simultaneous}.
In the current contribution, our main goal is to generalize the approach of \cite{amir2018simultaneous} and to demonstrate several (out of many) geometric controls that can be obtained. 
Nevertheless, we are motivated by a particular class of applications, namely structural assemblies---or in other words, structures that are manufactured in parts and then joined together by e.g.~hardware assembly, welding, riveting etc.
One specific example arose recently with the development of additive manufacturing (AM): 
The current sizes of printed parts are limited by the sizes of the machines.
This is particularly critical in metals, where most AM facilities can manufacture parts that are not longer than a few hundred millimeters. 
Consequently, for AM of larger structural components, the design should be divided into parts that are manufactured separately and assembled subsequently. 

The fact that the ultimate structural component is an assembly of several separate parts may impose certain limitations regarding the geometry and the material properties at the interfaces between the parts. 
Hence it is not expected that optimizing the component as a whole and subsequently defining interface regions for manufacturing as separate parts will lead to optimal performance.
Furthermore, one cannot impose limitations on interface regions without knowing their location, meaning that significant post-processing may be required.
The mixed method proposed herein aims to remove these obstacles:  
We optimize the location and shape of the interfaces between parts using an explicit geometric representation and employ standard density-based topology optimization elsewhere.
The outcomes are structures that are optimized with respect to the performance as an assembly while considering limitations at interfaces, without defining their shape and location a-priori.  
 
The remainder of the article is organized as follows: In Sec.~\ref{sec:mixed} we present the coupled parametrization that is based on density and geometric variables.
In Sec.~\ref{sec:formulation} we briefly discuss the general type of topology optimization problems considered.
Further details on how the mixed formulation is used to control specific entities and to impose limitations related to interfaces of components are provided in Sec.~\ref{sec:control}.
In Sec.~\ref{sec:sensan} we discuss the sensitivity analysis of the various constraints and in Sec.~\ref{sec:examples} we apply the proposed approach to several example applications.
Final conclusions are drawn in Sec.~\ref{sec:end}.


\section{Mixed parametrization of the design problem}
\label{sec:mixed}
In this section we present the parametrization of the topology optimization problem considered in this work.
We simultaneously optimize a fictitious density field of the distributed material assigned to each finite element, and the shape of the projection profile used to impose local controls (e.g.,~ geometric constraints) or material properties. Hence, the design variables are both fictitious density values used to interpolate the material properties between solid and void regions of the domain, and the geometric parameters that define the projection profile. In the current study, the latter are simply node positions of line segments, however many other geometric representations can be used.

\subsection{Density design variables}
We follow the popular density-based approach meaning that the material density is expressed in terms of design variables defined at the finite element level that are collected in the vector $\bm{\rho}$. That is, a density variable $\rho_{i}$ is associated to each finite element.
Therefore, we have $N_{ele}$ density design variables, where $N_{ele}$ is the number of finite elements adopted for the numerical approximation of the problem.
The density variables $\rho_{i}$ can assume $0-1$ values (i.e.~$\rho_{i}=\{0,1 \}$), defining the presence of void (i.e.~$\rho_{i}=0$) or solid material (i.e.~$\rho_{i}=1$) in the $i$-th element.
However, since for optimization we will use a gradient-based procedure all the variables need to be continuous. Hence, we relax the density variables' definition and we allow them to assume also the intermediate values between their upper and lower bounds: $0 \leq \rho_{i} \leq 1$.
In Sec. \ref{sec:formulation} we will provide further details on the methodology adopted to converge to nearly discrete optimized designs starting from a continuous problem formulation.

\subsection{Shape variables of the projection profiles}
\begin{figure}[h]
\centering
  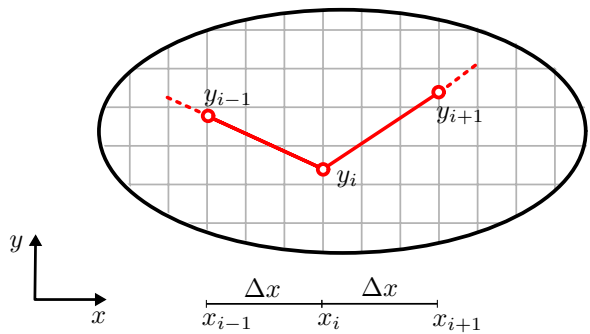
\caption{Geometric representation of a 1-D profile used for projection. In this case, the profile is developed along the horizontal direction. Thus, the $x$ coordinates of the nodes are fixed, whereas the $y$ coordinates are variable}     
\label{fig:ex_geom_var}  
\end{figure}

The projection of selected properties is performed considering one or several evolving 1-D profiles whose location in space is defined through additional geometric variables.
Each 1-D profile is a piece-wise linear geometric entity, composed of a chain of piece-wise linear segments.
The influence of each profile extends to a sub-domain identified by a strip along its development.
For example, in Fig.~\ref{fig:ex_geom_var} we show a portion of a horizontal profile made of two linear segments, and three nodes. 
In this case, we shall consider the coordinates $x_{i}$ of the nodes fixed, and the coordinates $y_{i}$ as additional variables of the problem.
In the eventuality of a profile with vertical direction of development, we shall consider the coordinates $y_{i}$ of the nodes fixed, and the coordinates $x_{i}$ as additional variables of the problems.
In this work, the geometric description of the profile used for projection is quite simple. In fact, it is made of linear segments composing a chain, and we allow for two cases: horizontal profile (i.e.~the $x_{i}$ coordinates of the profile nodes are fixed), and vertical profile (i.e.~the $y_{i}$ coordinates of the profile nodes are fixed).
Nevertheless, the approach is general and can accommodate more advanced line descriptions (such as splines) and in principle any explicit geometric entity.
Furthermore, it can be used in combination with state-of-the-art projection techniques \citep{guo2014doing,norato2015geometry}.

\subsection{Density-based parametrization}
\label{subsec:densityparam}
We consider a three-field density representation, i.e $\bm{\rho} \rightarrow \tilde{\bm{\rho}} \rightarrow \bar{\bm{\rho}}$ .
First, the density variable field $\bm{\rho}$ is regularized to avoid checkerboard patterns and mesh-dependent solutions though the well-known density filter \citep{bruns2001topology,bourdin2001filters}. The resulting filtered density field is $\tilde{\bm{\rho}}$ whose components are:
\begin{equation}\label{eq:densityfilter}
\tilde{\rho}_{i} = \frac{\sum_{j\in N_{i}} w(\Delta\textbf{x}_{ij})\rho_{j}}{\sum_{j\in N_{i}} w(\Delta\textbf{x}_{ij})}
\end{equation}
where $w(\Delta\textbf{x}_{ij})$ is a linear weight function:
\begin{equation}
w(\textbf{x}_{j}) = \max \bigl( r_{min} - \|\textbf{x}_{j}-\textbf{x}_{i}\|, 0\bigr)
\end{equation}
where $r_{min}$ is the specified filter radius, $\textbf{x}_{i}$ is the position of the centroid of the element $i$, and $\textbf{x}_{j}$ is the position of the centroid of the element $j$. If the element $j$ is in the neighborhood $N_{i}$ of the element $i$ (defined by the radius $r_{min}$) the weight $w(\textbf{x}_{j})$ has a positive value different from zero.
In a matrix form Eq.~\eqref{eq:densityfilter} can be stated as follows:
\begin{equation}
\tilde{\bm{\rho}} = \mathcal{D}\left(\textbf{H}_{s}\right)^{-1}\left(\textbf{H}\,\bm{\rho}\right)
\end{equation}
where $\textbf{H}$ is a $[N_{ele}\times N_{ele}]$ matrix, $\textbf{H}_{s}$ is a $[N_{ele}\times 1]$ vector, and $\mathcal{D}()$ is an operator that transforms a vector into a diagonal matrix and vice versa, similarly to the \texttt{diag()} MATLAB function. 
The entries of $\textbf{H}$ and $\textbf{H}_{s}$ are defined as follows:
\begin{equation}
\begin{split}
& H_{ij} = w(\Delta\textbf{x}_{ij}), \quad H_{s,i} = \sum_{j} H_{ij}
\end{split}
\end{equation}

Additionally, we project the filtered densities $\tilde{\bm{\rho}}$ into $\bar{\bm{\rho}}$ with smooth Heaviside functions to improve the convergence of the algorithm towards crisp solid-void material distributions in the final optimized designs \citep{guest2004achieving,xu2010volume}.
In particular, we introduce an eroded density vector $\bar{\bm{\rho}}^{ero}$ and a dilated density vector $\bar{\bm{\rho}}^{dil}$, in addition to the intermediate density vector $\bar{\bm{\rho}}^{int}$ \citep{sigmund2009manufacturing,wang2011projection}.
It should be noted that $\bar{\bm{\rho}}^{ero}$, $\bar{\bm{\rho}}^{int}$, and $\bar{\bm{\rho}}^{dil}$ depend explicitly on $\tilde{\bm{\rho}}$, which depends explicitly on $\bm{\rho}$.
In problems involving only stiffness and volume it is possible to consider the eroded density field 
to quantify the worst case for stiffness, and the dilated field to quantify the worst case for volume \citep{lazarov2016length}.
The above mentioned projected density vectors are defined as follows:
\begin{equation}\label{eq:rhoproject}
\begin{split}
&\bar{\rho}^{ero}_{i} = \frac{tanh\left(\beta_{HS}\eta_{ero}\right) + tanh\left(\beta_{HS}(\tilde{\rho}_{i}-\eta_{ero})\right)}{tanh\left(\beta_{HS}\eta_{ero}\right) + tanh\left(\beta_{HS}(1-\eta_{ero})\right)}, \\
&\bar{\rho}^{int}_{i} = \frac{tanh\left(\beta_{HS}\eta_{ero}\right) + tanh\left(\beta_{HS}(\tilde{\rho}_{i}-\eta_{ero})\right)}{tanh\left(\beta_{HS}\eta_{ero}\right) + tanh\left(\beta_{HS}(1-\eta_{ero})\right)}, \\
&\bar{\rho}^{dil}_{i} = \frac{tanh\left(\beta_{HS}\eta_{dil}\right) + tanh\left(\beta_{HS}(\tilde{\rho}_{i}-\eta_{dil})\right)}{tanh\left(\beta_{HS}\eta_{dil}\right) + tanh\left(\beta_{HS}(1-\eta_{dil})\right)},\\
& \text{with e.g. } \eta_{ero}=0.6,\; \eta_{int}=0.5,\; \eta_{dil}=0.4
\end{split}
\end{equation}
where $\beta_{HS}$ control the sharpness of the projection in the transition zone; and $\eta_{ero}$, $\eta_{int}$, and $\eta_{dil}$ are the projection threshold parameters.
It should be noted that the final design solution is represented by $\bar{\bm{\rho}}^{int}$, and in the current context it is used  during optimization only for updating the volume fraction for the volume constraint.
The field $\bar{\bm{\rho}}^{dil}$ is used to evaluate the volume constraint during the optimization process, and $\bar{\bm{\rho}}^{ero}$ is used to calculate the structural stiffness and to evaluate the compliance functional.
More details regarding the optimization problem formulation will be given in Sec.~\ref{sec:formulation}.

\subsection{Coupling between geometric and density variables}
\begin{figure}[h]
\centering
  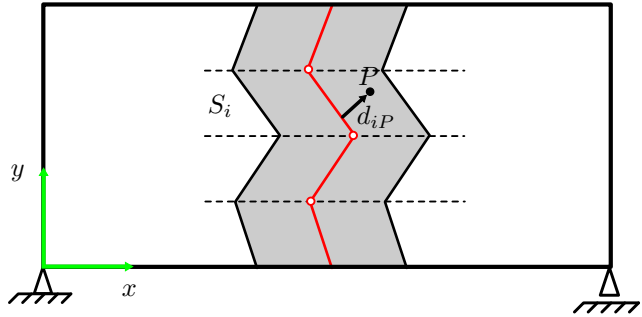
\caption{Graphic representation of the type of profiles considered. The red line represents the profile used for projection. The grey area indicates the portion of the domain affected by the profile}
\label{fig:1cut}
\end{figure}

We consider piece-wise linear profiles similar to the one shown in Fig.~\ref{fig:1cut} in red.
In this case the profile is made of four segments, and the nodes of the segments have variable coordinates.
The profile of Fig.~\ref{fig:1cut} develops along the vertical $y$ direction. The $y$ coordinates of the nodes are fixed, the $x$ coordinates of the nodes are variable.
Each point $P$ in space has a distance $d_{i,P}$ from the segment $i$ of the profile, measured perpendicularly from the segment.
We define also a maximum distance $\beta_{\phi}$, that delimits the region of influence of the profile colored in grey in Fig.~\ref{fig:1cut}.
Points whose distance is smaller than $\beta_{\phi}$ will be affected by the projection process while points outside this area will not be affected.
We consider also the case of a profile that develops along the horizontal direction $x$. In that case, the $x$ coordinates of the nodes are fixed, the $y$ coordinates of the nodes are variable. 
For a given configuration of the $i-th$ profile, and once each point $P$ has a distance parameter defined, the projection is defined through a Super-Gaussian function:
\begin{equation}
\phi_{i,P} = e^{-\frac{1}{2}\left(\frac{d^{2}_{i,P}}{\beta^{2}_{\phi}} \right)^{\mu_{\phi}}},\;\forall \, P
\end{equation}
where $\mu_{\phi}$ controls the sharpness of the Super-Gaussian projection.

\begin{figure*}[h]
    \centering
    \begin{subfigure}[t]{0.32\textwidth}
        \centering
        \includegraphics[width=1\columnwidth]{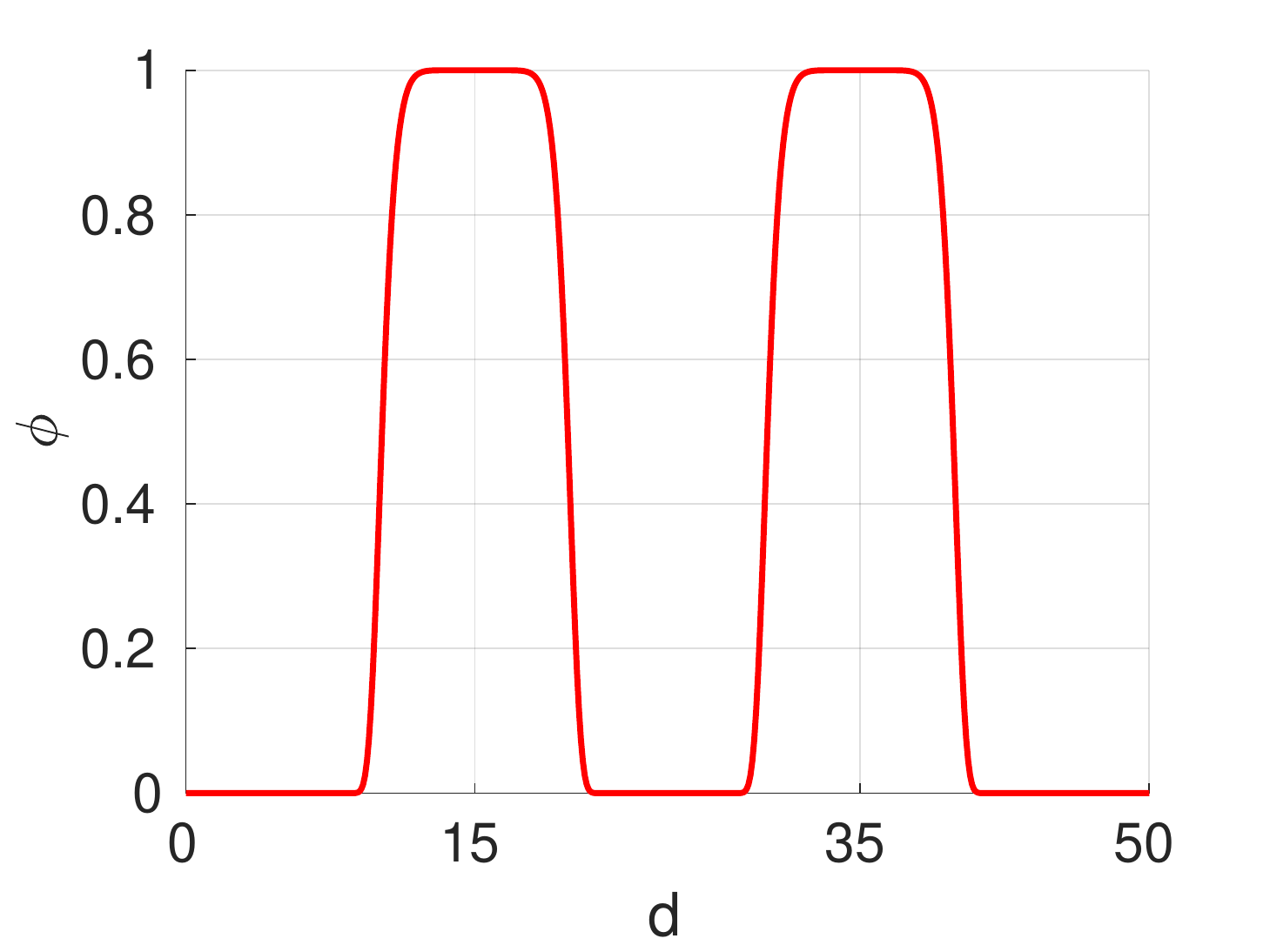}
        \caption{$x_{01}=15$ $x_{02}=35$.}
    \end{subfigure}%
    ~ 
    \begin{subfigure}[t]{0.32\textwidth}
        \centering
        \includegraphics[width=1\columnwidth]{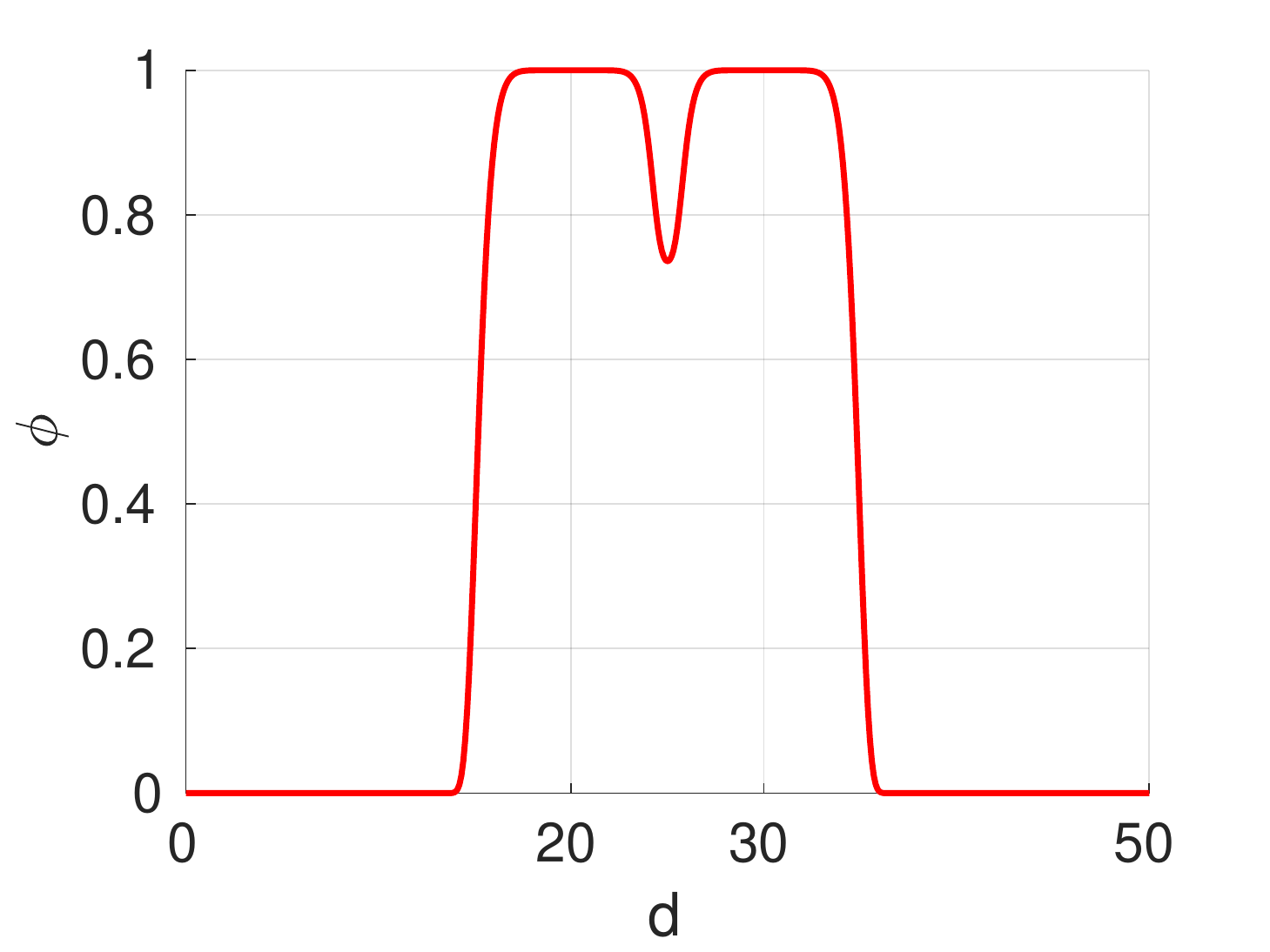}
        \caption{$x_{01}=20$ $x_{02}=30$.}
    \end{subfigure}
		~ 
    \begin{subfigure}[t]{0.32\textwidth}
        \centering
        \includegraphics[width=1\columnwidth]{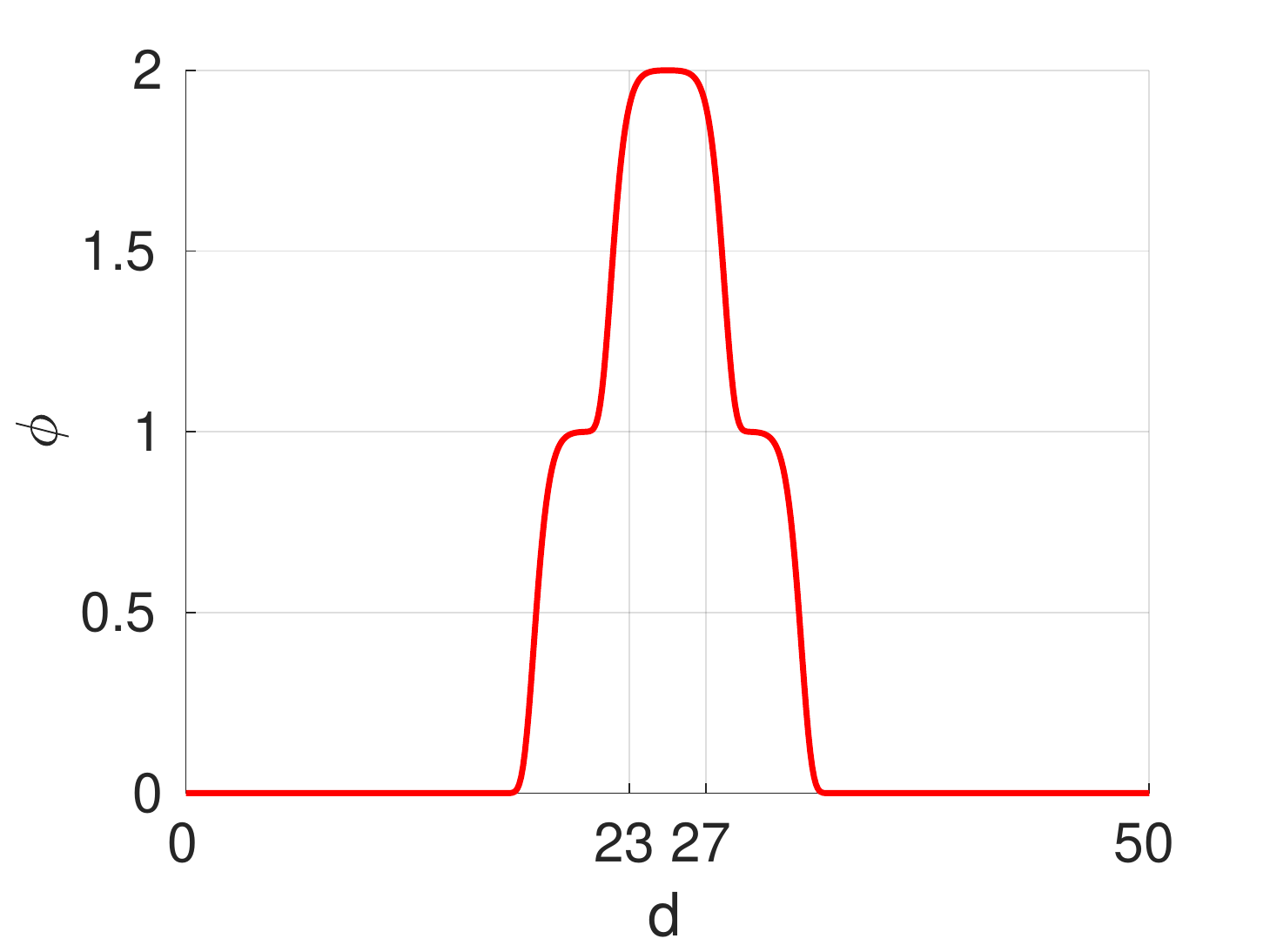}
        \caption{$x_{01}=23$ $x_{02}=27$.}
    \end{subfigure}
    \caption{Super-Gaussian projection in the case of two nearing vertical profiles. The profiles are centered in $x_{01}$ and $x_{02}$}
    \label{fig:phiSupGauss1}
\end{figure*}

In the case of two profiles $i$ and $j$ that are either parallel or perpendicular in terms of their direction of development, each point in space will have two distances with respect to two segments associated to each profile: $d_{i,P}$, $d_{j,P}$.
If we perform the projection summing up the two contributions we have the following:
\begin{equation}\label{eq:projsum}
\phi_{P} = e^{-\frac{1}{2}\left(\frac{d^{2}_{i,P}}{\beta^{2}_{\phi}}\right)^{\mu_{\phi}}}+ e^{-\frac{1}{2}\left(\frac{d^{2}_{j,P}}{\beta^{2}_{\phi}}\right)^{\mu_{\phi}}},\;\forall \, P
\end{equation}
The projection approach based on the sum of the projections of each profile in Eq.~\eqref{eq:projsum} results in the graphical behavior shown in Fig.~\ref{fig:phiSupGauss1}.
As it can be seen, when the two profiles are close to each other the points that lay in the vicinity of the two profiles are associated to projected variables $\phi$ whose values are bigger than one.
This is an undesired behavior that we wish to avoid.
We want the points close to two segments of two different profiles to have a projected variable $\phi$ at most equal to one.
The key for avoiding an overestimation of the projection, is to assign to each $i$-th point only the minimum of the two distances to the two profiles:
\begin{figure}[h]
\centering
  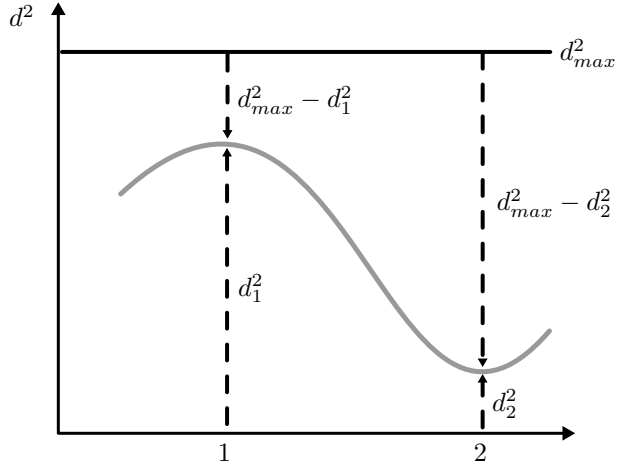
\caption{Graphic description on the procedure used to calculate the minimum of two distance measures}      
\end{figure}
\begin{equation}\label{eq:projmax}
\begin{split}
& \phi_{i} = e^{-\frac{1}{2}\left(\frac{\bar{d}^{2}_{i}}{\beta^{2}_{\phi}}\right)^{\mu_{\phi}}} \text{ for }i=1,\dots,N_{ele}\\
& \text{where } \forall\, i:\\
&\bar{d}^{2}_{i}= \min \left( [d^{2}_{i,1},\; d^{2}_{i,2}]\right)=\\
&\quad=d^{2}_{max} - \max\left( [d^{2}_{max}-d^{2}_{i,1},\; d^{2}_{max}-d^{2}_{i,2}]\right)
\end{split}
\end{equation}
The formulation of Eq.~\eqref{eq:projmax} allows to select only the minimum distance based on the maximum distance definition.
Thus, in a differentiable form Eq.~\eqref{eq:projmax} becomes:
\begin{equation}\label{eq:distprojmax}
\begin{split}
& \phi_{i} = e^{-\frac{1}{2}\left(\frac{\bar{d}^{2}_{i}}{\beta^{2}_{\phi}}\right)^{\mu_{\phi}}}\text{ for }i=1,\dots,N_{ele}\\
&\text{where } \forall\, i: \\
&\bar{d}^{2}_{i} = d^{2}_{max} - \frac{\left(d^{2}_{max}-d^{2}_{i,1}\right)^{q+1} + \left(d^{2}_{max}-d^{2}_{i,2}\right)^{q+1}}{\left(d^{2}_{max}-d^{2}_{i,1}\right)^{q} + \left(d^{2}_{max}-d^{2}_{i,2}\right)^{q}}
\end{split}
\end{equation}
In Eq.~\eqref{eq:distprojmax} $q$ is a large number, e.g. $10^6$.
\begin{figure*}[h]
    \centering
    \begin{subfigure}[t]{0.32\textwidth}
        \centering
        \includegraphics[width=1\columnwidth]{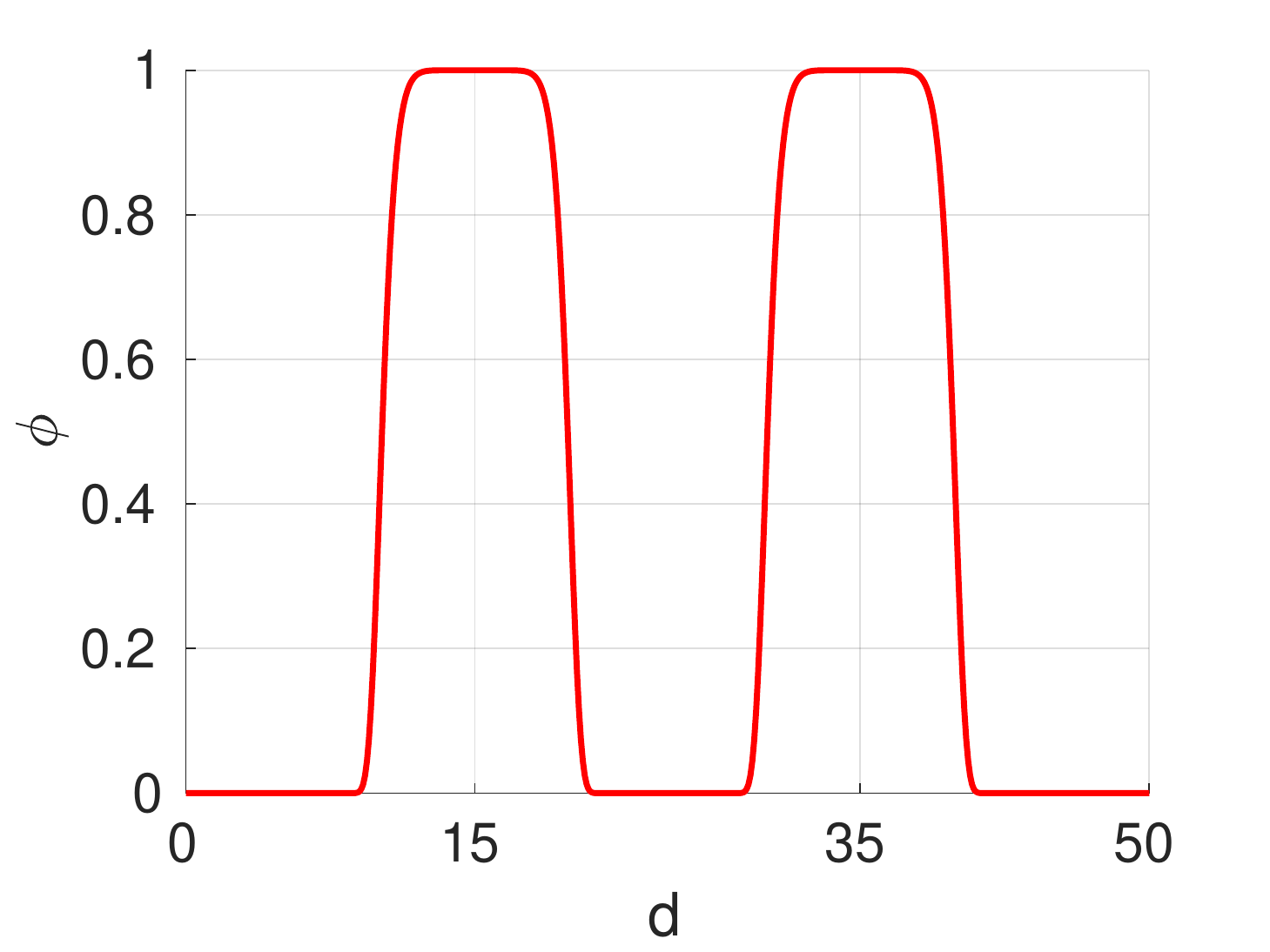}
        \caption{$x_{01}=15$ $x_{02}=35$.}
    \end{subfigure}%
    ~ 
    \begin{subfigure}[t]{0.32\textwidth}
        \centering
        \includegraphics[width=1\columnwidth]{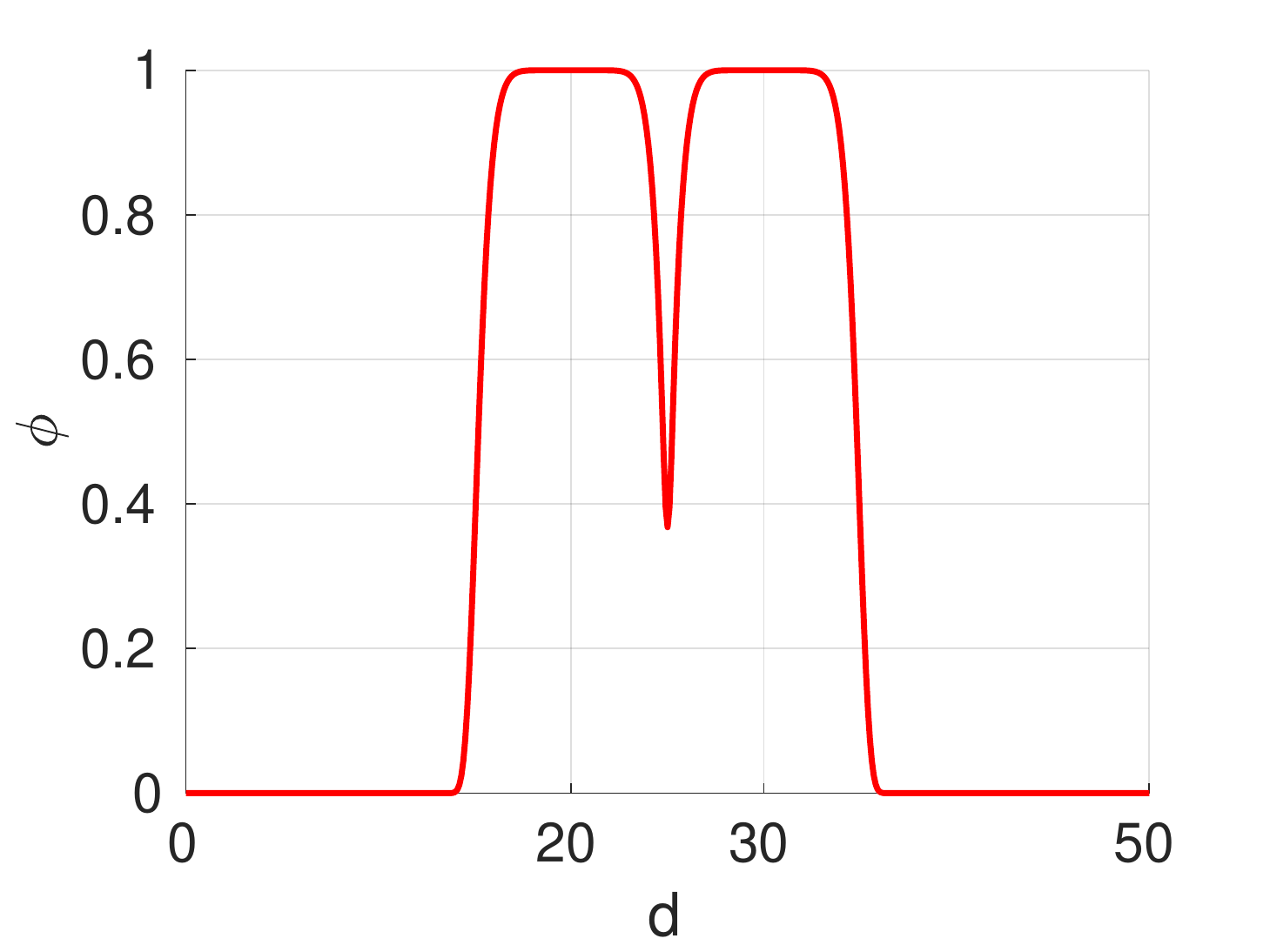}
        \caption{$x_{01}=20$ $x_{02}=30$.}
    \end{subfigure}
	~ 
    \begin{subfigure}[t]{0.32\textwidth}
        \centering
        \includegraphics[width=1\columnwidth]{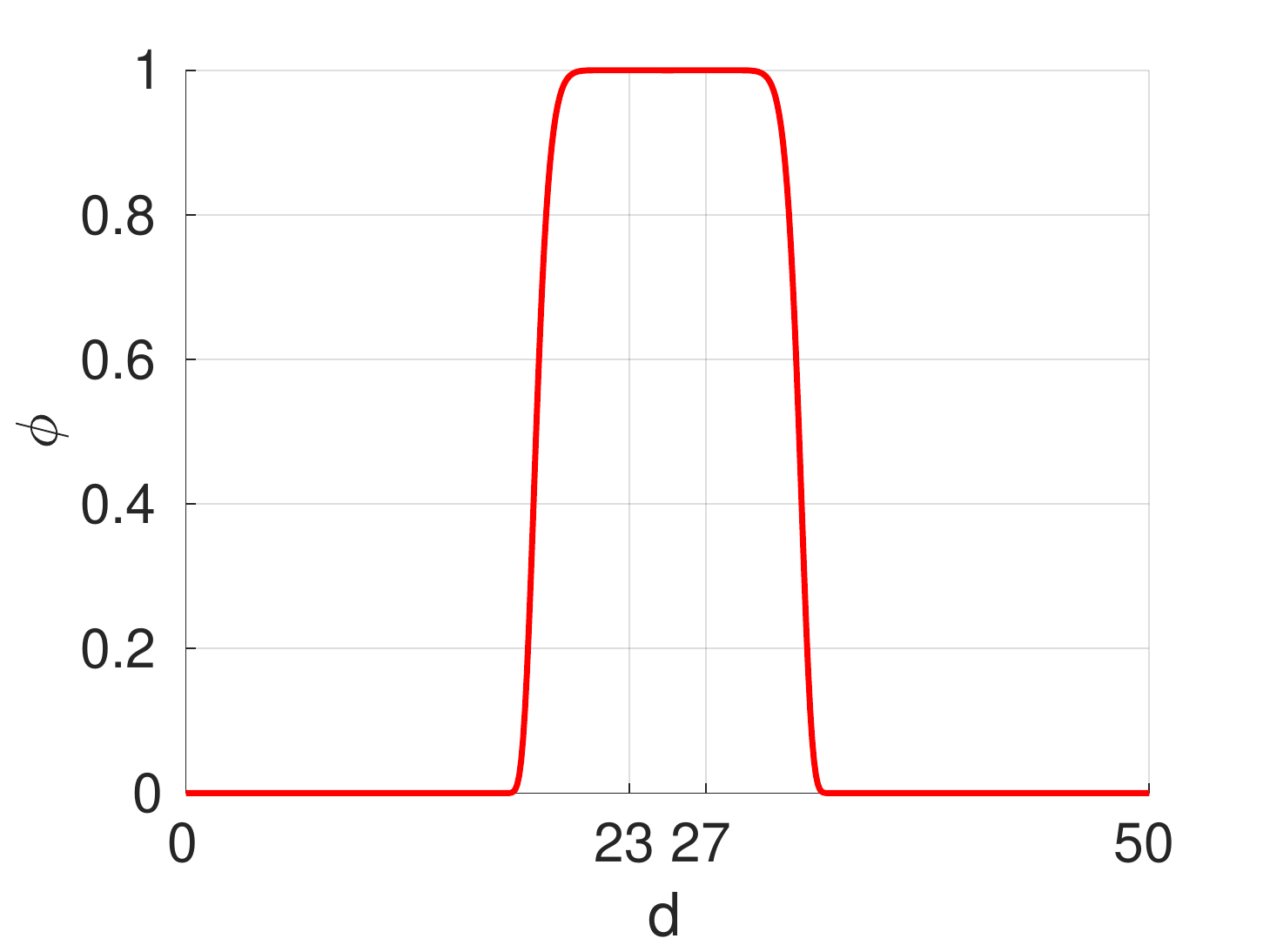}
        \caption{$x_{01}=23$ $x_{02}=27$.}
    \end{subfigure}
    \caption{Supergaussian projection in the case of two nearing vertical profiles, considering the minimum distance to the profiles for each element. The profiles are centered in $x_{01}$ and $x_{02}$}
    \label{fig:phiSupGauss2}
\end{figure*}
In Fig.~\ref{fig:phiSupGauss2} it can now be observed that when the two profiles are close to each other the points that lay in the vicinity of the two profiles are associated to projected variables $\phi$ whose values are at most equal to one.
 
\begin{figure*}[h]
    \centering
    \begin{subfigure}[t]{0.32\textwidth}
        \centering
        \includegraphics[width=1\columnwidth,trim=0.5in 0.3in 0.5in 0.30in,clip]{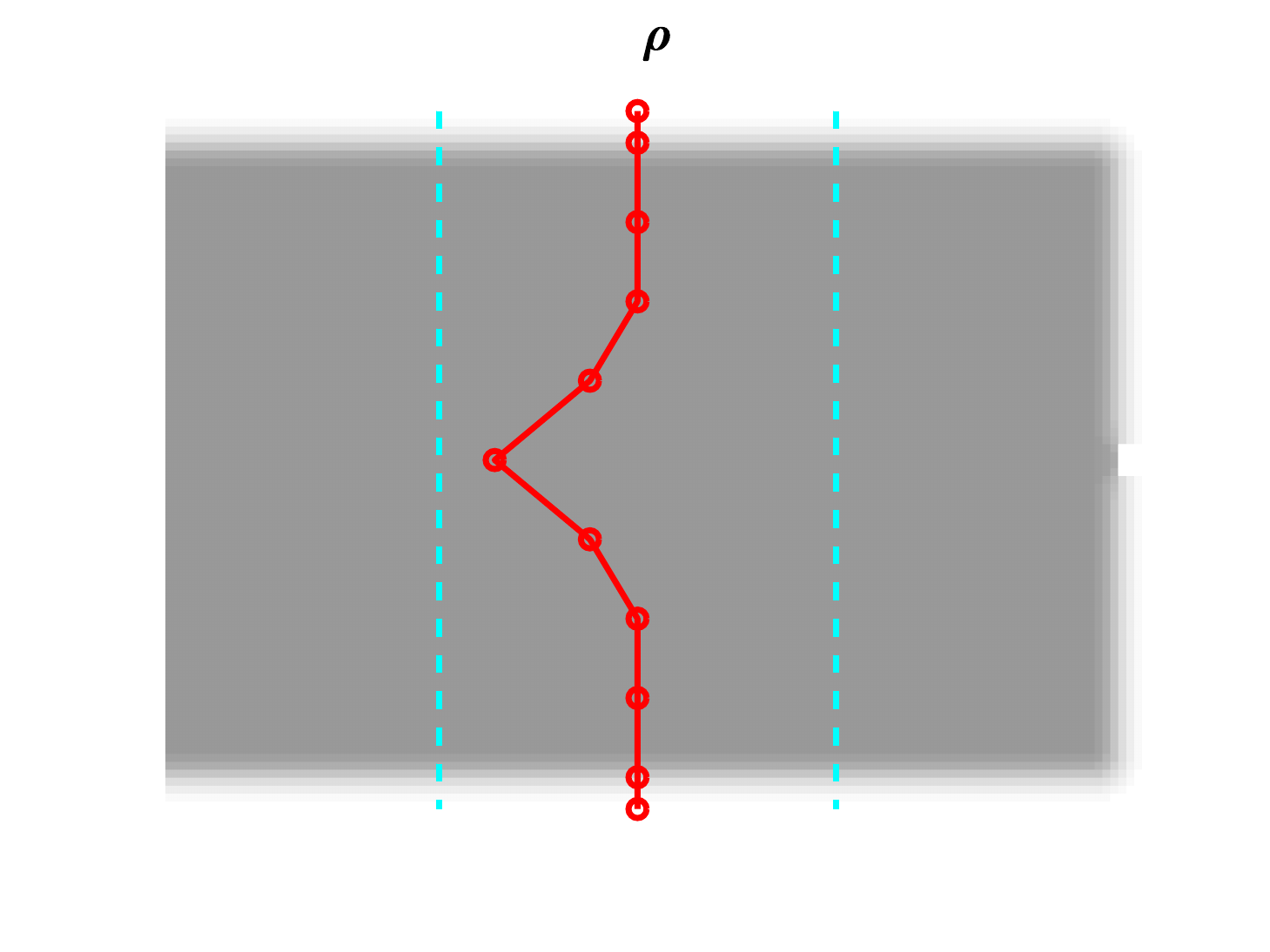}
        \caption{Graphic representation of the projection profile}
        \label{fig:phi_d_notfilt1}
    \end{subfigure}%
	~ 
    \begin{subfigure}[t]{0.32\textwidth}
        \centering
        \includegraphics[width=1\columnwidth,trim=0.5in 0.3in 0.5in 0.30in,clip]{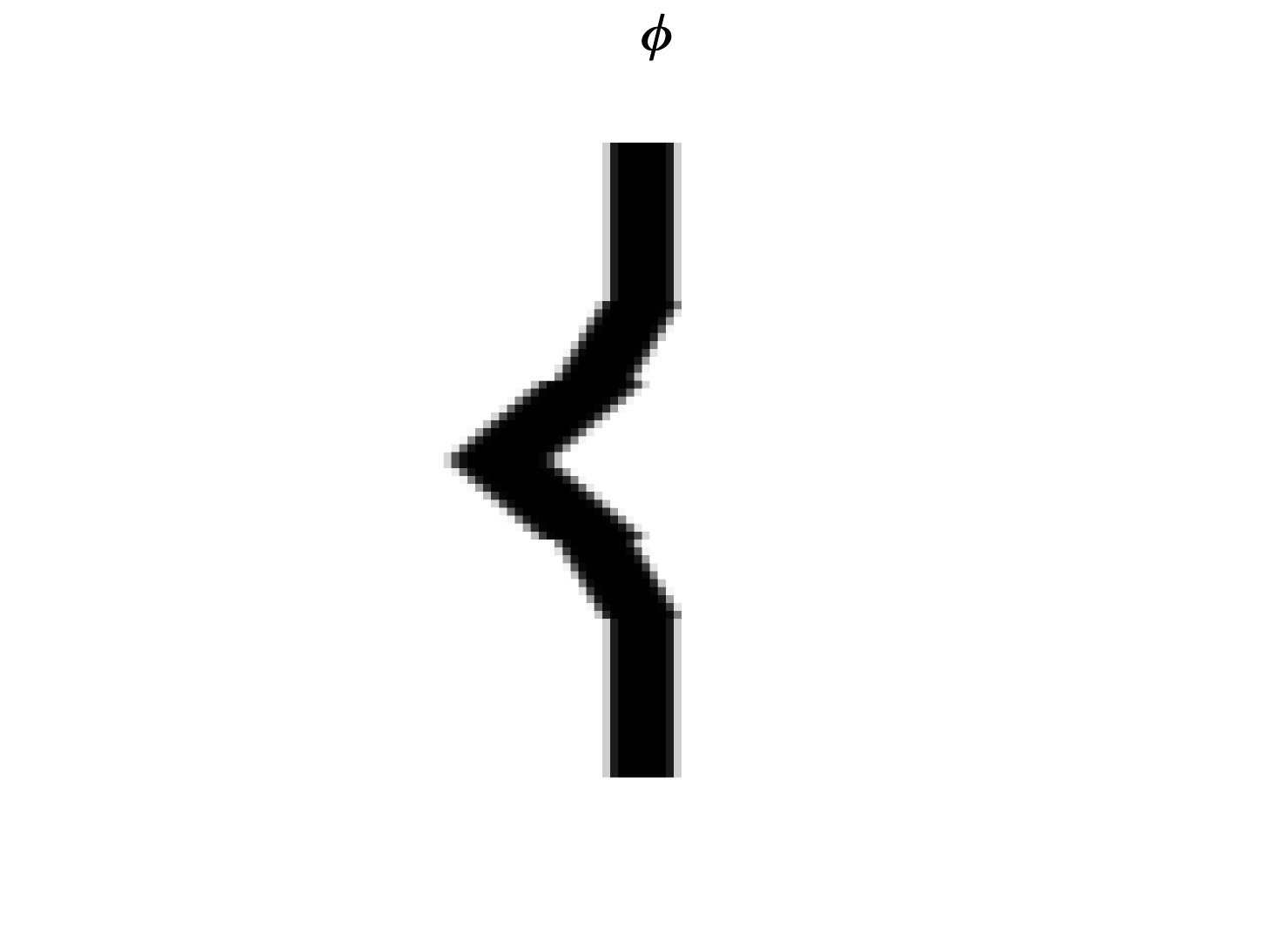}
        \caption{$\phi (\bar{d})$ - distance field not filtered}
        \label{fig:phi_d_notfilt2}
    \end{subfigure}%
	~ 
    \begin{subfigure}[t]{0.32\textwidth}
        \centering
        \includegraphics[width=1\columnwidth,trim=0.5in 0.3in 0.5in 0.30in,clip]{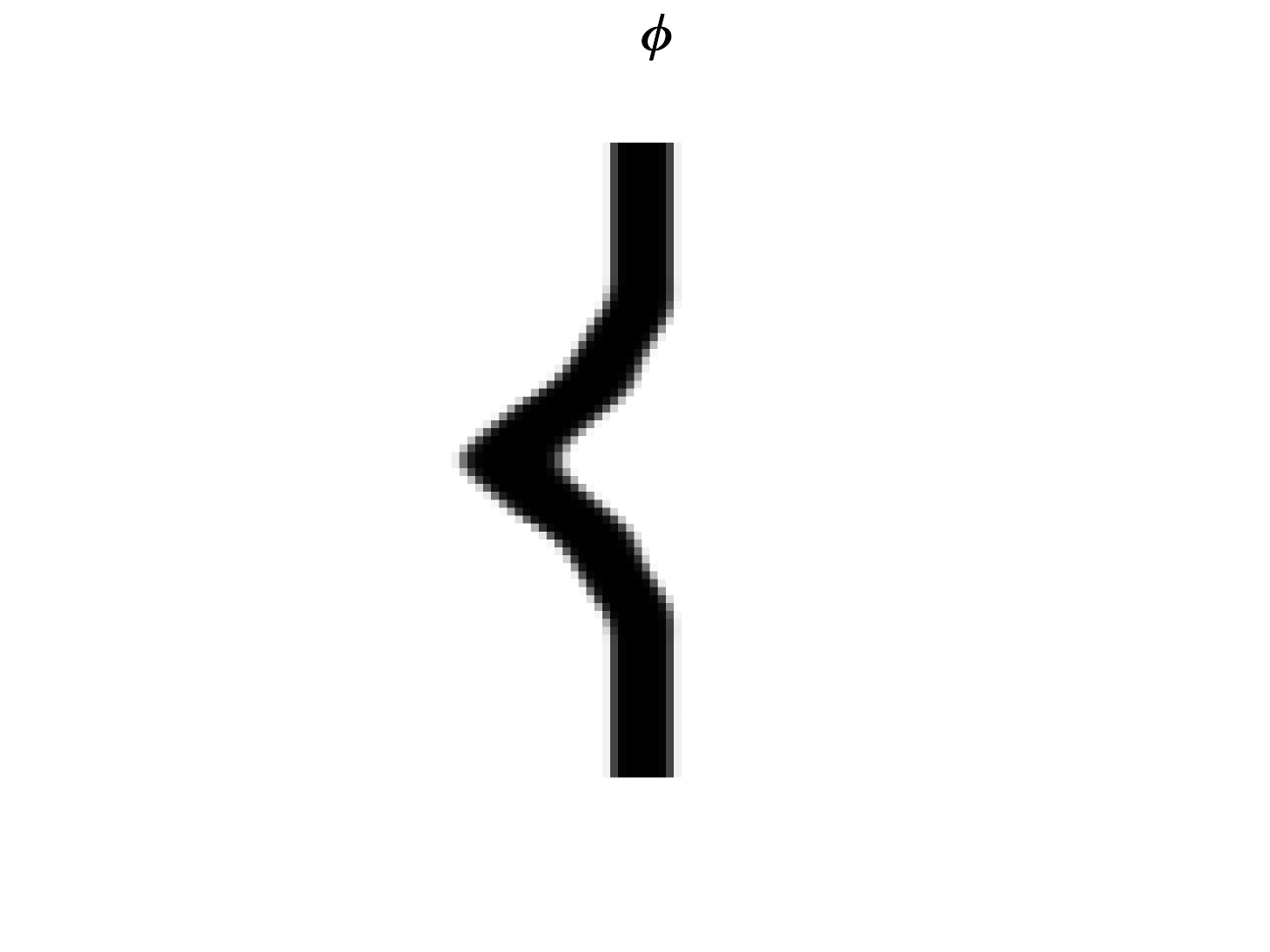}
       \caption{$\phi (\tilde{d})$ - filtered distance field}
       \label{fig:phi_d_filt3}
   \end{subfigure}
    \caption{Projection profiles without and with filtering of the distance field. In Eq. \eqref{eq:finalphi} $\beta_{fil}=4$, $\mu_{fil}=5$, and the radius of the filter of Eq. \eqref{eq:filterphi} is $r_{\phi} = 4$}
    \label{fig:phidfiltered4}
\end{figure*}

The projected quantity $\phi (\bar{d})$ of Eq. \eqref{eq:distprojmax} is characterized by sharp transitions between consecutive segments because of the piece-wise linear definition of the projection profile.
This can be seen in Fig.~\ref{fig:phi_d_notfilt2}, for the given profile shown in Fig.~\ref{fig:phi_d_notfilt1}. 
To smooth the corners of the projected strip, we filter the distance field $\bar{d}^{2}_{e}$,
\begin{equation}\label{eq:filterphi}
\tilde{d}^{2}_{i} = \frac{\sum_{j\in N_{i}} w(\Delta\textbf{x}_{ij})\bar{d}^{2}_{j}}{\sum_{j\in N_{i}} w(x_{j})}
\end{equation}
or in a matrix form:
\begin{equation}
\tilde{\textbf{d}}^{2} = \mathcal{D}\left(\textbf{H}_{s}^{\phi}\right)^{-1}\left(\textbf{H}^{\phi}\,\bar{\textbf{d}}^{2}\right)
\end{equation}
where $\textbf{H}^{\phi}$ is a $[N_{ele}\times N_{ele}]$ matrix, $\textbf{H}^{\phi}_{s}$ is a $[N_{ele}\times 1]$ vector.
The entries of $\textbf{H}^{\phi}$ and $\textbf{H}^{\phi}_{s}$ are defined as follows:
\begin{equation}
\begin{split}
& H_{ij}^{\phi} = w(\Delta\textbf{x}_{ij}), \quad H_{s,i}^{\phi} = \sum_{j} H_{ij}.
\end{split}
\end{equation}
The benefit of the filtering process applied to $\bar{d}$ can be observed in Fig.~\ref{fig:phi_d_filt3}. The figure highlights the smoother transitions of the projected quantity between consecutive profile segments.
Consequently, the final Super-Gaussian projection is:
\begin{equation}\label{eq:finalphi}
\phi_{i} = e^{-\frac{1}{2}\left(\frac{\tilde{d}^{2}_{i}}{\beta^{2}_{\phi}}\right)^{\mu_{\phi}}} \text{ for }i=1,\dots,N_{ele}
\end{equation}


\section{Problem formulation}
\label{sec:formulation}

In this section we present the formulation of the optimization problem based on the design variables and the parametrization described in the previous section. 
We provide also details regarding the governing equations of the problem and additional computational considerations.
In particular, the optimization problem considered herein can be formulated in general terms as follows:
\begin{equation}\label{eq:topoptprob}
\begin{split}
\underset{\bm{\rho},\textbf{x}}{\text{minimize :}} & \quad f(\bm{\rho},\,\textbf{x})\\
\text{subject to :} &\quad g_{k} (\bm{\rho},\textbf{x}) \leq 0,\quad k=0,...,m \\
\text{ :}& \quad 0 \leq \rho_{i} \leq 1,\quad i=1,...,N_{ele}\\
\text{ :}& \quad x_{lb} \leq x_{j} \leq x_{ub}, \quad j=1,...,N_{node}\\
\text{with :} &\quad \textbf{K}(\bm{\rho},\textbf{x})\, \textbf{u} = \textbf{f}\\
\end{split}
\end{equation}
where $\bm{\rho}$ and $\textbf{x}$ are the vectors of the design variables, corresponding to densities and nodal coordinates, respectively; $f(\bm{\rho},\textbf{x})$ is the objective function; $g_{k} (\bm{\rho},\textbf{x})$ are the $m$ inequality constraints of the problem; $N_{ele}$ is the number of finite elements used to discretize the problem; $N_{node}$ is the number of nodes defining the projection profile; and $x_{ub}$ and $x_{lb}$ are predefined upper and lower bounds for the variable geometric coordinates of the projection profiles' nodes.
It should be noted that the shape variables $x_{i}$ are mapped from a normalized domain $[0-1]$ to the actual domain as follows: $X_{i}=x_{i}\, N_{ele,x}\,a$, where $N_{ele,x}$ is the number of elements in the $x$ direction and $a$ is the size of each element.
In case of a vertical profile, the same applies to $y_{i}$, $Y_{i}$, and  $N_{ele,y}$.
Finally, $\textbf{K}(\bm{\rho},\textbf{x})$ is the stiffness matrix of the structural system considered and $\textbf{u}$ are the displacements of the deformed structure due to the external loads $\textbf{f}$.

As it has been mentioned in Sec.~\ref{subsec:densityparam}, we apply a `robust' topology optimization approach \citep{wang2011projection}. 
This improves the control over the minimum length scale, and for problems that consider only stiffness and volume, the robust approach results in considering the eroded design to compute the stiffness and the dilated design to compute the structural volume \citep{lazarov2016length}.
The objective function minimized in the topology optimization problem is the structural compliance evaluated considering the eroded density field, i.e.~$\bar{\bm{\rho}}^{ero}$. In particular, the compliance is defined as follows:
\begin{equation}
f(\bm{\rho},\textbf{x}) = \textbf{f}^{T} \, \textbf{u}(\bar{\bm{\rho}}^{ero},\textbf{x}).
\end{equation}
The displacement field is computed by solving the linear system of equations for the structural equilibrium. The structural stiffness matrix is computed based on the current value of the design variables, therefore:
\begin{equation}
\textbf{K}(\bar{\bm{\rho}}^{ero},\textbf{x}) \,\textbf{u}= \textbf{f}.
\end{equation}
The definition of Young's modulus $E$ of each element $i$ in the eroded layout is based on the Modifed SIMP interpolation scheme \citep{sigmund1997design}: 
\begin{equation}\label{eq:simp}
E_{i} = E_{min} + ( E_{max} - E_{min})\, \left( \bar{\rho}^{ero}_i \right)^{p_{E}}
\end{equation}
where $E_{min}$ is associated to an ersatz material that represents void and it has a relatively small value to avoid singularities in the stiffness matrix (e.g. $10^{-6}$); $E_{max}$ is the actual Young's modulus of the distributed material; $p_{E}$ is a penalization factor that for values bigger than one penalizes the intermediate values of $\bar{\rho}^{ero}_i$, making them uneconomical and thus driving the optimizer towards near discrete optimized designs; last, $\bar{\rho}^{ero}_i$ is defined according to Eq.~\eqref{eq:rhoproject}.

We consider also a volume constraint on the full structural domain:
\begin{equation}\label{eq:control0}
g_{0}(\bm{\rho}) =\frac{\sum_{i=1}^{N_{ele}} \bar{\rho}^{dil}_{i} v_{i}}{\sum_{i=1}^{N_{ele}} v_{i}} - g_{0,dil}^{*} \leq 0
\end{equation}
where $v_{i}$ is the volume of the $i$-th finite element; and $g_{0,dil}^{*}$ is the available solid volume fraction in the dilated layout.


\section{Control over projected entities}
\label{sec:control}
The topology optimization approach discussed herein is quite flexible as it can control several design features by combining projection- and density-based design variables in the problem formulation.
In this section we discuss several of the features that can be controlled.
These are a subset of the many possibilities and they have been selected to show the potential of the methodology presented. 
The control of selected features is expressed through optimization constraints $g_{k} (\bm{\rho},\textbf{x})$ in the optimization problem formulation \eqref{eq:topoptprob}, or by projecting specific mechanical properties.

The first example of a feature that can be controlled through a localized constraint is the total volume occupied by the structural material in the projection area. The possibility of controlling the amount of material in a specific area can be important in the manufacturing process of structural assemblies. In these cases, it may be desirable to reduce the amount of material that needs to be welded or connected at the interface of different sub-components of a structural assembly. It could also be relevant for structural assemblies where one wishes to reduce the structural complexity at the connecting interface of sub-components.
This type of control can be defined through a local volume constraint, affecting only the portion of the structure that is included in the projection area.
As the shape of the projection profile is also optimized together with the structural topology, this local constraint is imposed on a sub domain that changes during the optimization process depending on the shape of the projection profile.
In particular, we define the following constraint:
\begin{equation}\label{eq:control1}
g_{1}(\bm{\rho},\textbf{x}) =\frac{\sum_{i=1}^{N_{ele}} \bar{\rho}^{dil}_{i} v_{i}\, \phi_{i}(\textbf{x})}{\sum_{i=1}^{N_{ele}} v_{i}\,\phi_{i}(\textbf{x})} - g_{1,dil}^{*} \leq 0
\end{equation}
where $\bar{\rho}^{dil}_{i}$ is the projected density of the $i$-th element in the dilated layout; $v_{i}$ is the volume of the $i$-th element; $\phi_{i}(\textbf{x})$ is the projection function evaluated in correspondence to the $i$-th element, expressed explicitly as a function of the projection profile coordinates variables $\textbf{x}$; and $g_{1,dil}^{*}$ is the solid volume fraction allowed in the projection area in the dilated layout. 
Essentially, Eq.~\eqref{eq:control1} represents a volume constraint imposed only on a sub-domain identified by those elements for which $\phi=1$ (i.e.~the gray area of Fig.~\ref{fig:1cut}).

Next, we consider an additional constraint that controls the maximum length scale of the solid phase \citep{guest2009imposing,wu2018infill}, and we impose it only in the projection area .
To this end, we first introduce the vector $\hat{\bm{\rho}}$ whose components $\hat{\rho}_{i}$ describe the local material distribution in a neighborhood $\hat{N}_{i}$ of a element $i$:
\begin{equation}\label{eq:filtrhomaxlengh}
\hat{\rho}_{i} = \frac{\sum_{j \in \hat{N}_{i}} \bar{\rho}^{dil}_{j}}{\sum_{j \in \hat{N}_{i}} 1}
\end{equation}
or in a matrix form:
\begin{equation}
\hat{\bm{\rho}} = \mathcal{D}\left( \textbf{H}_{s}^{ls}\right)^{-1}\left( \textbf{H}^{ls}\,
\bar{\bm{\rho}}^{dil} \right)
\end{equation}
where $\textbf{H}^{ls}$ is a $[N_{ele}\times N_{ele}]$ matrix, $\textbf{H}^{ls}_{s}$ is a $[N_{ele}\times 1]$ vector.
The entries of $\textbf{H}^{ls}$ and $\textbf{H}^{ls}_{s}$ are defined as follows:
\begin{equation}
\begin{split}
& H_{ij}^{ls} = w^{ls}_{ij}, \quad H_{s,i}^{ls} = \sum_{j} H_{ij}^{ls}
\end{split}
\end{equation}
with
\begin{equation}
 w^{ls}_{ij}=
 \begin{cases}
    1,& \text{if } j \in \hat{N}_{i}\\
    0,              & \text{otherwise}
\end{cases}
\end{equation}
and we require that:
\begin{equation}\label{eq:aggregconst}
\hat{\rho}_{i} \phi_{i} \leq \alpha \quad \forall \, i=1, \dots, N_{ele}.
\end{equation}
Eq.~\eqref{eq:aggregconst} represents $N_{ele}$ constraints on the averaged percentage of solid material in a neighborhood of each element $i$ (i.e. $\hat{\rho}_{i}$), where the maximum allowed percentage of material is $\alpha$.
These constraints are equivalent to:
\begin{equation}\label{eq:aggregconst2}
\max_{i}(\hat{\rho}_{i} \phi_{i}) \leq \alpha \quad \text{for } \, i=1, \dots, N_{ele}
\end{equation}
The $\max$ function in Eq. \eqref{eq:aggregconst2} is not differentiable and therefore not suitable for numerical optimization approaches based on first-order information.
To reduce the number of constraints in \eqref{eq:aggregconst} from $N_{ele}$ to one in a differentiable manner, we aggregate them with a $p$-norm into a single constraint that approximates the maximum value of the set considered similarly to \cite{wu2018infill}:
\begin{equation}\label{eq:lengthscalefilter}
\left(\sum_{i=1}^{N_{ele}} \hat{\rho}_{i}^{p}\phi_{i}\right)^{1/p} \leq \left(\sum_{i=1}^{N_{ele}} \alpha^{p}\phi_{i}\right)^{1/p}
\end{equation}
In Eq. \eqref{eq:lengthscalefilter} the quantities $\phi_{i}$ are not raised to the power of $p$. This is done to account also for the elements on the perimeter of the projection area (where $0<\phi_{i}<1$) in the constraint \eqref{eq:lengthscalefilter} during the early stages of the optimization. However, as the optimization progresses and the projected quantities $\phi_{i}$ assume near discrete $0$-$1$ values, the fact that the quantities $\phi_{i}$ are not raised to the power of $p$ becomes less relevant.
After rearranging the terms we obtain the following constraint formulation:
\begin{equation}\label{eq:lengthscaleconst}
g_{2}(\hat{\rho},\textbf{x})=\left(\frac{\sum_{i=1}^{N_{ele}} \hat{\rho}_{i}^{p}\phi_{i}}{\sum_{i=1}^{N_{ele}} \phi_{i}}\right)^{1/p} -  \alpha \leq0
\end{equation}

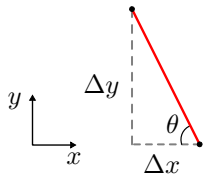
\begin{figure}[h]
\centering
\begingroup%
  \makeatletter%
  \providecommand\color[2][]{%
    \errmessage{(Inkscape) Color is used for the text in Inkscape, but the package 'color.sty' is not loaded}%
    \renewcommand\color[2][]{}%
  }%
  \providecommand\transparent[1]{%
    \errmessage{(Inkscape) Transparency is used (non-zero) for the text in Inkscape, but the package 'transparent.sty' is not loaded}%
    \renewcommand\transparent[1]{}%
  }%
  \providecommand\rotatebox[2]{#2}%
  \newcommand*\fsize{\dimexpr\f@size pt\relax}%
  \newcommand*\lineheight[1]{\fontsize{\fsize}{#1\fsize}\selectfont}%
  \ifx\svgwidth\undefined%
    \setlength{\unitlength}{77.96821895bp}%
    \ifx\svgscale\undefined%
      \relax%
    \else%
      \setlength{\unitlength}{\unitlength * \real{\svgscale}}%
    \fi%
  \else%
    \setlength{\unitlength}{\svgwidth}%
  \fi%
  \global\let\svgwidth\undefined%
  \global\let\svgscale\undefined%
  \makeatother%
  \begin{picture}(1,0.82410452)%
    \lineheight{1}%
    \setlength\tabcolsep{0pt}%
    \put(0,0){\includegraphics[width=\unitlength,page=1]{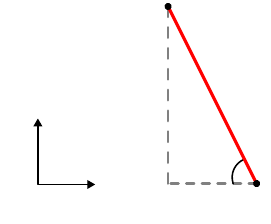}}%
    \put(0.78574498,0.20089312){\color[rgb]{0,0,0}\makebox(0,0)[lt]{\lineheight{0}\smash{\begin{tabular}[t]{l}$\theta$\end{tabular}}}}%
    \put(0.38636788,0.39231226){\color[rgb]{0,0,0}\makebox(0,0)[lt]{\lineheight{0}\smash{\begin{tabular}[t]{l}$\Delta y$\end{tabular}}}}%
    \put(0.67494701,0.00754009){\color[rgb]{0,0,0}\makebox(0,0)[lt]{\lineheight{0}\smash{\begin{tabular}[t]{l}$\Delta x$\end{tabular}}}}%
    \put(0.01998119,0.34081203){\color[rgb]{0,0,0}\makebox(0,0)[lt]{\lineheight{1.25}\smash{\begin{tabular}[t]{l}$y$\end{tabular}}}}%
    \put(0.30345899,0.04968185){\color[rgb]{0,0,0}\makebox(0,0)[lt]{\lineheight{1.25}\smash{\begin{tabular}[t]{l}$x$\end{tabular}}}}%
  \end{picture}%
\endgroup%

\caption{Generic segment of the projection profile inclined by an angle $\theta$ with respect to the horizontal direction $x$. In this case the segment belongs to a profile developed along the vertical $y$ direction, but the same idea applies to horizontal profiles}  
\label{fig:segmentslope}
\end{figure}

The next constraint that we wish to discuss is related to the maximum allowed slope of the segments composing the projection profile.
As it has been already mentioned, the geometric description of the projection profiles discussed herein relies on simple piece-wise linear segmented profiles.
Large variations of the inclination angles between contiguous segments can result in segmented projection profiles with undesired sharp corners (see Fig.~\ref{fig:phidfiltered4}). 
This issue has been mitigated with the introduction of a filtered distance of the elements with respect to their associated profile segment. 
Here, we discuss an additional precaution that can be taken to further regularize the projection profile: We consider a constraint on the slope of the profile segments.
If we consider a generic segment belonging to a vertical profile as shown in Fig.~\ref{fig:segmentslope}, $\Delta x$ is the difference between the $x$ coordinates of the segment nodes, $\Delta y$ is the difference between the $y$ coordinates of the segment nodes, and $\theta$ is the angle of the segment with respect to the horizontal direction.
In the case of a vertical profile, $\Delta x$ is variable whereas $\Delta y$ fixed. The opposite would be true for a horizontal profile.
A limit on the maximum slope $\theta_{max}$ is equivalent to imposing a limit on the maximum $\Delta x$ for a fixed $\Delta y$:
\begin{equation}
 \Delta x_{max}= \frac{\Delta y}{tan(\theta_{max})}
\end{equation}
Consequently, the constraint is formulated as follows:
 \begin{equation}\label{eq:g3max}
g_{3}(\textbf{x})= \max \left( \Delta x_{i}^{2}\right) \leq \Delta x_{max}^{2},\; \text{for } i=1,\dots,N_{node}-1
\end{equation}
The constraint \eqref{eq:g3max} is equivalent to $N_{node}$-$1$ linear constraints for the shape variables $x_{i}$. It has been defined as a single aggregated constraint to facilitate its use with the optimization algorithm chosen for this work, the Method of Moving Asymptotes \citep{svanberg1987method}, which is less suited for liner constraints, and in general large numbers of constraints.
It should be mentioned, however, that the constraints of Eq. \eqref{eq:g3max} could have been treated separately with other optimization algorithms more suited for larger number of constraints (e.g. Sequential Linear Programming, Sequential Quadratic Programming \citep{nocedal2006numerical}).
Subsequently, we replace Eq. \eqref{eq:g3max} with an approximated differentiable formulation based on a $p$-norm:
\begin{equation}\label{eq:constrg3}
g_{3}(\textbf{x})= \frac{1}{\Delta x_{max}^{2}}\left(\sum_{i=1}^{N_{node}-1} \left( \Delta x_{i}^{2}\right)^{p} \right)^{\frac{1}{p}} -1\leq 0.
\end{equation}
The constraint $g_{3}(\textbf{x})$ defined in Eq.~\eqref{eq:constrg3} limits the inclination of the projection profile segments, and leads to more regular optimized projection profiles. 

The projection variables $\phi_{i}$ allow also to assign different material properties to the elements included in the projection area. 
In particular, if we wish to assign a Young's modulus equal to $E_{\phi}$ to the elements in the projection area, we can do that through a material interpolation scheme based on the Modified SIMP:
\begin{equation}\label{eq:simp2}
\begin{split}
& E_{i} = E_{min} + ( E_{max} - E_{max}\, r_{E}\,\phi_{i} - E_{min})\, \bar{\rho}^{p_{E}}_{i,ero}\\
& r_{E} = (E_{max}-E_{\phi})/E_{max}
\end{split}
\end{equation}
where the parameter $r_{E}$ is introduced to prescribe a different Young's modulus to the elements inside the projection area, where $\phi_{i}=1$ and $E_{i}=E_{\phi}$.
The interpolation scheme defined in \eqref{eq:simp2} allows to assign the Young's modulus $E_{max}$ to the solid finite elements outside the projection area, and the Young's modulus $E_{\phi}$ to the solid elements inside the projection area. This could be used for example in the case of a structural assembly composed of separate welded components. In this case, $E_{\phi}$ would define the material property along the welded interface identified by the projection profile.

The last controlled feature that we discuss is the spatial variability of the length scale. We implement a density filter with varying radius, similarly to the one suggested in \cite{amir2018achieving}. This allows to obtain a spatially variable minimum or maximum length scale control.
More precisely, we consider a density filter definition whose radius varies according to the current relative position in the domain of the element considered and of the projection profile.
In order to obtain a variable minimum length scale control, we modify the well known expression for the density filter as follows:
 \begin{equation}\label{eq:densityfilt_var}
\tilde{\rho}_{i} = \frac{\sum_{j\in N_{i}} w(\Delta\textbf{x}_{ij},\phi_{i})\rho_{j}}{\sum_{j\in N_{i}} w(\Delta\textbf{x}_{ij},\phi_{i})}
\end{equation}
where $w(\Delta\textbf{x}_{ij},\phi_{i})$ is the modified weight function:
\begin{equation}\label{eq:varfiltweight1}
w(\Delta\textbf{x}_{ij},\phi_{i}) = \max \bigl( \bar{r}_{min}(\phi_{i}) - \|\textbf{x}_{j}-\textbf{x}_{i}\|, 0\bigr).
\end{equation}
In \eqref{eq:varfiltweight1}, the variable filter radius is defined as:
\begin{equation}\label{eq:varfiltweight2}
\bar{r}_{min}(\phi_{i}) = r_{min}(1 + \gamma\, \phi_{i})
\end{equation}
where $r_{min}$ is the assigned filter radius, and $\gamma$ the filter radius amplification factor. 
For example if $\gamma=1$, the filter radius is doubled in the projection area, i.e.~where $\phi=1$.
In a matrix form the filtering transformation \eqref{eq:densityfilt_var} reads as follows:
\begin{equation}
\tilde{\bm{\rho}} = \mathcal{D}\left(\textbf{H}_{s}(\bm{\phi})\right)^{-1}\left(\textbf{H}(\bm{\phi})\,\bm{\rho}\right)
\end{equation}
where $\textbf{H}$ is a $[N_{ele}\times N_{ele}]$ matrix, $\textbf{H}_{s}$ is a $[N_{ele}\times 1]$ vector. 
The entries of $\textbf{H}$ and $\textbf{H}_{s}$ are defined as follows:
\begin{equation}
\begin{split}
& H_{ij} = w(\Delta\textbf{x}_{ij},\phi_{i}), \quad H_{s,i} = \sum_{j} H_{ij}.
\end{split}
\end{equation}
Alternatively, it is possible to consider a different filter weight. 
In fact, we can replace the definition \eqref{eq:varfiltweight1} with a Gaussian weight as in \cite{amir2018achieving}:
\begin{equation}\label{eq:varfiltweight_gaus}
w_{G}(\Delta\textbf{x}_{ij},\phi_{i}) = exp \left(-\left(\frac{\|\textbf{x}_{j}-\textbf{x}_{i}\|}{\bar{r}_{min}(\phi_{i})/2}\right)^{n}\right).
\end{equation}
In Eq.~\eqref{eq:varfiltweight_gaus} $exp()$ is the exponential function, and $n$ is a predefined positive number. 
For a density filter with a similar effect of the filter \eqref{eq:densityfilt_var}, $n$ can be set to $2$.

\begin{figure*}[h]
    \centering
    \begin{subfigure}[t]{0.475\textwidth}
        \centering
        \includegraphics[width=0.99\columnwidth]{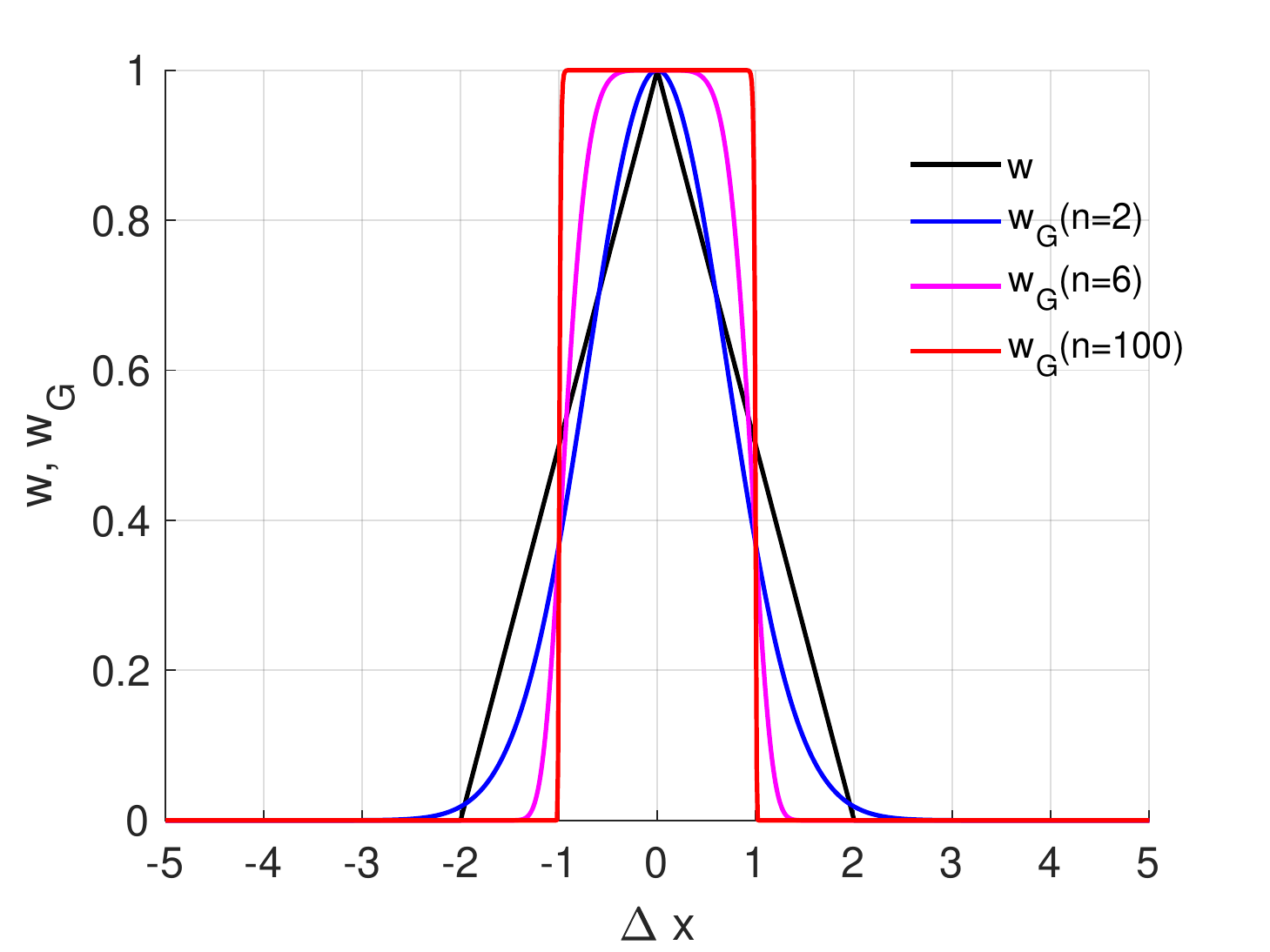}
        \caption{ }
				\label{subfig:filterweights1}
    \end{subfigure}%
    ~ 
    \begin{subfigure}[t]{0.475\textwidth}
        \centering
        \includegraphics[width=0.99\columnwidth]{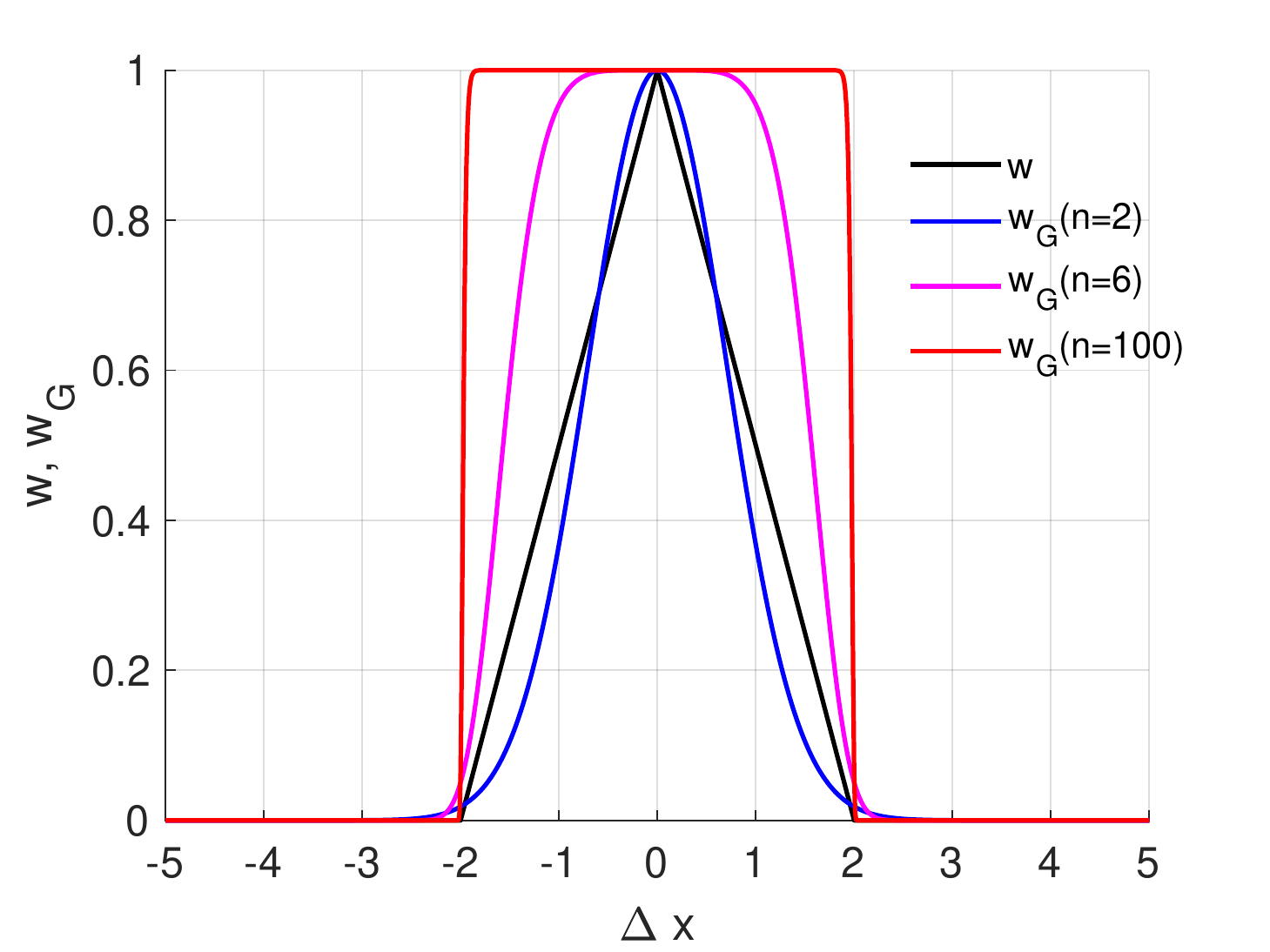}
        \caption{ }
				\label{subfig:filterweights2}
    \end{subfigure}
    \caption{Comparison of filter weights $w=\max ( r_{min} - x, 0)/r_{min}$ and $w_{G}(n)$ for $r_{min}=2$, and $n={2,6,100}$. (a) $w_{G}(\bar{r}_{min})$. (b) $w_{G}(\hat{r}_{min})$}
		\label{fig:filter_weights_compare}
\end{figure*}

It is possible to define also a variable maximum length scale control.
In regard to this, previously in Eq.~\eqref{eq:filtrhomaxlengh} the following filtering procedure has been considered:
\begin{equation}
\hat{\rho}_{i} = \frac{\sum_{j\in N_{i}} w(\Delta\textbf{x}_{ij}) \bar{\rho}^{dil}_{j}}{\sum_{j\in N_{i}} w(\Delta\textbf{x}_{ij})}
\end{equation}
where $w(\Delta\textbf{x}_{ij})$ was equivalent to:
\begin{equation}\label{eq:weightmaxlenghH}
w(\Delta\textbf{x}_{ij}) = \mathcal{H} \bigl( \max(\bar{r}_{min}(\phi_{i}) - \|\textbf{x}_{j}-\textbf{x}_{i}\|,0)\bigr)
\end{equation}
In Eq.~\eqref{eq:weightmaxlenghH}, $\mathcal{H}(x)$ is the Heaviside function and it returns $1$ for a non zero argument $x$, otherwise it returns $0$.
The filter weights of Eq.~\eqref{eq:weightmaxlenghH} can be closely approximated also by the Gaussian weight function.
By increasing the value of $n$ in Eq.~\eqref{eq:varfiltweight_gaus}, the weights $w_{G}$ become more homogeneous and approximately equal to $1$.
However, in Fig.~\ref{subfig:filterweights1} it can also be observed that for increasing values of $n$ the effective filter radius reduces up to approximately half of its initial value.
To preserve a meaningful definition of the filter weights with respect to the filter radius $r_{min}$, we propose the following modification of the assigned filter radius $r_{min}$ based on the given value of the exponent $n$:
\begin{equation}\label{eq:modifyrmin}
\tilde{r}_{min} = r_{min}  (1 + ( 1-2/n ) )
\end{equation}
such that
\begin{equation}\label{eq:finalwg}
\begin{split}
&w_{G}(\Delta\textbf{x}_{ij},\phi_{i}) = exp \left(-\left(\frac{\|\textbf{x}_{j}-\textbf{x}_{i}\|}{\hat{r}_{min}(\phi_{i})/2}\right)^{n}\right),\\
&\text{with: }
\hat{r}_{min}(\phi_{i},n) = \tilde{r}_{min}(1 + \gamma\, \phi_{i})
\end{split}
\end{equation}
It can be observed in Fig.~\ref{subfig:filterweights2} that by replacing $r_{min}$ with $\tilde{r}_{min}$ we preserve the desired length scale control also for increasing values of $n$.
Thus, we can now define a constraint for maximum length scale control with variable maximum feature size allowed:
\begin{equation}\label{eq:lengthscaleconst2}
g_{4}(\hat{\rho},\textbf{x})=\left(\frac{\sum_{i=1}^{N_{ele}} \hat{\rho}_{i}^{p}}{N_{ele}}\right)^{1/p} -  \alpha \leq0
\end{equation}
where
\begin{equation}
\hat{\rho}_{i} = \frac{\sum_{j\in N_{i}} w_{G}(\Delta\textbf{x}_{ij},\hat{r}_{min}(\phi_{i},n)) \bar{\rho}^{dil}_{j}}{\sum_{j\in N_{i}} w_{G}(\Delta\textbf{x}_{ij},\hat{r}_{min}(\phi_{i},n))},\; \forall i=1,\dots,N_{ele}
\end{equation}
and $n$ is a positive number higher than $2$ (e.g. $n= \{4,6\}$).

\section{Sensitivity analysis}
\label{sec:sensan}
The optimization approach adopted in this work relies on first-order information. Hence, the gradients of the objective and constraints functions need to be calculated. The dependency of the objective function on the optimization problem variables is implicitly defined through the equilibrium equations. 
Hence in this section we provide additional details regarding the adjoint sensitivity analysis considered to calculate the gradients of the objective function and some of the constraint functions.
In particular, some of the functions involved in the problem formulation depend on both the topological variables $\bm{\rho}$ and on the coordinates of the projection profile $\textbf{x}$, namely the constraints $g_{1}$, $g_{2}$, $g_{4}$, and the objective function $f$ (if the material interpolation scheme \eqref{eq:simp2} is considered). 
The dependency on both sets of variables originates from the fact that the projection profile configuration defines the subdomain on which some of the constraints are defined (i.e. $g_{1}$, $g_{2}$, and $g_{4}$), and the subdomain on which the solid elements are given the Young's modulus $E_{\phi}$, if Eq. \eqref{eq:simp2} is considered.
The constraint $g_{3}$ depends only on the shape variables $\textbf{x}$.
The explicit calculations of the gradients are provided in the following.

The gradient of the objective function is calculated with an adjoint method where we first define an augmented function $\hat{f}$:
\begin{equation}\label{eq:objgrad}
\begin{split}
& \hat{f}(\bm{\rho},\textbf{x}) = \textbf{f}^{T}\, \textbf{u} + \bm{\lambda}^{T}(\textbf{K}(\bm{\rho}_{ero},\textbf{x}) \,\textbf{u} - \textbf{f})
\end{split}
\end{equation}
It should be noted that when the equilibrium is satisfied (i.e. $\textbf{K}(\bm{\rho}_{ero},\textbf{x}) \,\textbf{u} - \textbf{f}$=\textbf{0}) then $f=\hat{f}$.
The derivative of $\hat{f}$ with respect to a generic variable $\xi$ is given by:
\begin{equation}\label{eq:objgrad2}
\begin{split}
& \frac{\partial \hat{f}}{\partial \xi} = \textbf{f}^{T}\,\frac{\partial \textbf{u}}{\partial \xi}  + \bm{\lambda}^{T}(\frac{\partial \textbf{K}}{\partial \xi}\,\textbf{u} + \textbf{K}\frac{\partial \textbf{u}}{\partial \xi})
\end{split}
\end{equation}
where we assumed that the force vector $\textbf{f}$ does not depend on the design variables. This does not imply any loss of generality, see for example design-dependent forces in \cite{amir2018simultaneous}.
In order to eliminate the unknown implicit derivatives $\frac{\partial \textbf{u}}{\partial \xi}$ from the gradient calculation, the following equation is solved:
\begin{equation}\label{eq:objgrad3}
\begin{split}
\textbf{K}^{T}\bm{\lambda} = -\textbf{f}.
\end{split}
\end{equation}
Once the adjoint vector $\bm{\lambda}$ has been calculated, it is possible to calculate the actual gradient:
\begin{equation}\label{eq:objgrad4}
\begin{split}
& \frac{\partial f}{\partial \xi} = \bm{\lambda}^{T}\,\frac{\partial \textbf{K}}{\partial \xi}\,\textbf{u}.
\end{split}
\end{equation}
We now specify the derivative $\frac{\partial \hat{f}}{\partial \xi}$ with respect to $\bm{\rho}$:
\begin{equation}\label{eq:derphi1}
\begin{split}
\frac{\partial f}{\partial \rho_{i}} &= \sum_{j\in N_{i}}\left( \frac{\partial \hat{f}}{\partial \bar{\rho}^{ero}_{j}}  \frac{\partial \bar{\rho}^{ero}_{j}}{\partial \tilde{\rho}_{j}} \right) \frac{\partial \tilde{\rho}_{j}}{\partial \rho_{i}}=\\
 &=\sum_{j\in N_{i}} H_{ij}\left( \frac{1}{H_{s,j}} \frac{\partial \hat{f}}{\partial \bar{\rho}^{ero}_{j}}  \frac{\partial \bar{\rho}^{ero}_{j}}{\partial \tilde{\rho}_{j}} \right)
 \end{split}
\end{equation}
where
\begin{equation}\label{eq:derphi2}
\begin{split}
 &\frac{\partial \hat{f}}{\partial \bar{\rho}^{ero}_{i}} = \bm{\lambda}^{T}\,\left(\frac{\partial \textbf{K}}{\partial \bar{\rho}^{ero}_{i}}\,\textbf{u}\right)\\
 &\frac{\partial \textbf{K}}{\partial \bar{\rho}^{ero}_{i}}= \opA_{i=1}^{N_{ele}} p_{E}(E_{max} - E_{min})(\bar{\rho}^{ero}_{i})^{p_{E}-1}\textbf{K}_{e,i}
 \end{split}
\end{equation}
In Eq.~\eqref{eq:derphi2}, $\opA$ is the matrix assembly operator, and $\textbf{K}_{e,i}$ is the stiffness matrix of each finite element normalized by its interpolated Young's modulus $E(\bar{\rho}^{ero}_{i})$.

In cases where the material interpolation scheme of Eq.~\eqref{eq:simp2} is considered, the stiffness matrix will depend also on the shape coordinates of the projection profile, and as a consequence also the objective function $f$. Hence:
\begin{equation}\label{eq:derphi3}
\begin{split}
\frac{\partial f}{\partial x_{i}} &= \sum_{k\in N_{j}} \left(\left(\frac{\partial \hat{f}}{\partial \phi_{k}} \frac{\partial \phi_{k}}{\partial \tilde{d}^{2}_{k}} \right)\frac{\partial \tilde{d}^{2}_{k}}{\partial \bar{d}^{2}_{j}}  \right)
\frac{\partial \bar{d}^{2}_{j}}{\partial x_{i}} =\\
&=\sum_{k\in N_{j}} \left(H_{jk}^{\phi}\left(\frac{1}{H_{s,k}^{\phi}}\frac{\partial \hat{f}}{\partial \phi_{k}} \frac{\partial \phi_{k}}{\partial \tilde{d}^{2}_{k}} \right)  \right)
\frac{\partial \bar{d}^{2}_{j}}{\partial x_{i}}
\end{split}
\end{equation}
where
\begin{equation}\label{eq:derphi4}
\begin{split}
 &\frac{\partial \hat{f}}{\partial \phi_{i}} = \bm{\lambda}^{T}\,\left(\frac{\partial \textbf{K}}{\partial \phi_{i}}\,\textbf{u}\right)\\
 &\frac{\partial \textbf{K}}{\partial \phi_{i}}= \opA_{i=1}^{N_{ele}} -r_{E}E_{max} (\bar{\rho}^{ero}_{i})^{p_{E}}\textbf{K}_{e,i}
 \end{split}
\end{equation}
In Eq.~\eqref{eq:derphi5}, the matrix $\textbf{H}^{\phi}$ and the vector $\textbf{H}_{s}^{\phi}$ are used to perform the filtering of Eq. \eqref{eq:filterphi} that transform the distance field $\bar{d}^{2}$ in $\tilde{d}^{2}$.

The functions that define the constraints $g_{i}$ for $i=0,\dots , m$ are formulated explicitly in terms of the variables of the problem, and for this reason the calculation of their gradient does not require a dedicated sensitivity analysis. 
In particular, the gradient of $g_{0}$ with respect to $\bm{\rho}$ is calculated as follows:
\begin{equation}
\begin{split}
\frac{\partial g_{0}}{\partial \rho_{i}} &= \sum_{j\in N_{i}}\left( \frac{\partial g_{0}}{\partial \bar{\rho}^{dil}_{j}}  \frac{\partial \bar{\rho}^{dil}_{j}}{\partial \tilde{\rho}_{j}}\right)\frac{\partial \tilde{\rho}_{j}}{\partial \rho_{i}}=\\
&=\sum_{j\in N_{i}} H_{ij}\left( \frac{1}{H_{s,j}} \frac{\partial g_{0}}{\partial \bar{\rho}^{dil}_{j}}  \frac{\partial \bar{\rho}^{dil}_{j}}{\partial \tilde{\rho}_{j}}\right)
\end{split}
\end{equation}
The gradient of $g_{1}$ with respect to $\bm{\rho}$ is similar to that of  $g_{0}$, and in particular:
\begin{equation}
\begin{split}
\frac{\partial g_{1}}{\partial \rho_{i}} &= \sum_{j\in N_{i}}\left(\frac{\partial g_{1}}{\partial \bar{\rho}^{dil}_{j}}   \frac{\partial \bar{\rho}^{dil}_{j}}{\partial \tilde{\rho}_{j}}\right) \frac{\partial \tilde{\rho}_{j}}{\partial \rho_{i}}=\\
&=\sum_{j\in N_{i}} H_{ij}\left( \frac{1}{H_{s,j}} \frac{\partial g_{1}}{\partial \bar{\rho}^{dil}_{j}}  \frac{\partial \bar{\rho}^{dil}_{j}}{\partial \tilde{\rho}_{j}}\right)
\end{split}
\end{equation}
Moreover, $g_{1}$ is also a function of $\textbf{x}$ and the associated gradient is defined as follows:
\begin{equation}\label{eq:derphi5}
\begin{split}
\frac{\partial g_{1}}{\partial x_{i}} &= \sum_{k\in N_{j}} \left(\left(\frac{\partial g_{1}}{\partial \phi_{k}} \frac{\partial \phi_{k}}{\partial \tilde{d}^{2}_{k}} \right)\frac{\partial \tilde{d}^{2}_{k}}{\partial \bar{d}^{2}_{j}}  \right)
\frac{\partial \bar{d}^{2}_{j}}{\partial x_{i}} =\\
&=\sum_{k\in N_{j}} \left(H_{jk}^{\phi}\left(\frac{1}{H_{s,k}^{\phi}}\frac{\partial g_{1}}{\partial \phi_{k}} \frac{\partial \phi_{k}}{\partial \tilde{d}^{2}_{k}} \right)  \right)
\frac{\partial \bar{d}^{2}_{j}}{\partial x_{i}}
\end{split}
\end{equation}
The gradient of $g_{2}$ with respect to $\bm{\rho}$ is calculated as follows:
\begin{equation}
\begin{split}
&\frac{\partial g_{2}}{\partial \rho_{i}} =  \sum_{j\in N_{i}}\left(\sum_{k\in N_{j}}\left(  \frac{\partial g_{2}}{\partial \hat{\rho}_{k}}\right)\frac{\partial\hat{\rho}_{k}}{\partial \bar{\rho}^{dil}_{j}} \frac{\partial \bar{\rho}^{dil}_{j}}{\partial \tilde{\rho}_{j}}\right)\frac{\partial \tilde{\rho}_{j}}{\partial \rho_{i}}\\
&=\sum_{j\in N_{i}}H_{ij}\left(\frac{1}{H_{s,j}}\sum_{k\in N_{j}}H_{jk}^{ls}\left( \frac{1}{H_{s,k}^{ls}} \frac{\partial g_{2}}{\partial \hat{\rho}_{k}}\right) \frac{\partial \bar{\rho}^{dil}_{j}}{\partial \tilde{\rho}_{j}}\right)
\end{split}
\end{equation}
 The constraint $g_{2}$ depends also on the shape variables $\textbf{x}$ and its gradient with respect to them is calculated as follows:
\begin{equation}
\begin{split}
\frac{\partial g_{2}}{\partial x_{i}} &=  \sum_{k\in N_{j}} \left(\left(\frac{\partial g_{2}}{\partial \phi_{k}} \frac{\partial \phi_{k}}{\partial \tilde{d}^{2}_{k}} \right)\frac{\partial \tilde{d}^{2}_{k}}{\partial \bar{d}^{2}_{j}}  \right)
\frac{\partial \bar{d}^{2}_{j}}{\partial x_{i}} =\\
&=\sum_{k\in N_{j}} \left(H_{jk}^{\phi}\left(\frac{1}{H_{s,k}^{\phi}}\frac{\partial g_{2}}{\partial \phi_{k}} \frac{\partial \phi_{k}}{\partial \tilde{d}^{2}_{k}} \right)  \right)
\frac{\partial \bar{d}^{2}_{j}}{\partial x_{i}}
\end{split}
\end{equation}
The constraint $g_{3}$ depends solely on the shape variables $\textbf{x}$. Hence the gradient $\frac{\partial g_{3}}{\partial x_{i}}$ is calculated directly from Eq.~\eqref{eq:constrg3}.

If a filter with variable radius is considered, similarly for example to Eq.~\eqref{eq:densityfilt_var}, the derivative with respect to $x_{i}$ becomes more articulated.
In the numerical examples we will also consider variable minimum and maximum length scale controls.
In the following we provide example sensitivities to show the effect of filters with variable radius on the sensitivity.
If we assume that a variable minimum length scale control is considered, the derivatives of the objective function with respect to the projection shape coordinates $x_{i}$ becomes:
\begin{equation}\label{eq:der_fHvar}
\begin{split}
\frac{\partial f}{\partial x_{i}} =& \text{Eq.\eqref{eq:derphi5}}+\\
&+\sum_{k\in N_{j}} \left(\left(\frac{\partial \hat{f}}{\partial \bar{\rho}^{ero}_{k}}  \frac{\partial \bar{\rho}^{ero}_{k}}{\partial \tilde{\rho}_{k}} \frac{\partial \tilde{\rho}_{k}}{\partial \phi_{k}} \frac{\partial \phi_{k}}{\partial \tilde{d}^{2}_{k}} \right)\frac{\partial \tilde{d}^{2}_{k}}{\partial \bar{d}^{2}_{j}}  \right)
\frac{\partial \bar{d}^{2}_{j}}{\partial x_{i}} \\
=&\text{Eq.\eqref{eq:derphi5}}+\\
&+\sum_{k\in N_{j}} \left(H_{jk}^{\phi}\left(\frac{1}{H_{s,k}^{\phi}}\frac{\partial \hat{f}}{\partial \bar{\rho}^{ero}_{k}}  \frac{\partial \bar{\rho}^{ero}_{k}}{\partial \tilde{\rho}_{k}} \frac{\partial \tilde{\rho}_{k}}{\partial \phi_{k}} \frac{\partial \phi_{k}}{\partial \tilde{d}^{2}_{k}} \right)  \right)
\frac{\partial \bar{d}^{2}_{j}}{\partial x_{i}}.
\end{split}
\end{equation}
In Eq.~\eqref{eq:der_fHvar}:
\begin{equation}\label{eq:der_fHvar2}
\begin{split}
& \frac{\partial \tilde{\rho}_{i}}{\partial \phi_{i}}=
\frac{(d\textbf{H}\, \bm{\rho})_{i}/H_{s,i} - (\textbf{H}\,\bm{\rho})_{i}\,dH_{s,i}}{H_{s,i}^{2}}\\
& dH_{ij}=\frac{\partial  w_{G}(\Delta\textbf{x}_{ij},\phi_{i}) }{\partial  \phi_{i}}, \; dH_{s,i}=\sum_{j} dH_{ij}
\end{split}
\end{equation}
where $w_{G}$ was defined in Eq.~\eqref{eq:finalwg}.

Similarly, the gradient of $g_{4}$ with respect to $\bm{\rho}$ is:
\begin{equation}
\begin{split}
&\frac{\partial g_{4}}{\partial \rho_{i}} =  \sum_{j\in N_{i}}\left(\sum_{k\in N_{j}}\left(  \frac{\partial g_{4}}{\partial \hat{\rho}_{k}}\right)\frac{\partial\hat{\rho}_{k}}{\partial \bar{\rho}^{dil}_{j}} \frac{\partial \bar{\rho}^{dil}_{j}}{\partial \tilde{\rho}_{j}}\right)\frac{\partial \tilde{\rho}_{j}}{\partial \rho_{i}}\\
&=\sum_{j\in N_{i}}H_{ij}\left(\frac{1}{H_{s,j}}\sum_{k\in N_{j}}H_{jk}^{ls}\left( \frac{1}{H_{s,k}^{ls}} \frac{\partial g_{4}}{\partial \hat{\rho}_{k}}\right) \frac{\partial \bar{\rho}^{dil}_{j}}{\partial \tilde{\rho}_{j}}\right)
\end{split}
\end{equation}
and the gradient of $g_{4}$ with respect to $\textbf{x}$ is:
\begin{equation}
\begin{split}
&\frac{\partial g_{4}}{\partial x_{i}} =  
\sum_{k\in N_{j}} \left(\left(\frac{\partial g_{4}}{\partial \phi_{k}} \frac{\partial \phi_{k}}{\partial \tilde{d}^{2}_{k}} \right)\frac{\partial \tilde{d}^{2}_{k}}{\partial \bar{d}^{2}_{j}}  \right)
\frac{\partial \bar{d}^{2}_{j}}{\partial x_{i}}+\\
&+\sum_{k\in N_{j}} \left(\left(\frac{\partial g_{4}}{\partial \bar{\rho}^{ero}_{k}}  \frac{\partial \bar{\rho}^{ero}_{k}}{\partial \tilde{\rho}_{k}} \frac{\partial \tilde{\rho}_{k}}{\partial \phi_{k}} \frac{\partial \phi_{k}}{\partial \tilde{d}^{2}_{k}} \right)\frac{\partial \tilde{d}^{2}_{k}}{\partial \bar{d}^{2}_{j}}  \right)
\frac{\partial \bar{d}^{2}_{j}}{\partial x_{i}}\\
&=\sum_{k\in N_{j}} \left(H_{jk}^{\phi}\left(\frac{1}{H_{s,k}^{\phi}}\frac{\partial g_{4}}{\partial \phi_{k}} \frac{\partial \phi_{k}}{\partial \tilde{d}^{2}_{k}} \right)  \right)
\frac{\partial \bar{d}^{2}_{j}}{\partial x_{i}} +\\
&+\sum_{k\in N_{j}} \left(H_{jk}^{\phi}\left(\frac{1}{H_{s,k}^{\phi}}\frac{\partial g_{4}}{\partial \bar{\rho}^{ero}_{k}}  \frac{\partial \bar{\rho}^{ero}_{k}}{\partial \tilde{\rho}_{k}} \frac{\partial \tilde{\rho}_{k}}{\partial \phi_{k}} \frac{\partial \phi_{k}}{\partial \tilde{d}^{2}_{k}} \right)  \right)
\frac{\partial \bar{d}^{2}_{j}}{\partial x_{i}}
\end{split}
\end{equation}
in which the results of Eq.~\eqref{eq:der_fHvar2} still hold.

\section{Numerical examples}
\label{sec:examples}
In this section we discuss several numerical examples that show the capability of the proposed mixed projection- and density-based topology optimization approach.
The optimization algorithm is based on a iterative nested approach, where we use the Method of Moving Asymptotes \citep{svanberg1987method}. 
The tight coupling between density and geometric shape variables makes it quite challenging to obtain a smooth convergence towards final optimized designs. Hence, during the optimization process we consider conservative moving limits equal to 0.2 for the density variables (i.e.~$\rho$) and to 0.005 for the geometric variables that define the shape of the projection profile (i.e.~$x$).

In the following numerical examples we consider the continuation schemes shown in Table \ref{tab:contscheme} with steps of $50$ iterations.
In Table \ref{tab:contscheme}, $p_{E}$ is the penalization parameter of the SIMP interpolation scheme (Eq.~\eqref{eq:simp}); $\mu_{fil}$ defines the sharpness of the Super-Gaussian function used for the projection (Eq.~\eqref{eq:finalphi}); and $\beta_{HS}$ defines the sharpness of the projection of the design variables $\tilde{\rho}$ in the dilated, intermediate, and eroded layouts (Eq.~\eqref{eq:rhoproject}).
In all the examples, $E_{max}$ is equal to $1$, and $E_{min}$ is equal to $10^{-6}$ to avoid singularities in the stiffness matrix. The Poisson's ratio considered is $\nu=0.3$.
Moreover, $\eta_{ero}=0.6$ and $\eta_{dil}=0.4$.
During the optimization runs, every $25$ iterations the allowable volume fractions of the dilated layout are updated such that eventually the volume fraction of the intermediate density meets its allowable value.

\begin{figure}[H]
\centering
  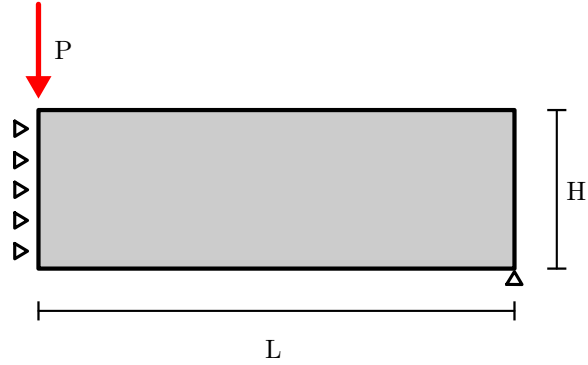
\caption{MBB beam structural scheme, $H/L=1/3$}   
\label{fig:mbb_scheme}
\end{figure}

\begin{figure}[H]
\centering
  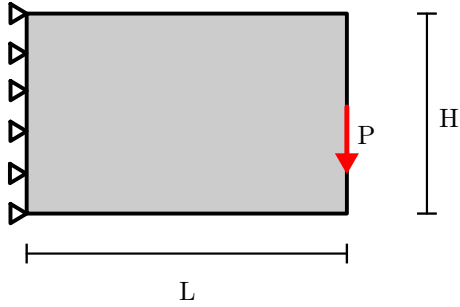
\caption{Short cantilever structural scheme, $H/L=2/3$}   
\label{fig:canti_scheme}
\end{figure}

In the following numerical applications we consider two structural systems: The MBB beam with height H and length L with a ratio of $H/L=1/3$ shown in Fig.~\ref{fig:mbb_scheme}, and a short cantilever with a height to length ratio equal to $H/L=2/3$ shown in Fig.~\ref{fig:canti_scheme}.
We first provide reference results for the two structural systems that are used as a base for a systematic comparison on the different design cases that will be discussed.
In both cases we consider a ``domain extension approach'' discussed in \cite{clausen2017filter} to avoid boundary effects. In the figures, the dashed black line represents the outer boundaries of the domain, including the added padding. 
The blue dashed line represents the blueprint layout, where the actual design is located. 
To obtain the reference results we set $\eta_{ero}=0.75$ and $\eta_{dil}=0.25$.

\begin{figure}[H]
\centering
  \includegraphics[width=0.6\columnwidth,trim=0.5in 1.1in 0.5in 1.1in,clip]{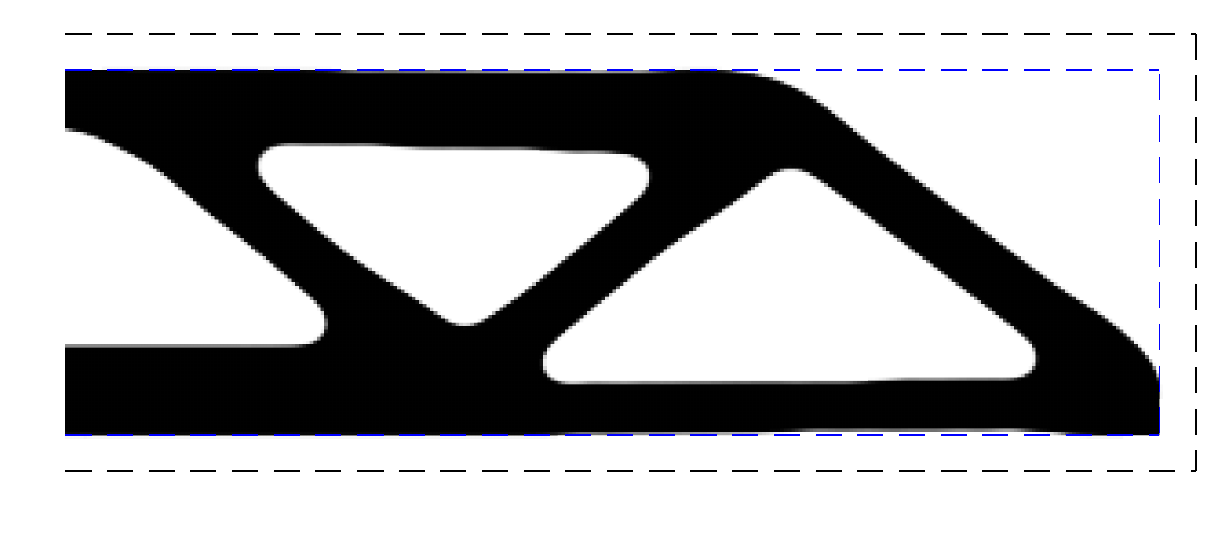}
\caption{MBB beam. Reference results: Compliance $f= 196.44$, and solid volume fraction $V=0.4009$ }   
\label{fig:mbb_refernce}
\end{figure}

\begin{figure}[H]
\centering
  \includegraphics[width=0.4\columnwidth,trim=0.5in 0.5in 0.5in 0.48in,clip]{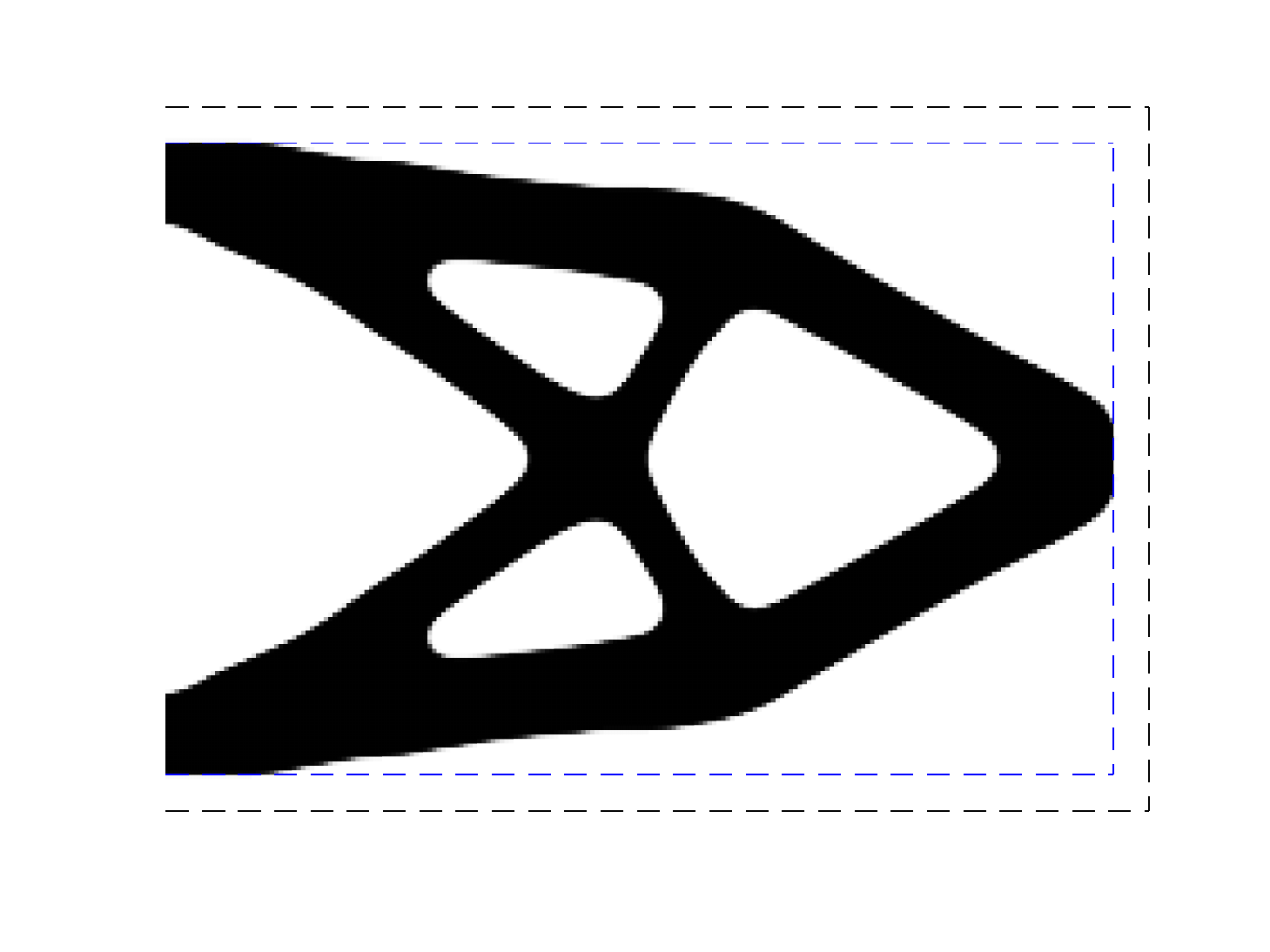}
\caption{Clamped beam. Reference results: Compliance $f=39.89$, and solid volume fraction $V=0.3505$}
\label{fig:clamped_refernce}
\end{figure}

The MBB beam is discretized with $100\times 300$ finite elements, considering a density filter radius $r=10$ elements, a constant SIMP penalization $p_{E}=1$, and a volume constraint of $40\%$. The optimization ran for $550$ iterations. The optimized topology is shown in Fig.~\ref{fig:mbb_refernce} and its final compliance is $f= 196.44$.
The short cantilever beam is discretized with $140\times 210$ finite elements, considering a density filter radius $r=8$ elements, a constant SIMP penalization $p_{E}=1$, and a volume constraint of $35\%$. The optimization ran for $550$ iterations. The optimized topology is shown in Fig.~\ref{fig:clamped_refernce} and its final compliance is $f= 39.89$. In both cases, the load $P=1$.

\setlength{\tabcolsep}{3pt}
\begin{table*}[htbp]
  \centering\scriptsize
  \caption{Continuation schemes for the numerical examples. Each step consists of $50$ optimization iterations}
    \begin{tabular}{c|ccc|ccc|ccc|ccc|ccc}
    
    & \multicolumn{3}{c}{Reference } & \multicolumn{3}{c}{Sec. \ref{sec:ex1}, \ref{sec:ex2}} & \multicolumn{3}{c}{Sec. \ref{sec:ex3-1}} & \multicolumn{3}{c}{Sec. \ref{sec:ex4.1}} & \multicolumn{3}{c}{Sec. \ref{sec:ex4.2},\ref{sec:ex4.3}}   \\
\hline
    Step &   $p_{E}$     & $\mu_{fil}$  & $\beta_{HS}$   & $p_{E}$     & $\mu_{fil}$  & $\beta_{HS}$   & $p_{E}$     & $\mu_{fil}$  & $\beta_{HS}$   & $p_{E}$     & $\mu_{fil}$  & $\beta_{HS}$   & $p_{E}$     & $\mu_{fil}$  & $\beta_{HS}$ \\
    1     & 1.00  & 1.00  & 1.00  & 1.00  & 1.25  & 1.00  & 1.00  & 1.50  & 2.00  & 1.00  & 1.25  & 1.50  & 1.25  & 1.25  & 1.00 \\
    2     & 1.00  & 1.41  & 1.46  & 1.50  & 1.77  & 1.46  & 1.50  & 2.12  & 2.93  & 1.50  & 1.77  & 2.52  & 1.50  & 1.77  & 1.41 \\
    3     & 1.00  & 2.00  & 2.14  & 2.00  & 2.50  & 2.14  & 2.00  & 3.00  & 4.29  & 2.00  & 2.50  & 4.24  & 2.00  & 2.50  & 2.00 \\
    4     & 1.00  & 2.83  & 3.14  & 2.50  & 3.54  & 3.14  & 2.50  & 4.24  & 6.28  & 2.50  & 3.54  & 7.14  & 2.50  & 3.54  & 2.83 \\
    5     & 1.00  & 4.00  & 4.59  & 3.00  & 5.00  & 4.59  & 3.00  & 5.00  & 9.19  & 3.00  & 5.00  & 12.00 & 3.00  & 5.00  & 4.00 \\
    6     & 1.00  & 5.00  & 6.73  & 3.00  & 5.00  & 6.73  & 3.00  & 5.00  & 13.45 & 3.00  & 5.00  & 20.18 & 3.00  & 5.00  & 5.66 \\
    7     & 1.00  & 5.00  & 9.85  & 3.00  & 5.00  & 9.85  & 3.00  & 5.00  & 19.70 & 3.00  & 5.00  & 33.94 & 3.00  & 5.00  & 8.00 \\
    8     & 1.00  & 5.00  & 14.42 & 3.00  & 5.00  & 14.42 & 3.00  & 5.00  & 28.84 & 3.00  & 5.00  & 57.08 & 3.00  & 5.00  & 11.31 \\
    9     & 1.00  & 5.00  & 21.11 & 3.00  & 5.00  & 21.11 & 3.00  & 5.00  & 32.00 & 3.00  & 5.00  & 96.00 & 3.00  & 5.00  & 16.00 \\
    10    & 1.00  & 5.00  & 30.91 & 3.00  & 5.00  & 30.91 & 3.00  & 5.00  & 32.00 & 3.00  & 5.00  & 100.00 & 3.00  & 5.00  & 22.63 \\
    11    & 1.00  & 5.00  & 32.00 & 3.00  & 5.00  & 32.00 & 3.00  & 5.00  & 32.00 & 3.00  & 5.00  & 100.00 & 3.00  & 5.00  & 32.00 \\
    12    &    &   &   &    &    &   &    &    &    & 3.00  & 5.00  & 100.00 &    &    &   \\
    \hline
    \end{tabular}%
  \label{tab:contscheme}%
\end{table*}%

Table \ref{tab:optsettings} lists additional settings for the optimization algorithms adopted in the numerical examples.
In particular: $r_{min}$ is the density filter radius for the minimum length scale control; $r_{\phi}$ is the filter radius used to filter the distance field of the elements with respect to the projection profile; $r_{max}$ is the radius of the maximum length scale control; $\beta_{fil}$ is the distance from the projection profile in each direction and it defines the area of domain affected by the projection; $g_{0}^{*}$ is the allowable volume fraction for the constraint $g_{0}$, and $g_{0,dil}^{*}$ (Eq.~\eqref{eq:control0}) is equal to $1.05\,g_{0}^{*}$ ; $\alpha$ is the volume fraction allowed in the maximum length scale control in each bubble with radius $r_{max}$; $\gamma$ is the amplification factor for $r_{min}$ or $r_{max}$ used to obtain variable length scale controls. 
\begin{table}[htbp]
  \centering\scriptsize
  \caption{Settings adopted in the optimization algorithms of the numerical examples}
    \begin{tabular}{lccccccc}
    \hline
     Sec.     &  \ref{sec:ex1}   & \ref{sec:ex2}   & \ref{sec:ex3-11} & \ref{sec:ex3-12} & \ref{sec:ex4.1} & \ref{sec:ex4.2} & \ref{sec:ex4.3} \\
    $r_{min}$  & 10    & 10    & 3     & 3     & 2-10  & 3     & 3-6 \\
    $r_{\phi}$  & 10    & 10    & 3     & 3     & 2     & 3     & 3 \\
    $r_{max}$  & n/a   & n/a   & 5     & 5     & n/a   & 7-14  & 20 \\
    $\beta_{fil}$  & 20    & 20    & 10-5  & 10-5  & 15    & 20    & 20 \\
    $g_{0}^{*}$   & 0.4   & 0.4   & 0.35  & 0.35  & 0.4   & 0.4   & 0.4 \\
    $\alpha$ & n/a   & n/a   & 0.5   & 0.5   & n/a   & 0.6   & 0.6 \\
    $\gamma$ & n/a   & n/a   & n/a   & n/a   & 4     & 1     & 1 \\
    \hline
    \end{tabular}%
  \label{tab:optsettings}%
\end{table}%

All the following examples are obtained solving Problem \eqref{eq:topoptprob} with at least two constraints.
The first is the total volume constraint $g_{0}$.
The second is $g_{3}$ defined in Eq.~\eqref{eq:constrg3}, and it is used in all cases to obtain more regular shapes of the optimized projection profile, by controlling the slopes of the segments composing the projection profile. In particular, we consider an exponent for the $p$-norm in Eq.~\eqref{eq:constrg3} equal to $10$, and a maximum allowed inclination angle of $60$ degrees.
All the numerical analyses were performed on a Linux machine with $8$ Gb of RAM and a dual-core Intel $i7$ CPU at $2.00$ GHz.

\subsection{Example 1: MBB beam with local volume constraint}
\label{sec:ex1}
\begin{figure}[H]
\centering
  \includegraphics[width=0.6\columnwidth,trim=0.5in 1.1in 0.5in 1.1in,clip]{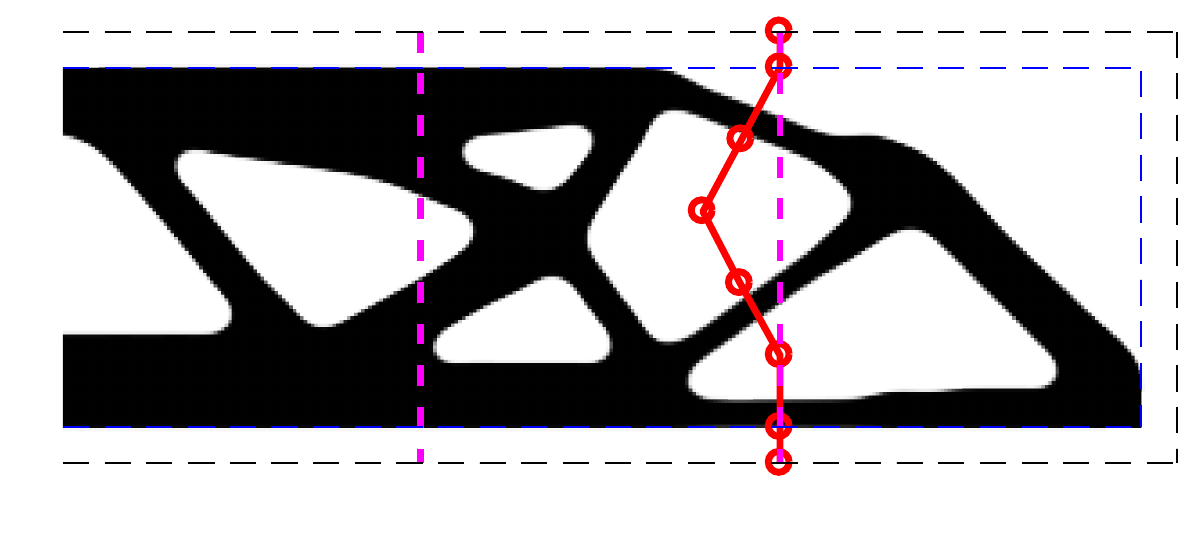}
\caption{Optimized MBB beam of Sec. \ref{sec:ex1} with local volume constraint. Final compliance $f= 202.13$, and solid volume fraction $V=0.4021$} 
\label{fig:ex1final}
\end{figure}

In this example we optimize the MBB beam solving problem \eqref{eq:topoptprob} with two volume constraints.
The first is the total volume constraint $g_{0}$. The second is a local volume constraint applied on the projection area $g_{1}$, which was discussed in Sec. \ref{sec:control}.
The allowed volume fraction of the intermediate density layout associated to $g_{0}$ is $40\%$, and the volume fraction associated to $g_{1}$ is $25\%$.
As it was already mentioned before, we consider also a third constraint to obtain more regular shapes of the optimized projection profile, namely $g_{3}$ which has been defined in Eq.~\eqref{eq:constrg3}.
The projection profile is composed of $5$ segments, each spanning $20$ elements in the vertical directions.
The $6$ nodes at the segments ends have fixed $y$ coordinates, and variable $x$ coordinates.
The geometric variables $x_{i}$ with $i=1,...,6$ are initially set to $0.5$ and their bounds are $x_{lb}=0.33$ and $x_{ub}=0.67$. These bounds correspond to stretches of $50$ elements in the negative and positive horizontal directions.
They are represented by  dashed lines colored in magenta in Fig. \ref{fig:ex1final}.
The motivation for defining this allowable region for the interface stems from manufacturing limitations: we consider a case where the whole beam cannot be manufactured in one part, and must be split into two parts whose maximum length is 2/3 of the total length.
Then, one would seek the optimal interface location that minimizes compliance while considering properties or limitations on the interface region.

Regarding the parameters that characterize the projection, in this example we consider a radius of the filter \eqref{eq:filterphi} $r_{\phi}=10$ elements, and a distance from the projection profile $\beta_{fil}=20$ elements on each side used to define the area on which the local volume constraint $g_{1}$ is applied. Moreover, we initialize the parameter $\mu_{fil}$ of Eq.~\eqref{eq:finalphi} to $1.25$, and increase it by steps of $\sqrt{2}$ at each continuation step up to a maximum value of $\mu_{fil}^{max}=5$ as shown in Table \ref{tab:contscheme}.
The parameters that define the robust approach described in Sec. \ref{subsec:densityparam} are set as follows: $\eta_{ero}=0.6$ and $\eta_{dil}=0.4$.
With this numerical example we want to simulate the search for an optimized topology with a volume constraint modified locally in the projection area.
This requirement may translate the need for reducing the structural elements' size at the interface of different parts manufactured separately and subsequently assembled.
\begin{figure}[H]
\centering
  \includegraphics[width=0.6\columnwidth,trim=0.5in 1.1in 0.5in 1.1in,clip]{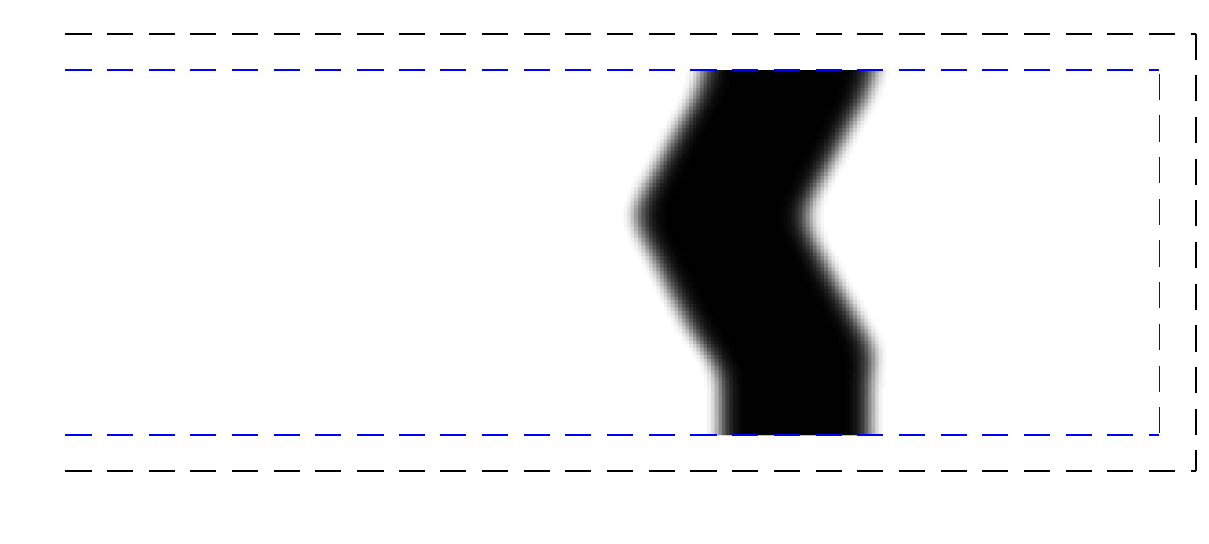}
\caption{Optimized projection area of the MBB beam of Sec. \ref{sec:ex1} with local volume constraint} 
\label{fig:piex1}     
\end{figure}
The final optimized design has a compliance $f=202.13$ and a volume of the intermediate layout of $V^{int}=40.21\%$. 
The optimized topology is shown in Fig.~\ref{fig:ex1final}, where each red circle represents a node of the profile whose horizontal coordinate $x$ is one of the geometric variables. Fig.~\ref{fig:piex1} shows the optimized shape of the projection profile and its associated projection area defined by the variables $\phi_{i}$.
As can be observed in Fig.~\ref{fig:ex1final}, by imposing a reduced amount of material in a sub-portion of the design domain we obtain an optimized design with thinner members in that portion of the domain, and hence a more flexible structure with poorer performance compared to the reference results. 
More precisely, compared to the reference design of Fig.~\ref{fig:mbb_refernce}, the obtained design shows a loss in performance of approximately $2.9\%$, even though both structures have about $40\%$ final solid volume fraction.
At the same time, the amount of material in the interface region of the result of Fig.~\ref{fig:ex1final} is smaller compared to the reference result. 
This implies that the effort invested in assembly (for example, total welding energy) is expected to reduce compared to the reference result.

\subsection{Example 2: MBB beam with a localized Modified SIMP}
\label{sec:ex2}
In this example we optimize the MBB beam solving problem \eqref{eq:topoptprob} with the local Modified SIMP defined in Eq.~\eqref{eq:simp2}, considering $r_{E}=0.5$. 
This implies that the material distributed in the projection area has a Young's modulus that is half of $E_{max}$.
The two constraints $g_{0}$ and $g_{3}$ are also considered in the optimization analysis for this example.
Moreover, the projection profile is made of $5$ segments spanning $20$ elements in the vertical directions.
The $6$ nodes at the segments ends have fixed $y$ coordinates, and variable $x$ coordinates.
The geometric variables $x_{i}$ with $i=1,...,6$ are initially set to $0.5$ and their bounds are $x_{lb}=0.33$ and $x_{ub}=0.67$. These bounds correspond to stretches of $50$ elements in the negative and positive horizontal directions.
All the remaining parameters are set according to Table \ref{tab:contscheme} and Table \ref{tab:optsettings}.
This example shows the possibility of optimizing a structure with an interface between substructures also simultaneously optimized, taking into account the presence of welded material with different mechanical properties connecting the different substructures.
It should be noted that the interface between the substructures is defined by the variable shape of the projection profile.

\begin{figure}[h]
\centering
  \includegraphics[width=0.6\columnwidth,trim=0.5in 1.1in 0.5in 1.1in,clip]{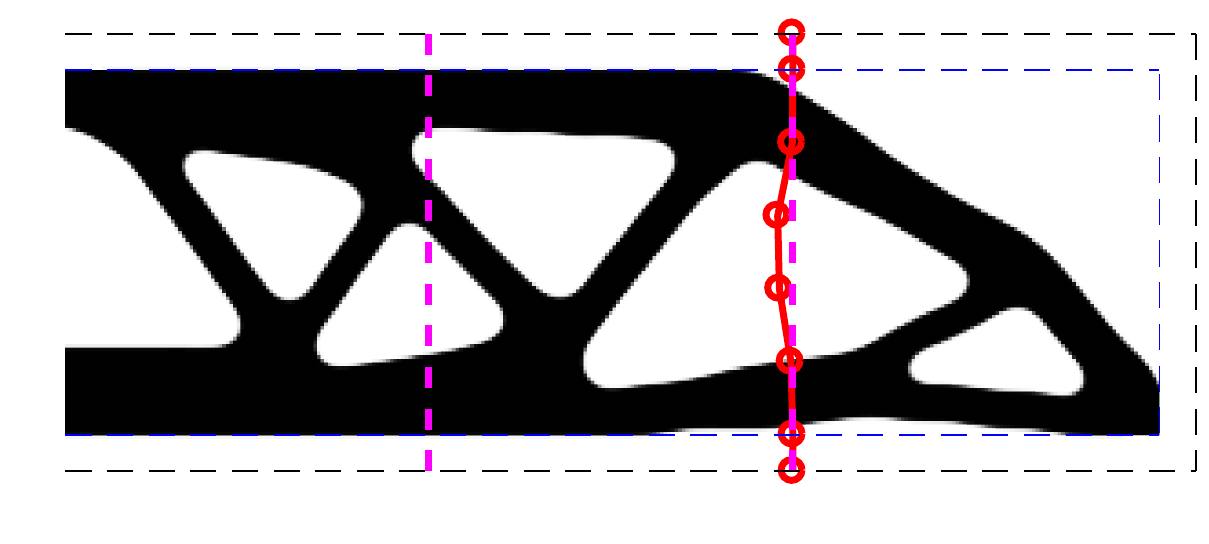}
\caption{MBB beam of Sec. \ref{sec:ex2} with local Modified SIMP. Final compliance $f=  207.13$, and solid volume fraction $V=0.4009$} 
\label{fig:ex2rho}
\end{figure}
\begin{figure}[h]
\centering
  \includegraphics[width=0.6\columnwidth,trim=0.5in 1.1in 0.5in 1.1in,clip]{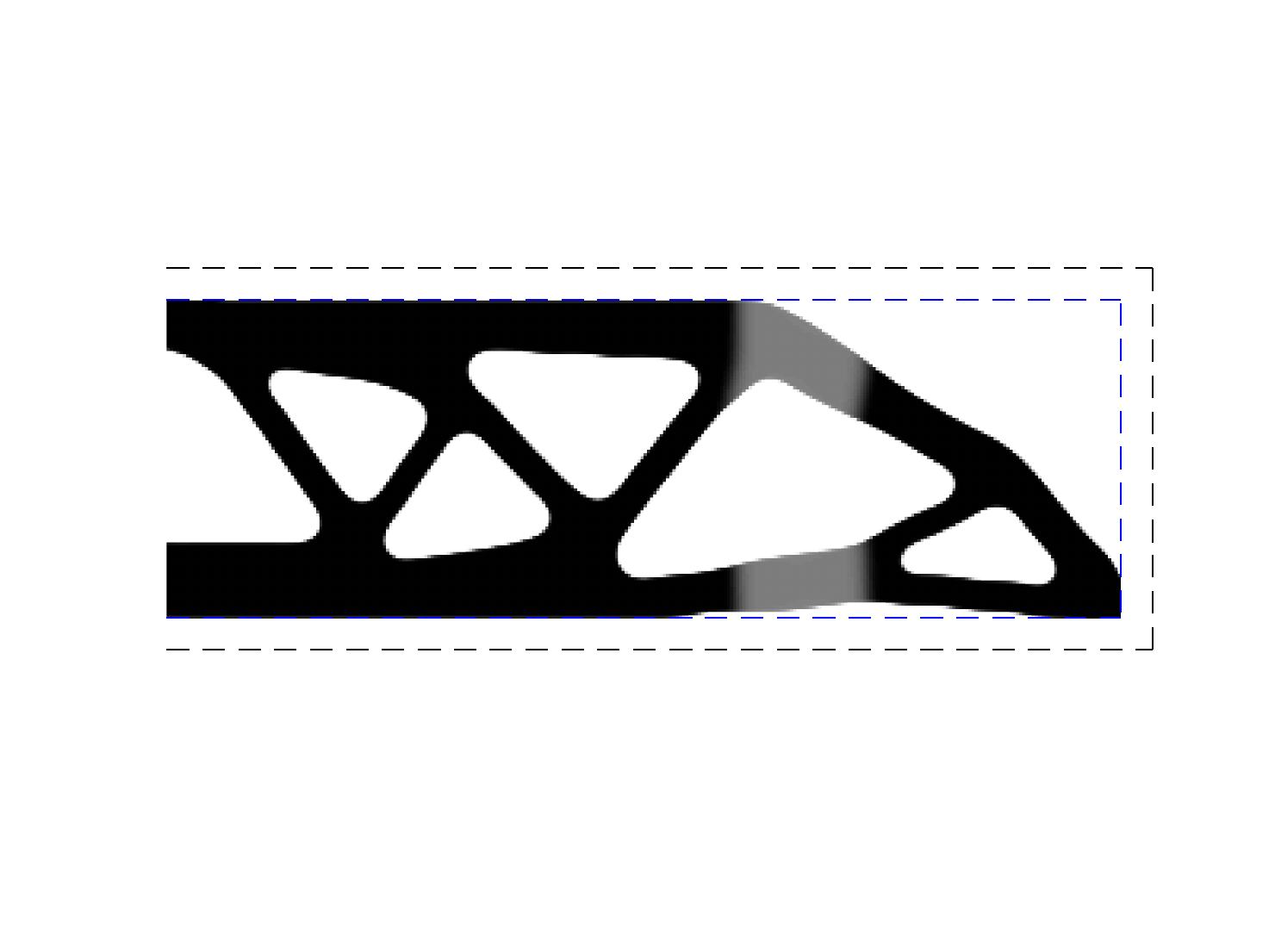}
\caption{MBB beam of Sec. \ref{sec:ex2} with local Modified SIMP. Plot of the Young's modulus $E(\bm{\rho},\bm{\phi})$ associated to the intermediate density field} 
\label{fig:ex2young}
\end{figure}
\begin{figure*}[h]
    \centering
    \begin{subfigure}[t]{0.475\textwidth}
        \centering
        \includegraphics[width=0.95\columnwidth,trim=0.5in 1.1in 0.5in 1.1in,clip]{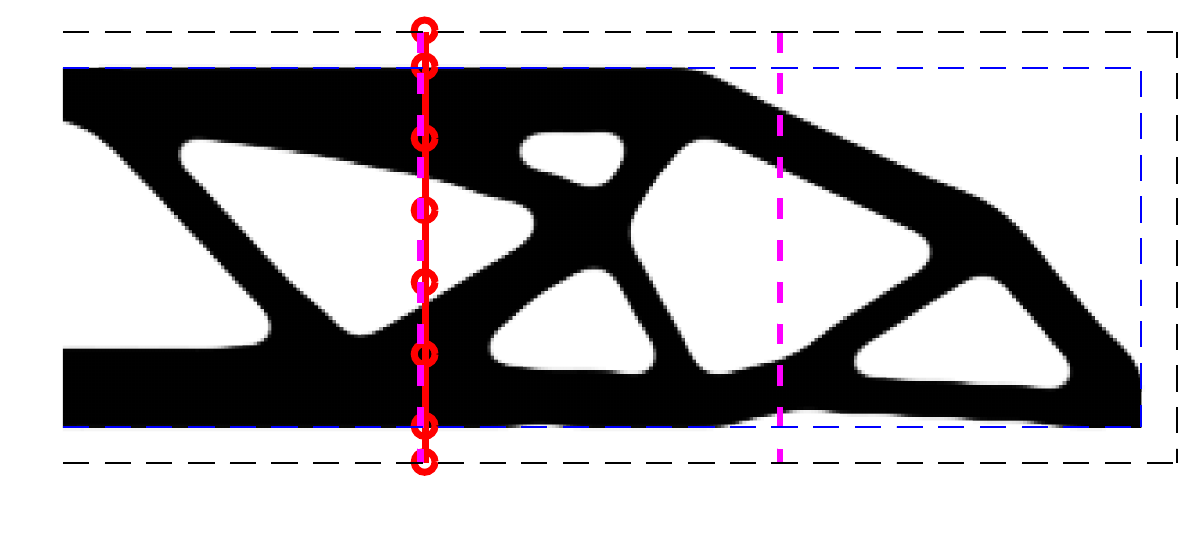}
        \caption{}
				\label{subfig:ex2rho_1}
    \end{subfigure}%
    ~ 
    \begin{subfigure}[t]{0.475\textwidth}
        \centering
        \includegraphics[width=0.95\columnwidth,trim=0.5in 1.1in 0.5in 1.1in,clip]{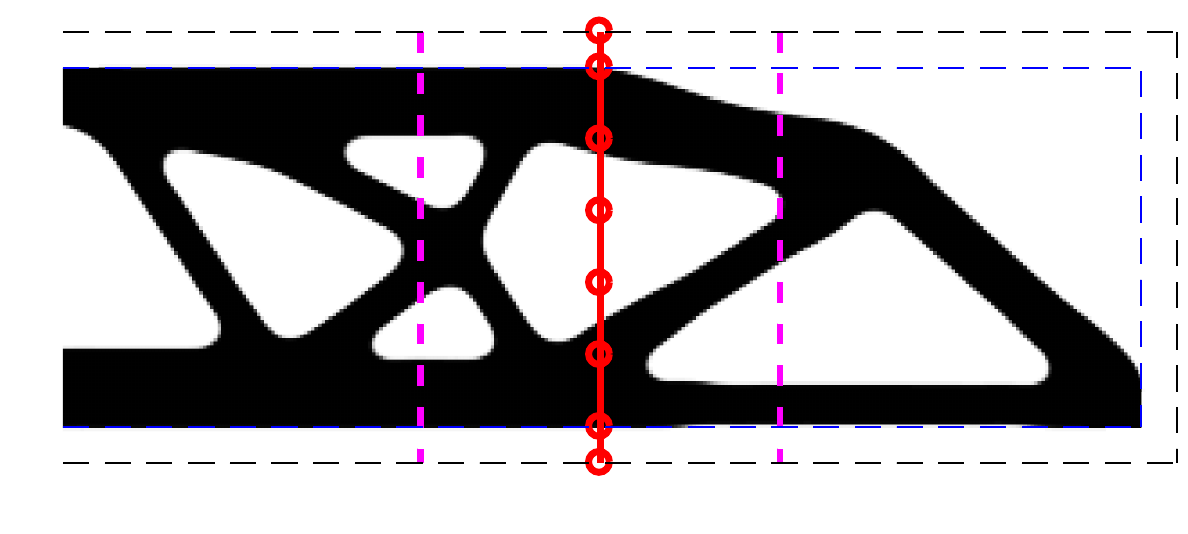}
        \caption{}
				\label{subfig:ex2rho_2}
    \end{subfigure}
    \caption{MBB beam of Sec. \ref{sec:ex2} with local Modified SIMP. (a) Fixed $x_{i}=x_{lb}$. Final compliance $f=  220.70$, and intermediate solid volume fraction $V=0.4008$. (b) Fixed $x_{i}=0.5$. Final compliance $f=  212.43$, and solid volume fraction $V=0.4008$ }
		\label{fig:ex2young_II}
\end{figure*}

Fig.~\ref{fig:ex2rho} shows the optimized topology and the optimized configuration of the projection profile.
The optimized structure has a compliance $f=207.13$ which is $5.4\%$ higher than the reference design of Fig.~\ref{fig:mbb_refernce}. 
The final solid volume of the intermediate layout is $V=40.09\%$.
Fig.~\ref{fig:ex2young} shows the plot of the Young's modulus of each solid element of the optimized structure. 
The black elements have a Young modulus equal to $E_{max}$. 
The gray elements are inside the area of influence of the projection profiles, and have a reduced Young's modulus equal to $E_{\phi}=0.5E_{max}$.
It is interesting to observe that the optimizer attempts to place the projection profile in the optimized configuration perpendicularly to the intersected structural elements.
We presume that by doing this, the optimizer tries to reduce the extent of poor material assigned to the structural elements.
Moreover, in Fig~\ref{fig:ex2rho} it is possible to also observe that the shape of the projection profile is very close to the upper bound of the variables $x_{i}$, namely $x_{ub}$, and almost straight.
To further investigate the solution identified by the optimizer we run two additional optimization analyses where we keep the shape variables $x_{i}$ fixed.
In one case we set $x_{i}=x_{lb}$ for $i=1,\dots,6$ and we obtain the final optimized topology shown in Fig.~\ref{subfig:ex2rho_1}. 
In the second case we set $x_{i}=0.5$ for $i=1,\dots,6$ which leads to the final optimized topology shown in Fig.~\ref{subfig:ex2rho_2}.
The performances of the optimized designs shown in  Fig.~\ref{fig:ex2young_II} are both inferior to the performance of our optimized design shown in Fig.~\ref{fig:ex2rho}.
This can explain why the optimizer placed the projection profile close to the upper bound $x_{lb}$ as shown in Fig.~\ref{fig:ex2rho}, where presumably the negative effect of the weak material is minimized.

\subsection{Example 3: Short cantilever with a localized maximum length scale control}
\label{sec:ex3-1}
We consider now the optimization of a short cantilever with two projection profiles. 
In particular, we consider two cases: one case in which the two profiles are vertical, and another case in which one profile is horizontal and another one is vertical.
Besides the usual total volume constraint $g_{0}$ and the constraint on the slope of the profiles' segments $g_{3}$, in this case we consider also the localized volume constraint for local maximum length scale control $g_{2}$ which has been introduced in Sec.~\ref{sec:control}. 
For the numerical example of this section, the moving limit for the density variables is set to $0.1$. 
The moving limit for the geometric variables is set here to $0.005$. 
This parameter modification was required because during initial numerical experiments the algorithm showed some difficulties in converging towards near discrete final designs.
In fact, in this example the portion of domain on which additional controls are imposed is more extended, and this increases the difficulty in converging towards optimized and feasible design solutions.
The allowable volume fraction of the intermediate layout is initially set to $35\%$. 
The allowable volume fraction for the maximum length scale control (the parameter $\alpha$ in Eq.~\eqref{eq:lengthscaleconst2}) is set to $50\%$. 
In both the following examples, the exponent $q$ of the max approximation of Eq.~\eqref{eq:distprojmax} is set to $10^{6}$.
The distance from the projection profile is set initially to $\beta_{fil}=10$, and decreased to $5$ at the first continuation scheme step. 

\subsubsection{Short cantilever with two vertical profiles}
\label{sec:ex3-11}
The first case considered is the short cantilever with two vertical profiles.
The settings for the continuation scheme and the optimization algorithm are shown in Table \ref{tab:contscheme} and Table \ref{tab:optsettings}.
The two vertical profiles are made of $7$ segments spanning $20$ elements.
The $8$ nodes at the segments ends have fixed $y$ coordinates, and variable $x$ coordinates.
The geometric variables of the two profiles $x^{1}_{i}$ and $x^{2}_{i}$ have been initially set to $0.25$ and $0.75$. 
Their bounds are defined by gaps of $30$ elements with respect to the original positions such that $x^{1}_{lb}=0.11$ and $x^{1}_{ub}=0.40$, and 
similarly $x^{2}_{lb}=0.61$ and $x^{2}_{ub}=0.90$.
\begin{figure*}[h]
    \centering
    \begin{subfigure}[t]{0.45\textwidth}
        \centering
        \includegraphics[width=0.85\columnwidth,trim=0.5in 0.35in 0.5in 0.35in,clip]{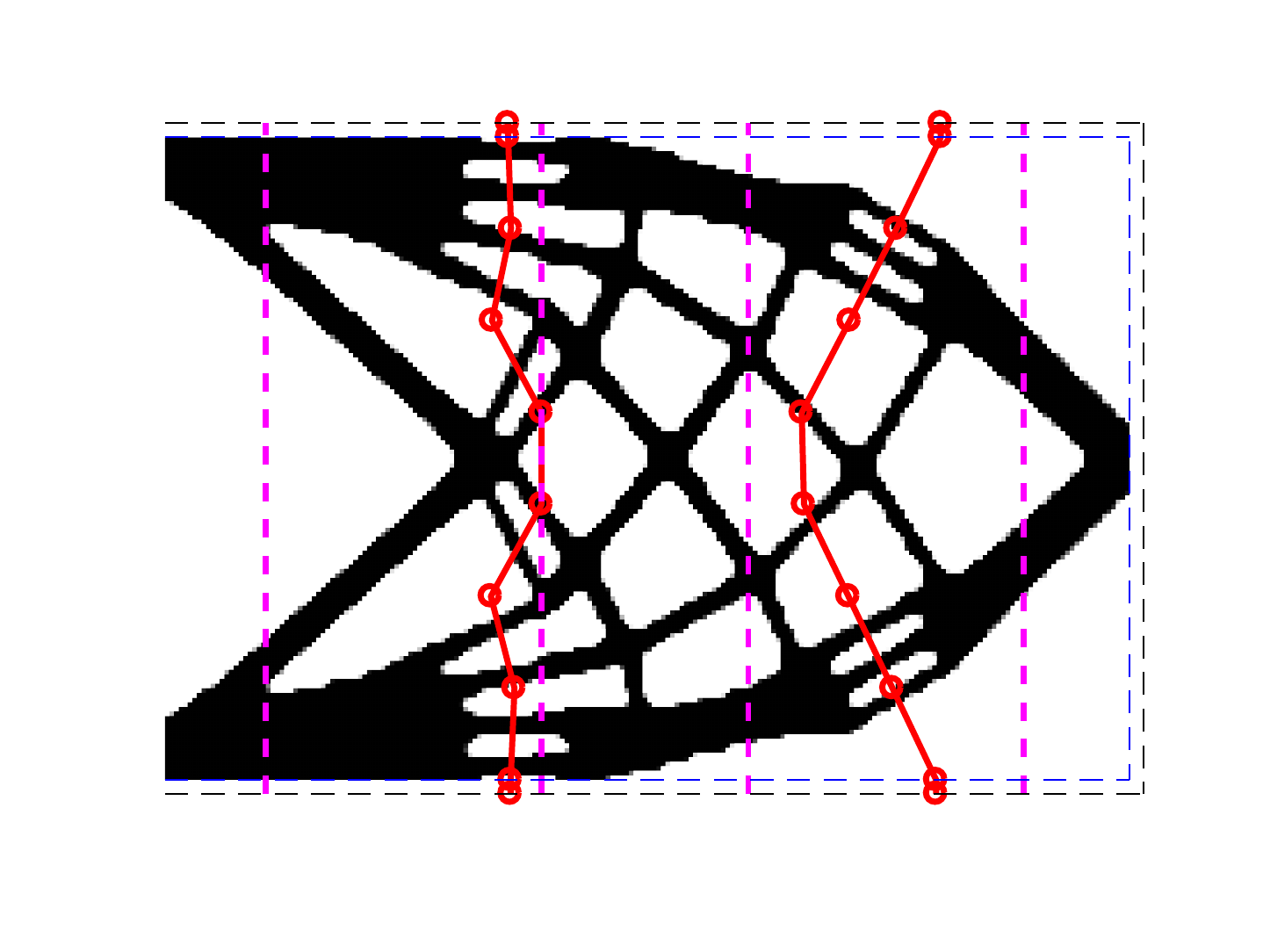}
        \caption{ }
				\label{subfig:ex3cantiVHrho}
    \end{subfigure}%
    ~ 
    \begin{subfigure}[t]{0.45\textwidth}
        \centering
        \includegraphics[width=0.85\columnwidth,trim=0.5in 0.35in 0.5in 0.35in,clip]{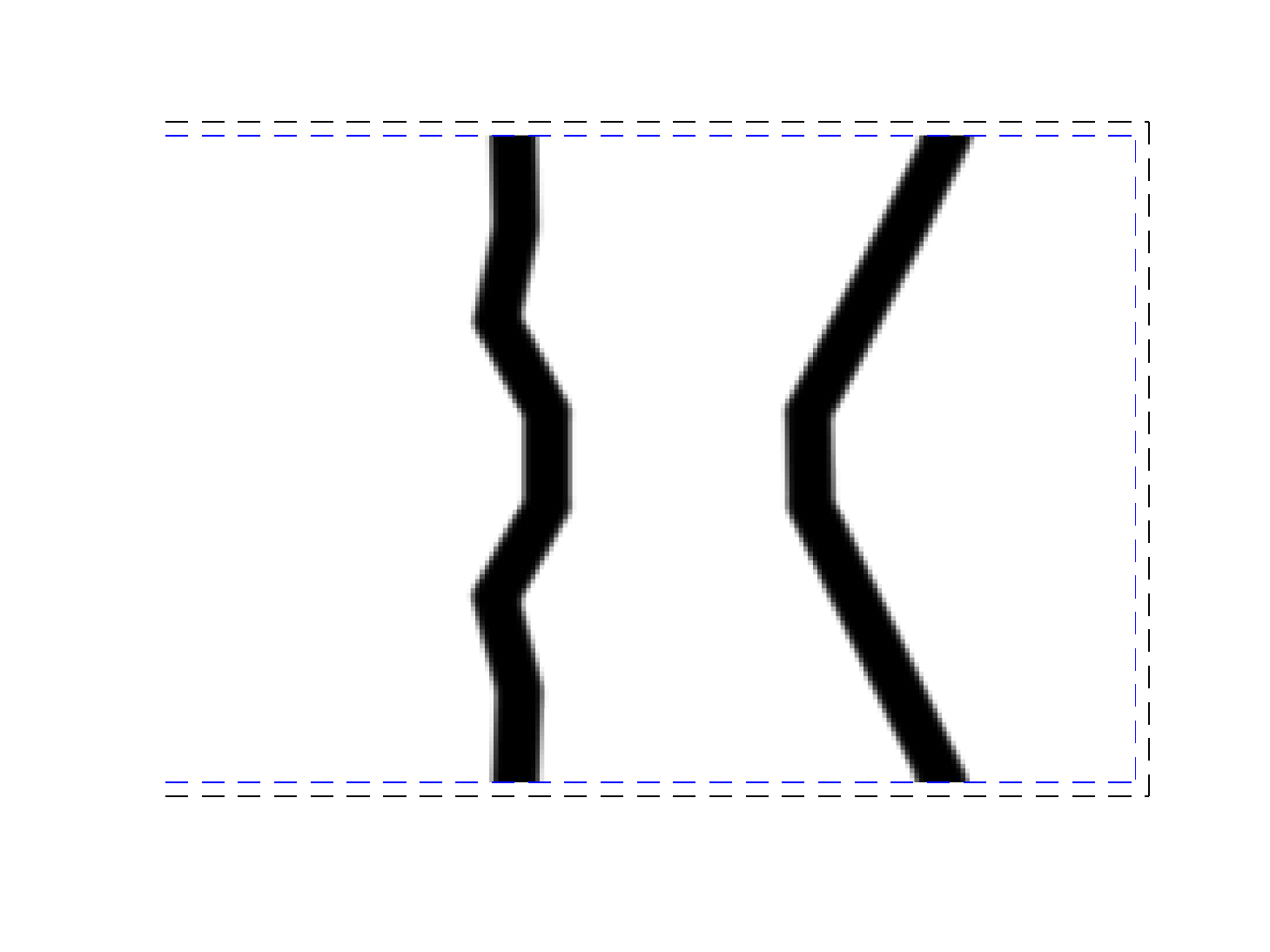}
        \caption{ }
				\label{subfig:ex3cantiVHphi}
    \end{subfigure}
    \caption{Short cantilever with two vertical projection profiles. (a) Optimized topology and configuration of the two profiles. (b) Final shape of the projection areas for which $\phi=1$. Final compliance $f=44.84$, and solid volume fraction $V=0.3501$}
		\label{fig:ex3_1}
\end{figure*}
Fig.~\ref{fig:ex3_1} shows the final optimized topology obtained after $550$ optimization iterations, and the final shape and configuration of the two projection profiles.
Fig.~\ref{fig:ex3_1} also displays the upper and lower bounds of the shape variables that are represented by dashed lines colored in magenta.
The optimized structure has a compliance $f=44.84$ ($12.4\%$ higher than the reference design of Fig.~\ref{fig:clamped_refernce}), and a final solid volume of the intermediate layout $V=35.01\%$.
The shape nodal variables are represented also in this case by red circles.
In particular, in Fig.~\ref{subfig:ex3cantiVHrho} it is possible to observe smaller structural features in the vicinity of the profiles as a direct consequence of the imposed local maximum length scale control $g_{3}$ defined in Eq.~\eqref{eq:lengthscaleconst}. 
This specific feature control may represent for example the requirement of having smaller elements that need to be connected in correspondence of an interface of different parts assembled. 
From a visual perspective, this topological layout resembles recent results achieved with graded porosity \citep{schmidt2019structural}. 
However, the underlying formulations are quite different. Primarily, we designate the region for imposing maximum length scale using an explicit geometry, whereas in \cite{schmidt2019structural} the authors propose manual control, auxiliary density field control or physics-based control.

\begin{figure*}[h]
    \centering
    \begin{subfigure}[t]{0.45\textwidth}
        \centering
        \includegraphics[width=0.85\columnwidth,trim=0.5in 0.35in 0.5in 0.35in,clip]{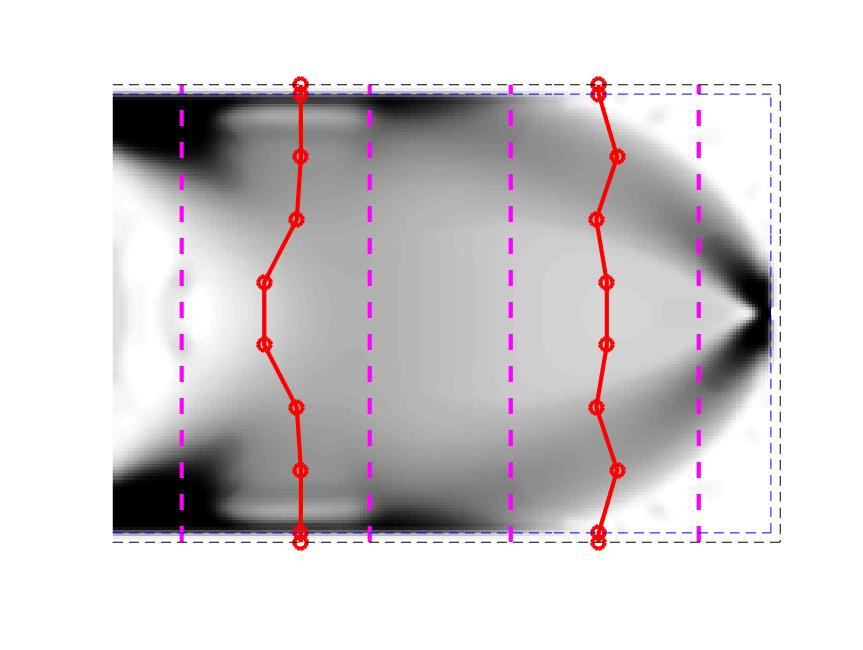}
        \caption{Iter $= 10$}
    \end{subfigure}%
    ~ 
    \begin{subfigure}[t]{0.45\textwidth}
        \centering
        \includegraphics[width=0.85\columnwidth,trim=0.5in 0.35in 0.5in 0.35in,clip]{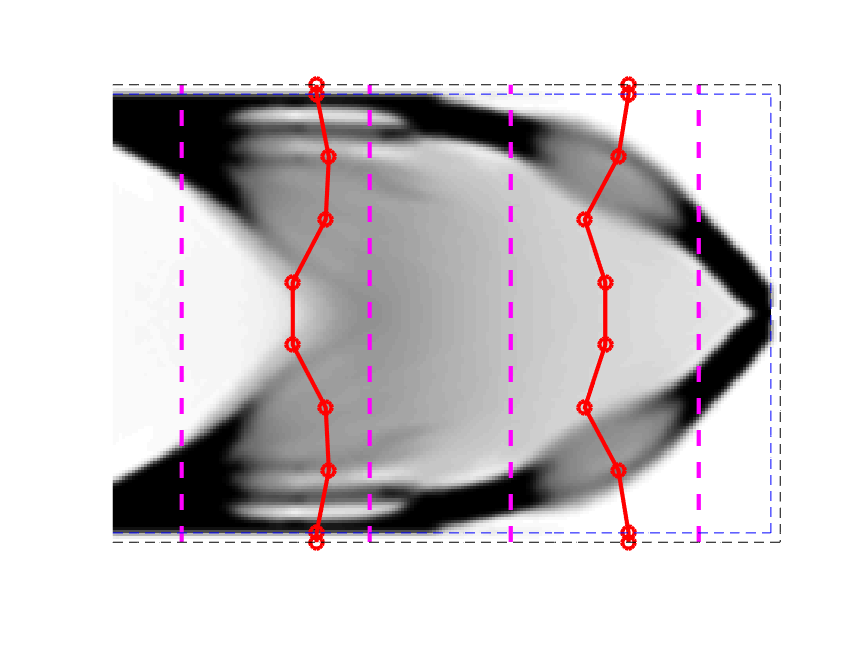}
        \caption{Iter $= 25$}
    \end{subfigure}
     ~
    \begin{subfigure}[t]{0.45\textwidth}
        \centering
        \includegraphics[width=0.85\columnwidth,trim=0.5in 0.35in 0.5in 0.35in,clip]{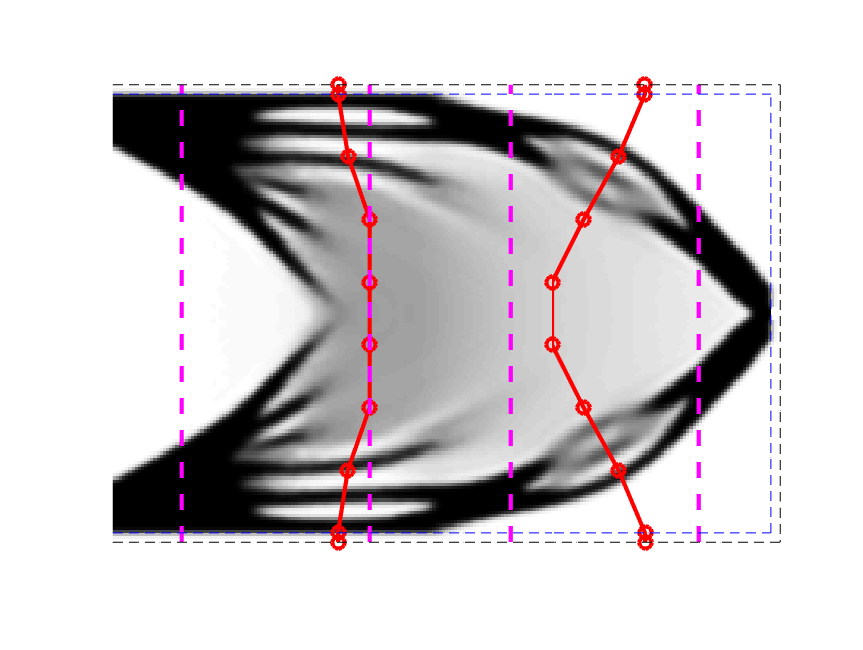}
        \caption{Iter $= 50$}
    \end{subfigure}%
    ~ 
    \begin{subfigure}[t]{0.45\textwidth}
        \centering
        \includegraphics[width=0.85\columnwidth,trim=0.5in 0.35in 0.5in 0.35in,clip]{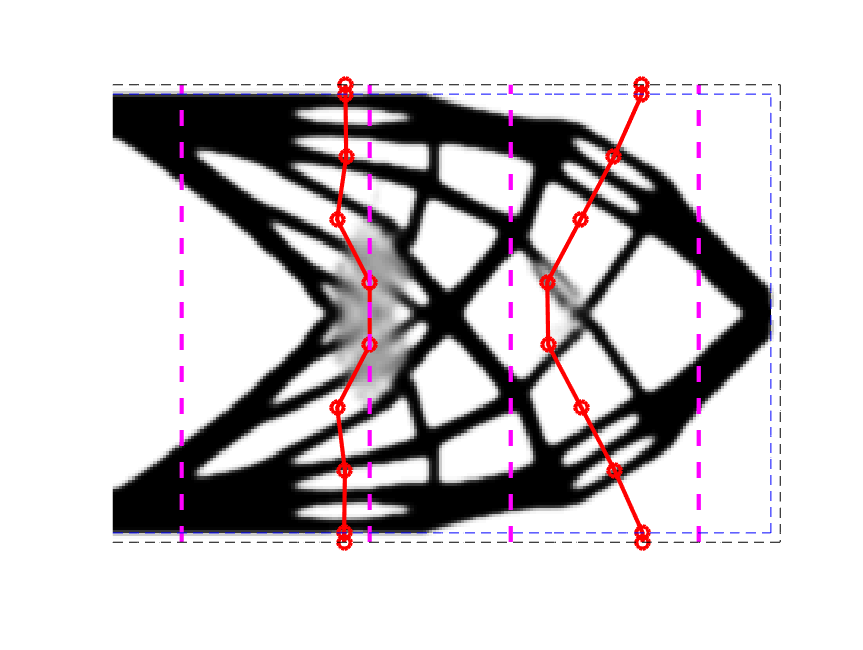}
        \caption{Iter $= 100$}
    \end{subfigure}
    \caption{Evolution of the topology during the optimization of the cantilever beam of Sec. \ref{sec:ex3-11}}
		\label{fig:evoluteex3}
\end{figure*}

Fig.~\ref{fig:evoluteex3} shows four intermediate design stages obtained in the initial $100$ optimization iterations.
It is possible to observe that initially the algorithm quickly identifies the topology outside of the projection areas.
Within the projection areas, the convergence towards a final near discrete design has been observed to be more challenging and slow due to the interplay between shape and density variables.

\subsubsection{Short cantilever with orthogonal projection profiles}
\label{sec:ex3-12}
\begin{figure*}[h]
    \centering
    \begin{subfigure}[t]{0.45\textwidth}
        \centering
        \includegraphics[width=0.85\columnwidth,trim=0.5in 0.35in 0.5in 0.35in,clip]{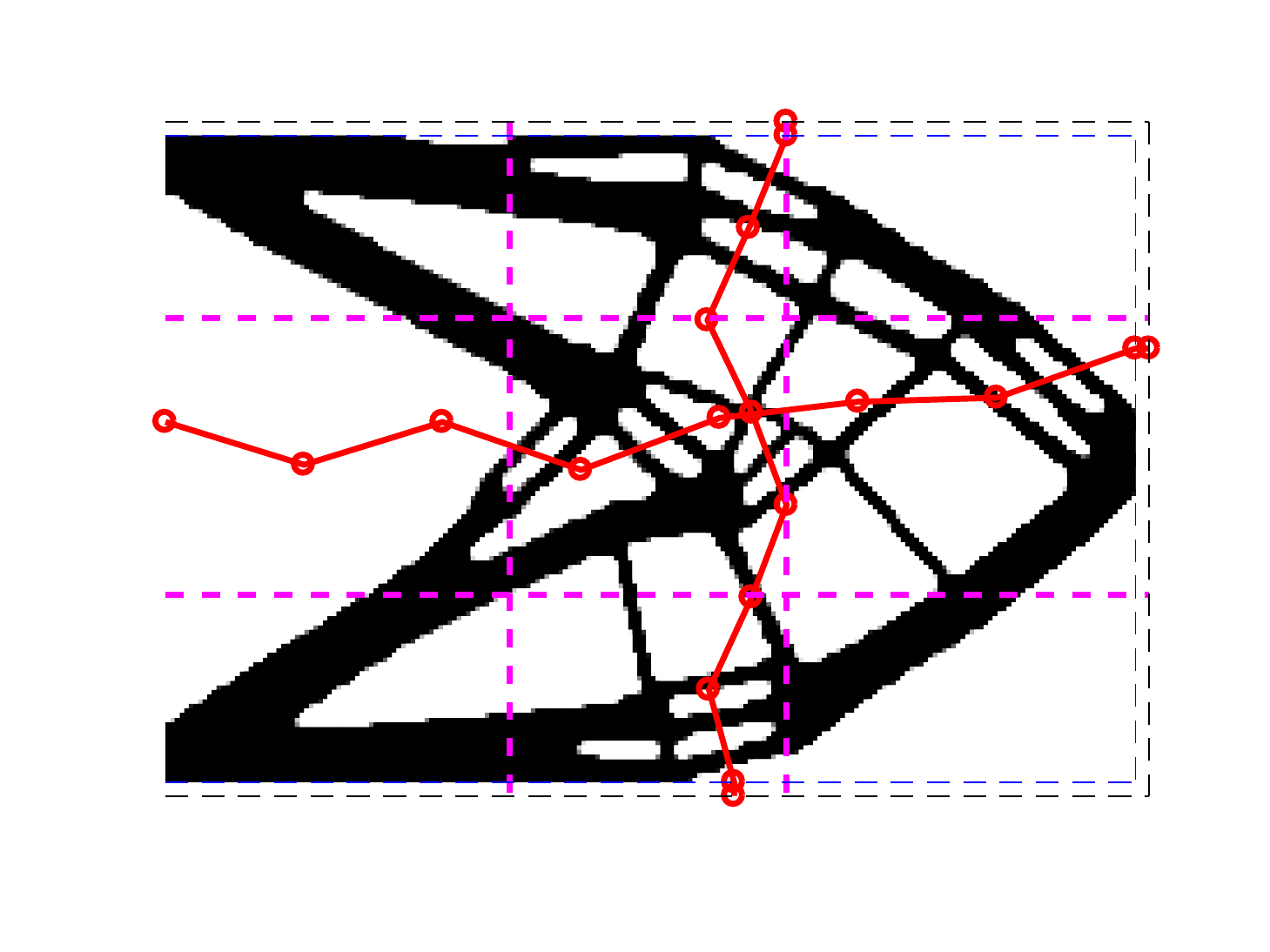}
        \label{subfig:cantiVHrho}
        \caption{}
    \end{subfigure}%
    ~ 
    \begin{subfigure}[t]{0.45\textwidth}
        \centering
        \includegraphics[width=0.85\columnwidth,trim=0.5in 0.35in 0.5in 0.35in,clip]{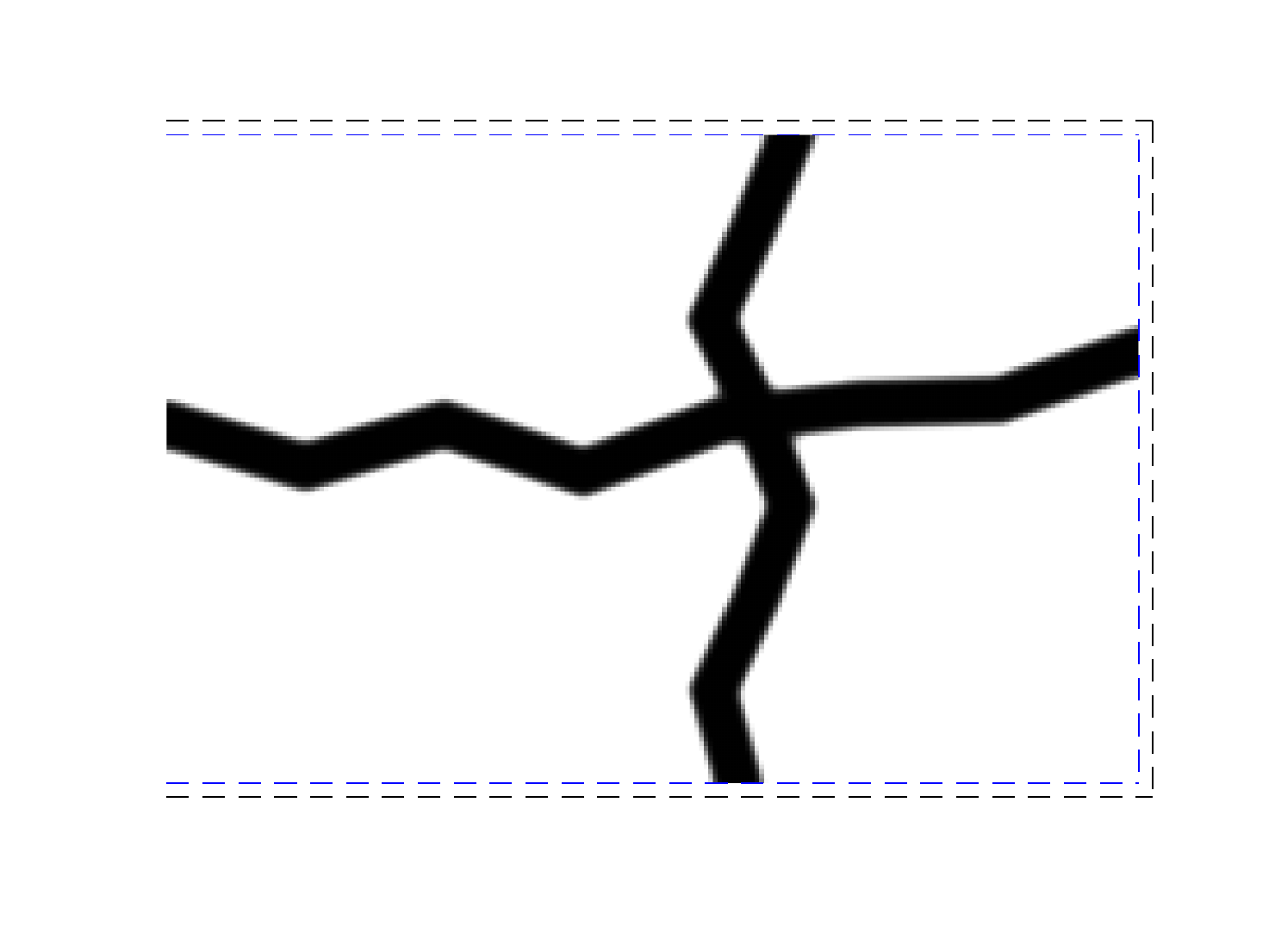}
        \label{subfig:cantiVHphi}
        \caption{}
    \end{subfigure}
    \caption{Short cantilever with two projection profiles, one vertical and one horizontal, of Sec.~\ref{sec:ex3-12}. (a) Optimized topology and configuration of the two profiles. (b) Final shape of the projection areas for which $\phi=1$. Final compliance $f=47.38$, and solid volume fraction $V=0.3502$. }
		\label{fig:orthogex3}
\end{figure*}

The short cantilever with orthogonal projection profiles has been optimized using the same settings for the parameters as in the previous example with two vertical profiles (i.e. Sec. \ref{sec:ex3-11}).
The vertical profile is made of $7$ segments spanning $20$ elements.
The horizontal profile is made of $7$ segments spanning $30$ elements.
The $8$ nodes at the vertical segments ends have fixed $y$ coordinates, and variable $x$ coordinates.
The $8$ nodes at the horizontal segments ends have fixed $x$ coordinates, and variable $y$ coordinates.
The variable geometric coordinates of the nodes of the two profiles $y^{1}_{i}$ and $x^{2}_{i}$ have been both initially set to $0.5$.
Their bounds are defined by gaps of $30$ elements with respect to the original positions such that $y^{1}_{lb}=0.29$ and $y^{1}_{ub}=0.71$, 
and similarly $x^{2}_{lb}=0.36$ and $x^{2}_{ub}=0.64$.
Fig.~\ref{fig:orthogex3} shows the final topology obtained after $550$ optimization iterations, and the corresponding shape of the optimized projection area.
The design solution obtained in this case is far from being symmetric. 
The reason could be that initially the horizontal profile is passing exactly through the point of application of the concentrated load.
As a consequence the maximum length scale control is imposed initially in the point of application of the load, which significantly weakens the structure.
And in fact, in the numerical experiments it has been observed that the optimizer moved the position of the profile away from the point of application of the load from the very beginning of the optimization analysis, thus converging towards a non symmetric final design.

In the two examples discussed in this section, it is possible to observe that the local volume constraint effectively leads towards final optimized designs with a maximum element length scale imposed on the projection areas. In fact, in these areas the optimized designs are characterized by thinner beam-like elements. 
This demonstrates the capability of the proposed procedure to control the maximum thickness of members in the interface of assemblies, as could be required by certain manufacturing considerations.


\subsection{Example 4: Variable minimum/maximum length scale control}
In this section we present three examples with different types of controls that impose variable minimum or maximum length scales.
More precisely, in the first example we perform the topology optimization of a short cantilever with a spatially variable minimum length scale. We adopt the density filter of Eq.~\eqref{eq:densityfilt_var} with weights defined according to Eq.~\eqref{eq:varfiltweight_gaus}. Through this filter formulation, a minimum length scale defined by the filter radius $\bar{r}_{min}$ is imposed in the domain inside the projection area (where $\phi=1$), and a minimum length scale defined by the radius $r_{min}$ otherwise. The ratio between $\bar{r}_{min}$ and $r_{min}$ is $5$.
The second example considers a spatially variable maximum length scale. For this purpose, we consider the constraint defined in Eq.~\eqref{eq:lengthscaleconst2}. This constraint is imposed on all the design domain. However, the radius used for imposing the maximum length scale varies, and it is doubled inside the projection area. In this way we allow for thicker features inside the projection area.
The third example shows a different approach for achieving a variable maximum length scale control. In particular, we consider the maximum length scale constraint of Eq.~\eqref{eq:lengthscaleconst2} with fixed radius $\hat{r}_{min}$ (i.e. $\gamma=0$). 
At the same time, we consider also a variable density filter for the minimum length scale given in Eq.~\eqref{eq:densityfilt_var} with filter weights defined in Eq.~\eqref{eq:varfiltweight_gaus} for $n=2$ and a filter radius doubled in the projection area. The density filter radius for the minimum length scale is kept smaller than that of the maximum length scale in order to be able to impose the desired feature size within the length scale allowed by the maximum length scale control.

\subsubsection{Variable minimum length scale}
\label{sec:ex4.1}
\begin{figure}[h]
\centering
  \includegraphics[width=0.6\columnwidth,trim=0.5in 1.1in 0.5in 1.1in,clip]{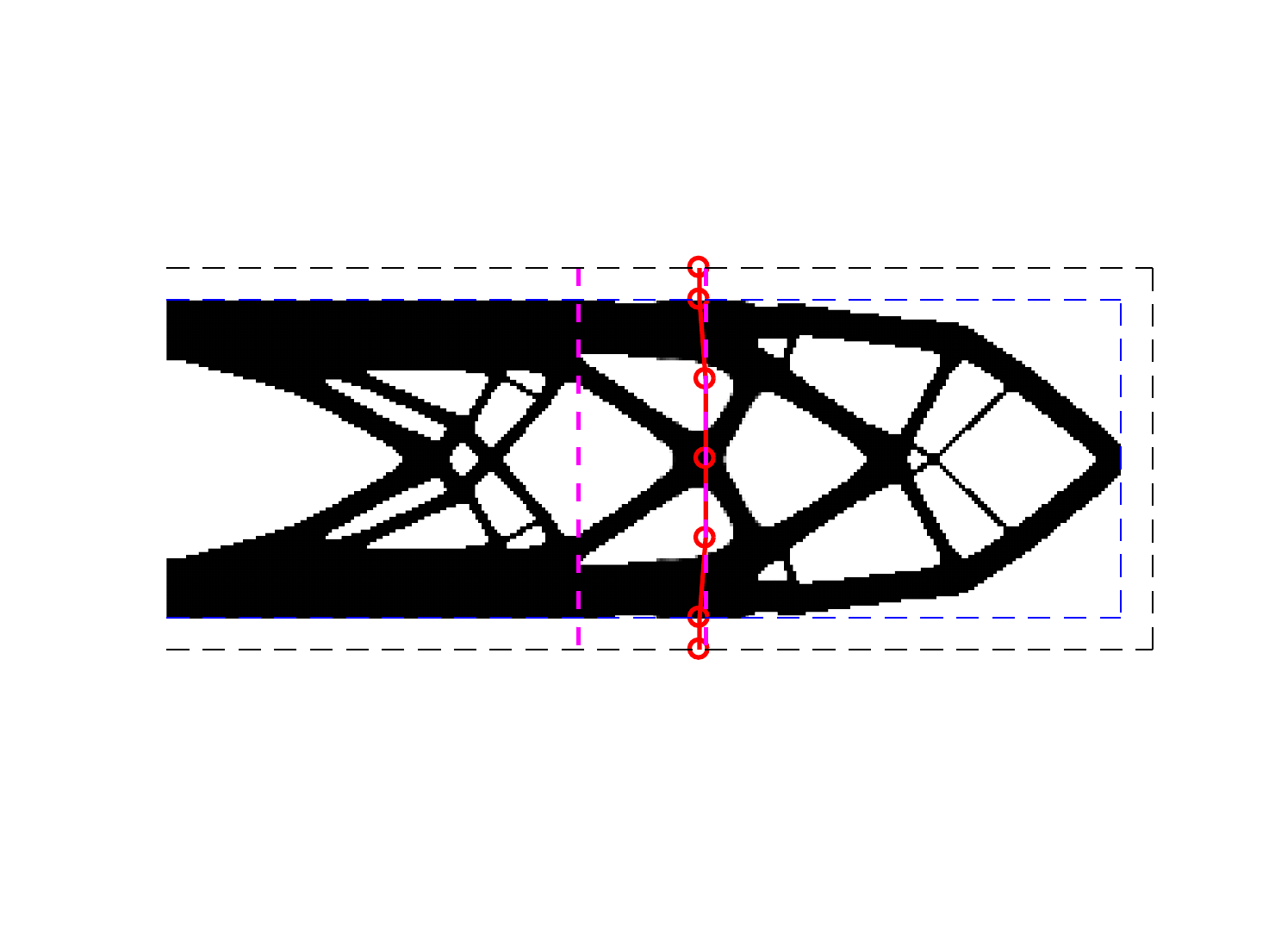}
\caption{Cantilever beam with variable minimum length scale control of Sec. \ref{sec:ex4.1}. Optimized topology and configuration of the projection profile. Final compliance $f=  170.44$, and solid volume fraction $V=0.4001$} 
\label{fig:ex4_1rho}
\end{figure}

\begin{figure}[h]
\centering
  \includegraphics[width=0.6\columnwidth,trim=0.5in 1.1in 0.5in 1.1in,clip]{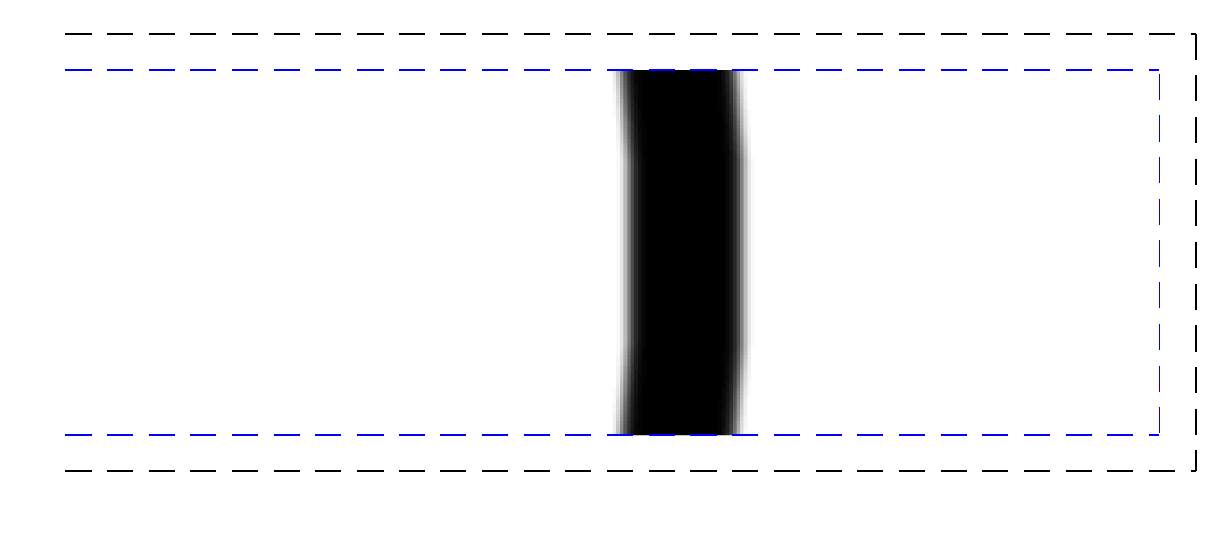}
\caption{Cantilever beam with variable minimum length scale control of Sec. \ref{sec:ex4.1}. Final shape of the projection area for which $\phi=1$} 
\label{fig:ex4_1phi}
\end{figure}

Here we consider a long cantilever beam as shown in Fig.~\ref{fig:canti_scheme} with a $H/L$ ratio of $1/3$.
We considered a longer geometry in the horizontal direction of the design domain to allow for a more distinct transition between the different minimum length scales imposed by the different filter radii considered during the optimization.
The structure is discretized with $100 \times 300$ finite elements with a density filter radius $r_{min}=2$ and a filter radius for the projection profile $r_{\phi}=2$.
For the minimum length scale filter, we adopted the filter weights defined in Eq.~\eqref{eq:finalwg} with a scaling factor of the filter radius in the projection area $\gamma =4$. As a consequence, for the $i$-th element included in the projection profile (i.e.~$\phi_{i}=1$) the radius of the minimum length scale is $\bar{r}_{min}=5\, r_{min}$.
We consider an allowable volume fraction for the total volume of the intermediate layout equal to $40\%$.
The parameters that define the continuation scheme are set as in the previous examples, with the exception of $\beta_{fil}=15$ elements, and $\beta_{HS}$ initialized to $1.5$ and increased by steps of $2^{0.75}$ up to $\beta_{HS,max}=100$.
In this example the final value $\beta_{HS,max}$ is significantly higher than that of the other examples because of the large minimum length scale radius $\bar{r}_{min}$ considered in the projection area.
A larger filter radius, in fact, requires a more sharp projection in Eq. \eqref{eq:rhoproject} in order to transform the filtered density filed into a near discrete topology.
The optimization process ran for $600$ iterations with steps of the continuation scheme of $50$ iterations. 
The projection profile is divided into four segments spanning $25$ elements each. 
Thus a total of five nodes with variable horizontal coordinate $x$ are also considered as deign variables.
The nodes' $x$ coordinates are initially set to $0.5$, and bounded with gaps of $20$ elements. 
That is, $x_{ub}=0.56$ and $x_{lb}=0.43$.
The final optimized topology and projection profile are shown in Fig.~\ref{fig:ex4_1rho}.
Fig.~\ref{fig:ex4_1phi} shows the final projection area in black, for which $\phi=1$.
In Fig.~\ref{fig:ex4_1rho} it is possible to observe a clear distinction of length scales between the areas of the domain inside and outside the projection area. Outside we observe thinner structural components in accordance with the regular filter radius, and inside only thick members and thick joints exist in accordance with the enlarged filter radius.
\subsubsection{Variable maximum length scale}
\label{sec:ex4.2}
Next, we consider a short cantilever beam as shown in Fig.~\ref{fig:canti_scheme} with a $H/L$ ratio of $1/2$.
The structure is discretized with $120 \times 240$ finite elements with a density filter radius $r_{min}=3$ elements and a filter radius for the projection profile set to $r_{\phi}=3$ elements.
The radius $r_{max}$ for the maximum length scale is set to $7$ elements outside the projection area. We consider an amplification factor $\gamma=1$, such that the $r_{max}$ is doubled inside the projection area.
The allowable volume fraction of the intermediate layout is initially set to $40\%$. The allowable volume fraction for the maximum length scale control (the parameter $\alpha$ in Eq.~\eqref{eq:lengthscaleconst2}) is set to $60\%$.
The projection area is defined in terms of distance from the profile and is set to $\beta_{fil}=20$ elements.
The exponent $p$ of the SIMP material interpolation is initialized to $1.25$ and increased to 1.5 at the first continuation scheme step. After, it is increased up to $3$ with steps of $0.5$.
The remaining parameter settings are listed in Tables \ref{tab:contscheme} and \ref{tab:optsettings}.
We consider a single vertical projection profile divided into six segments spanning $20$ elements each. 
Thus a total of seven nodes with variable horizontal coordinate $x$ are also considered as deign variables.
The nodes $x$ cooridnates are initially set to $0.5$, and bounded with gaps of $30$ elements. 
Hence, their upper bound is $x_{ub}=0.625$ and the lower bound is $x_{lb}=0.375$.
The optimization is run for $550$ iterations with steps of the continuation scheme of $50$ iterations.

\begin{figure}[h]
\centering
  \includegraphics[width=0.6\columnwidth,trim=0.5in 1.0in 0.5in 0.8in,clip]{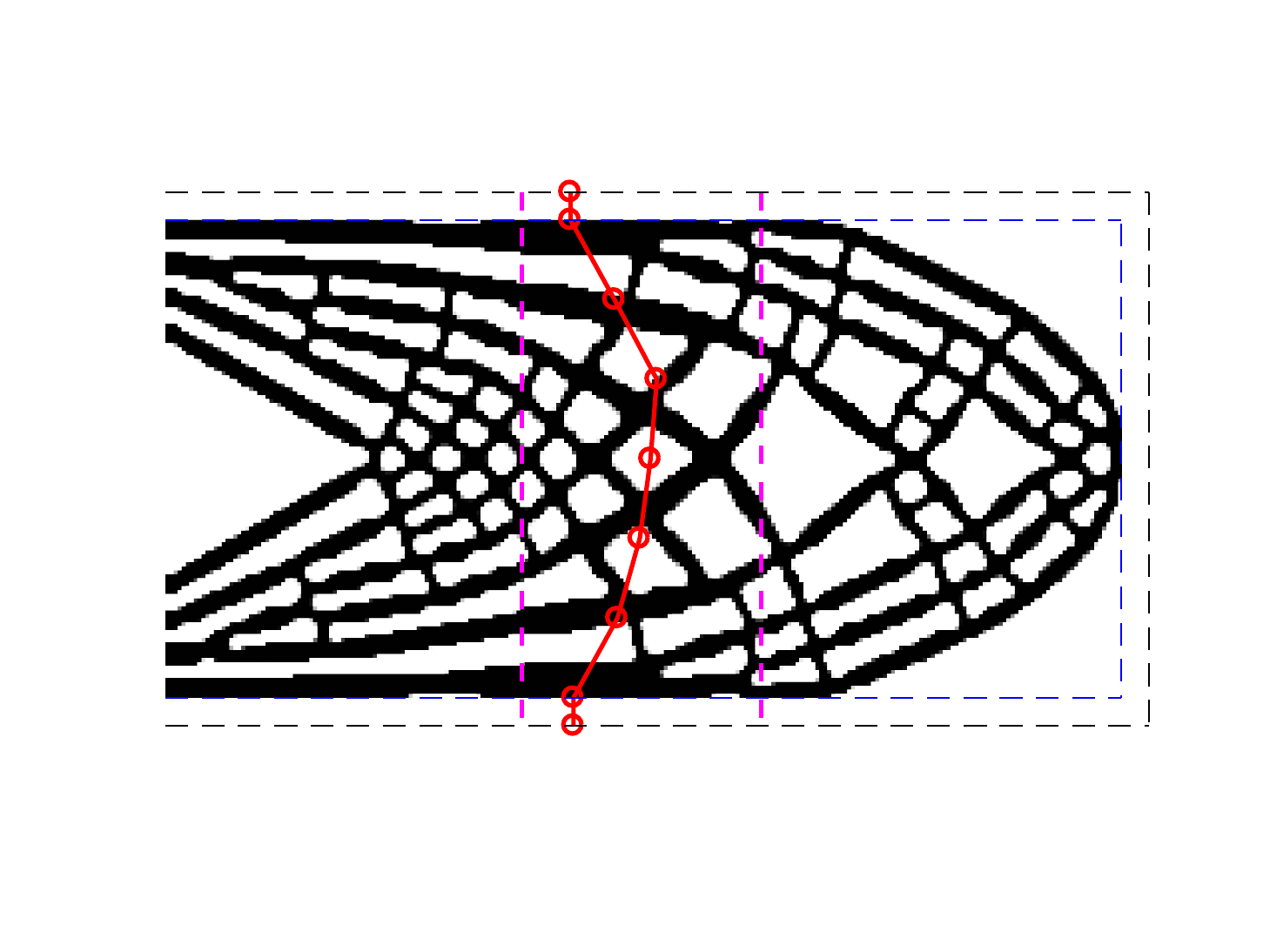}
\caption{Cantilever beam with variable maximum length scale control of Sec. \ref{sec:ex4.2}. Optimized topology and configuration of the projection profile. Final compliance $f=  87.56$, and solid volume fraction $V=0.3820$} 
\label{fig:ex4_2rho}
\end{figure}
\begin{figure}[h]
\centering
  \includegraphics[width=0.6\columnwidth,trim=0.5in 1.0in 0.5in 0.8in,clip]{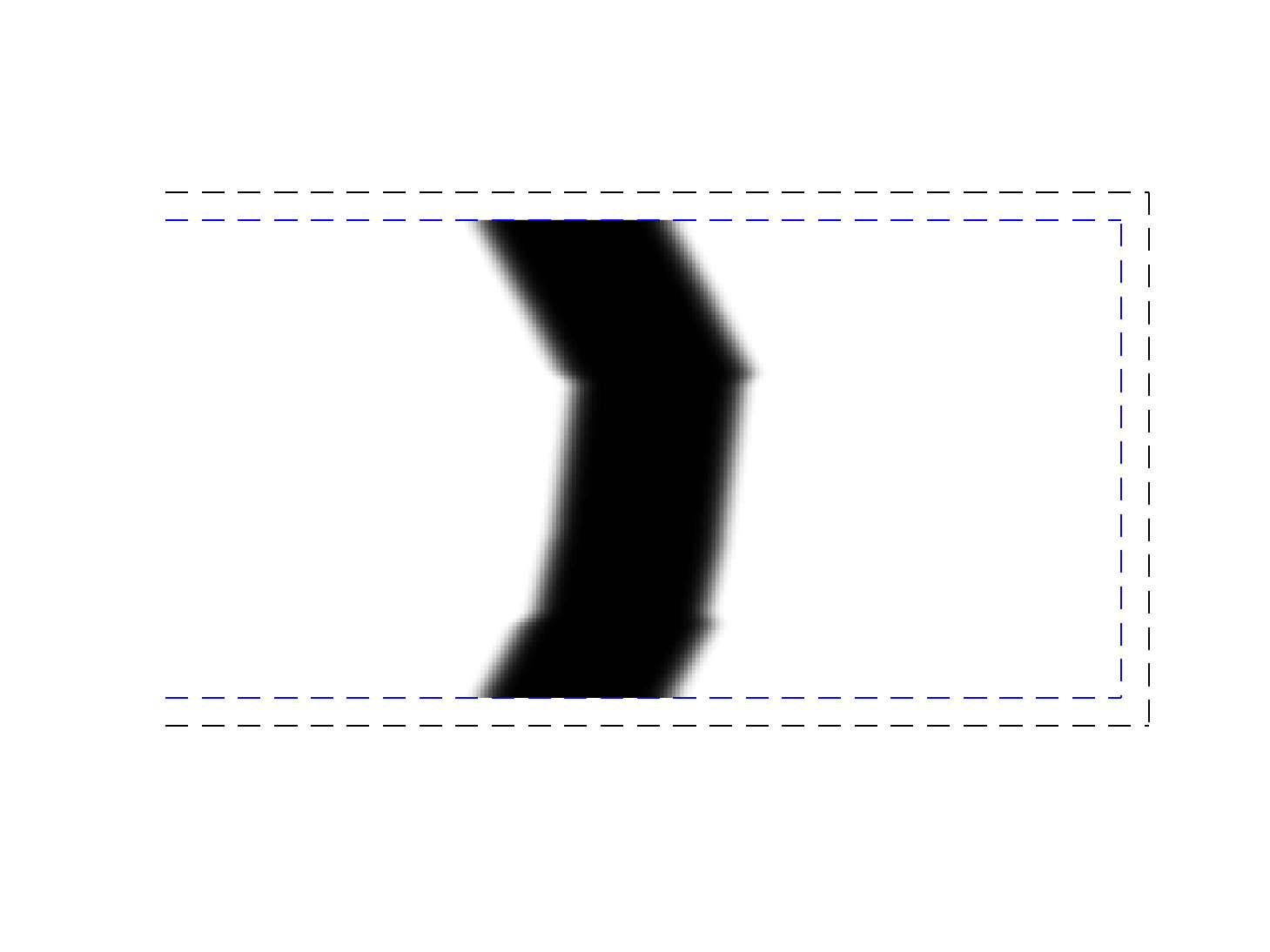}
\caption{Clamped beam of Sec. \ref{sec:ex4.2}. Final shape of the projection area  used to impose the variable maximum length scale control for which $\phi=1$} 
\label{fig:ex4_2phi}
\end{figure}

The compliance of the final optimized structure is $f=87.56$, and it is associated to a blueprint structural layout that occupies $38.2\%$ of the design domain.
The final optimized topology and projection profile are shown in Fig.~\ref{fig:ex4_2rho}. 
Fig.~\ref{fig:ex4_2phi} shows the final projection area in black, for which $\phi=1$.
In the final topology it is possible to observe a clear distinction between length scale outside and inside the projection area. Outside, the structure is characterized by thinner and more branched elements, resembling a porous medium.
Inside the projection area instead, the number of branching elements is reduced and the elements are thicker.
It should be mentioned that the problem formulation discussed in this section would probably benefit from even higher resolutions of the mesh. In fact the maximum length scale control imposed requires high resolutions in order to clearly identify the small features.
This of course would imply a higher computational cost and more time to perform the optimization. 
From an engineering design standpoint, this example shows how one can impose a maximum length scale throughout the design domain, while enlarging it in the assembly interface region in order to reduce the number of member connections.
\subsubsection{Maximum length scale control through a variable minimum length scale}
\label{sec:ex4.3}
\begin{figure}[h]
\centering
  \includegraphics[width=0.6\columnwidth,trim=0.5in 1.0in 0.5in 0.8in,clip]{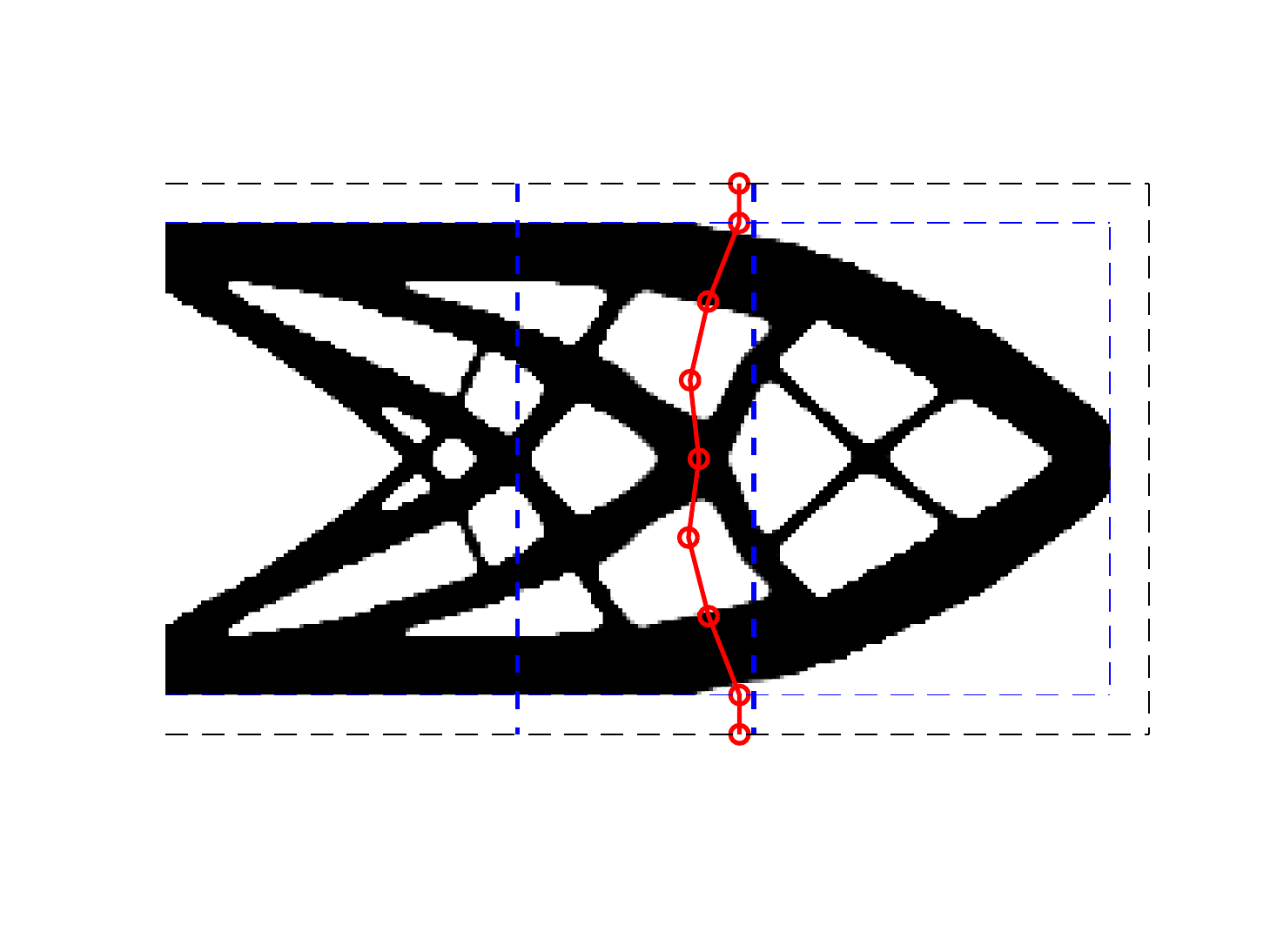}
\caption{Cantilever beam with maximum length scale and variable minimum length scale controls of Sec. \ref{sec:ex4.3}. Optimized topology and configuration of the projection profile. Final compliance $f=  64.93$, and solid volume fraction $V=0.4012$} 
\label{fig:ex4_3rho}
\end{figure}

Also in this example we consider the short cantilever beam shown in Fig.~\ref{fig:canti_scheme} which has a $H/L$ ratio of $1/2$.
The structure is discretized with $120 \times 240$ finite elements with a filter radius for the projection profile set to $r_{\phi}=3$ elements.
The radius $r_{max}$ for the maximum length scale is set to $20$ element.
The density filter radius is set to $r_{min}=3$ elements outside the projection area. We consider an amplification factor $\gamma=1$, such that $r_{min}$ is doubled inside the projection area.
The allowable volume fraction of the intermediate layout is initially set to $40\%$. The allowable volume fraction for the maximum length scale control (the parameter $\alpha$ in Eq.~\eqref{eq:lengthscaleconst2}) is set to $60\%$.
The remaining parameter settings are listed in Tables \ref{tab:contscheme} and \ref{tab:optsettings}.
We consider a single vertical projection profile divided into six segments spanning $20$ elements each. 
Thus a total of seven nodes with variable horizontal coordinate $x$ are also considered as design variables.
The nodes $x$ coordinates are initially set to $0.5$, and bounded with gaps of $30$ elements. 
That is, the upper bound is $x_{ub}=0.625$ and the lower bound is $x_{lb}=0.375$ and they are shown in Fig.~\ref{fig:ex4_3rho} with the magenta color.
The optimization is run for $550$ iterations with steps of the continuation scheme of $50$ iterations.
\begin{figure}[h]
\centering
  \includegraphics[width=0.6\columnwidth,trim=0.5in 1.0in 0.5in 0.8in,clip]{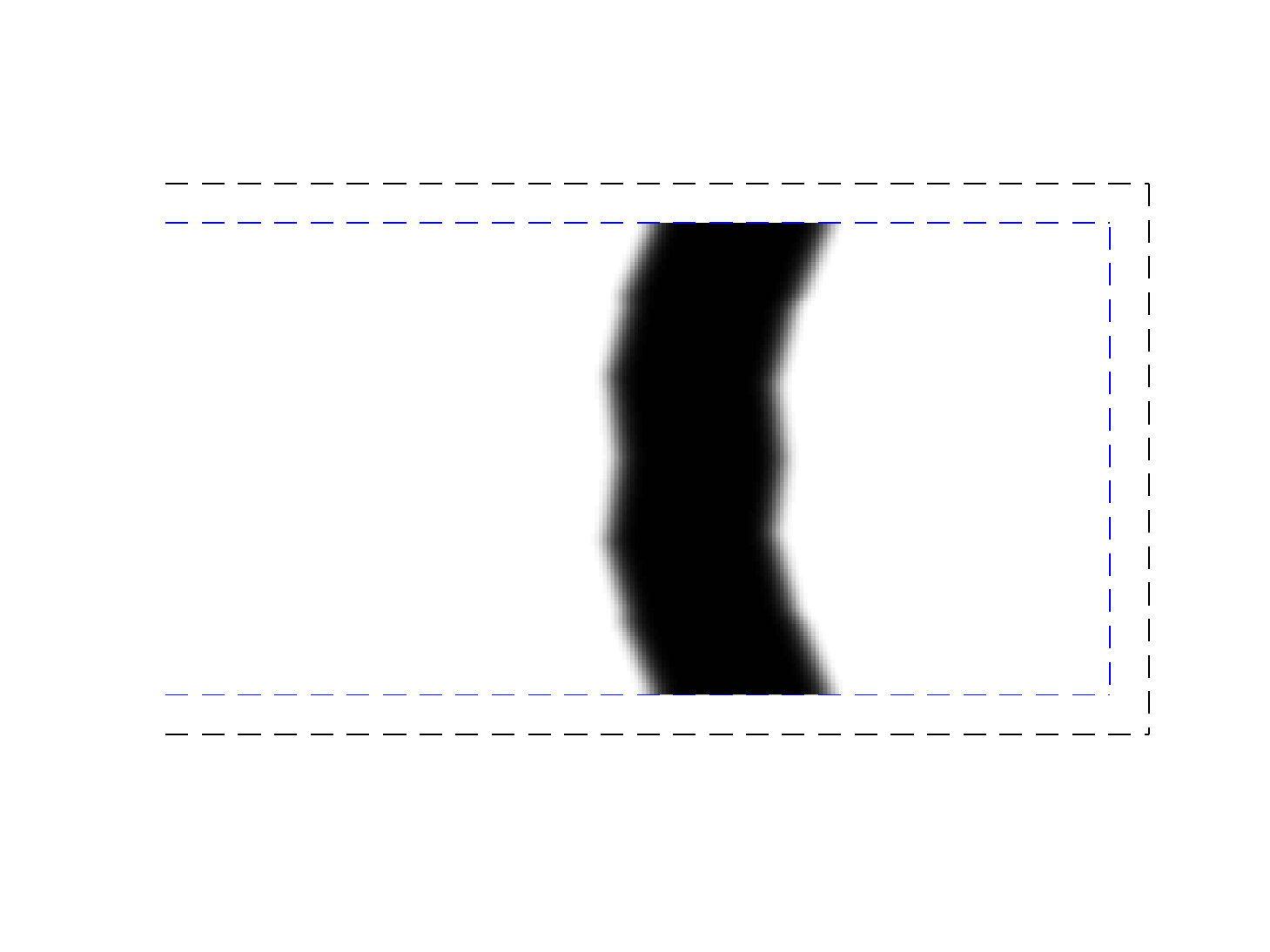}
\caption{Cantilever beam of Sec. \ref{sec:ex4.3}. Final shape of the projection area used to impose the maximum length scale through the variable minimum length scale control for which $\phi=1$} 
\label{fig:ex4_3phi}
\end{figure}
The compliance of the final optimized structure is $f=64.93$, and it is associated to a blueprint structural layout that occupies $40.12\%$ of the design domain.
The final optimized topology and projection profile are shown in Fig.~\ref{fig:ex4_3rho}. 
Also in this case, similarly to the results of Sec.~\ref{sec:ex4.1}, it is possible to observe an increase of the imposed length scale moving from the domain outside to the domain inside the projection area.
As expected, the results fulfill the expectation of elements with bigger size inside or in the vicinity of the projection area.
Fig. \ref{fig:ex4_3phi} shows the final projection area in black, for which $\phi=1$.
In a practical engineering prospective, this example, together with the examples of Sec. \ref{sec:ex4.2} and Sec. \ref{sec:ex4.2}, shows an additional way of imposing a maximum length scale throughout the design domain, which is enlarged in the assembly interface region in order to reduce the number of connections between structural elements.

\section{Conclusion}
\label{sec:end}
We presented a topology optimization approach that blends together projection- and density-based formulations.
The resulting formulation relies on an explicit geometric representation, parametrized by shape variables, that allows to impose selective controls over the design in specific regions, while using an implicit density representation elsewhere in the design domain.
The two representations are tied together via projection functions that couple the respective shape and density variables. 

The mixed formulation is rather general, for two main reasons: 1) Any type of explicit geometric entity can be utilized to define the domain where specific control is desired; and 2) Various geometric and response constraints can be imposed on the regions defined by the projected geometry.
In the current study, we are motivated by the optimal design of structural assemblies in 2-D, hence the geometric entities are piece-wise linear segments that define interfaces (or ``cuts'') between parts that are manufactured separately and subsequently assembled or joined together. 
Furthermore, the constraints represent several design considerations that may rise when optimizing an assembled part. 
Examples are: degraded material properties along the interface; total material volume along the interface; maximum member thickness in the interface region, while no limitation is imposed elsewhere; spatial variation of length scale, e.g.~imposing different minimum or maximum length scales near the interface and elsewhere. 
The examples in Sec. \ref{sec:examples} clearly demonstrate the capability to impose such restrictions in specific regions that are not predefined.
In other words, the overall topology and the shape of the interface between parts are found simultaneously, giving the optimization the freedom to choose where to impose restrictions while maintaining the full freedom of a density-based parametrization elsewhere.


\section*{Acknowledgements}
This work has been carried out as part of AATiD –- Advanced Additive Titanium Development Consortium. The authors wish to thank the Israeli Innovation Authority and the industrial partners for their generous financial support.

%


\bibliographystyle{spbasic}      


\end{document}